\documentclass[preprint,12pt]{elsarticle}

\usepackage{amssymb}
\usepackage{amsmath}
\usepackage{siunitx}
\usepackage[T1]{fontenc}
\usepackage{physics}
\usepackage{hyperref}
\hypersetup{pdftitle={ },
	pdfsubject={  },
	pdfauthor={J.M. Scherer},
	pdfkeywords={ }
	bookmarks=false,
	pdfpagemode=UseNone,
	linkcolor=blue,
	citecolor=blue,
	colorlinks=true,
	bookmarksopen=false,
	bookmarksopenlevel=1,
	urlcolor=blue}

\usepackage[lofdepth,lotdepth]{subfig}

\journal{ } 

\begin{document}

\begin{frontmatter}



\title{Deformation~\textcolor{black}{band} patterns and dislocation structures in finite strain crystal viscoplasticity} 


\author{Jean-Michel Scherer\corref{cor1}} 
\cortext[cor1]{Corresponding author.} 
\ead{jean-michel.scherer@minesparis.psl.eu} 

\address{Mines Paris, Universit{é} PSL, Centre des Mat{é}riaux (MAT), UMR7633 CNRS, \'Versailles, 78000, France}

\date{March 1$^{st}$, 2025} 

\begin{abstract}

Deformation band patterning in single crystals is investigated using a finite strain crystal viscoplasticity model based on the evolution of dislocation densities. In the presence of strong latent hardening and weak rate dependence, the deformation organizes into laminate microstructures consisting of single-slip regions separated by dislocation walls. The influence of material and numerical parameters on the nucleation and morphology of these patterns is analyzed in 2D single crystals under plane strain compression. Pattern formation is also observed in 3D single-crystal cylinders subjected to tension, where the characteristic size of deformation microstructures is found to depend on mesh size and boundary conditions in the absence of an intrinsic material length scale. To address this limitation, strain gradient plasticity is introduced, providing a length scale that governs the size of the patterns. Finally, we demonstrate that deformation patterns and dislocation structures also emerge in 2D and 3D polycrystals, highlighting the generality of the phenomenon.


\end{abstract}


\begin{highlights}
	\item A finite strain crystal viscoplasticity model based on dislocation density evolution is presented and extended to incorporate strain gradient effects.
	\item Deformation patterns naturally emerge under conditions of strong latent hardening and minimal rate dependence.
	\item Lamination patterns lead to the formation of dislocation structures, including sub-grain boundaries.
	\item Strain gradient plasticity introduces a material length scale that governs the characteristic size of deformation microstructures.  
	\item In both 2D and 3D polycrystals, deformation patterns and dislocation structures develop when the spatial resolution is sufficiently fine.
\end{highlights}

\begin{keyword}
	crystal plasticity \sep
	finite strain viscoplasticity \sep
	deformation patterning \sep
	dislocation structures \sep
	latent hardening \sep
	lamination \sep
	strain gradient plasticity \sep
	finite element method


\end{keyword}

\end{frontmatter}



\section{Introduction}
\label{sec:introduction}

Crystal plasticity provides a continuum mechanics framework for describing the collective behavior of dislocations and twinning at both the single and polycrystal scales. Numerous models have been developed in the literature to capture various aspects of crystal plasticity, including the kinematics of plastic slip~\citep{mandel1982definition} and twinning~\citep{van1978simulation,kalidindi1998incorporation}, strain hardening~\citep{meric1991single,kocks2003physics,madec2003role}, temperature and strain-rate dependence~\citep{lim2015physically,breumier2020high}, and size effects~\citep{gurtin2000plasticity,han2005mechanism,cordero2012grain} in single crystals.

A common characteristic of these models is the additive decomposition of the plastic velocity gradient into the sum of contributions from each active slip (or twin) system. Due to the symmetry of crystal lattices (e.g., FCC, BCC, HCP), the number of independent slip systems often exceeds the number of independent components in the plastic velocity gradient tensor. Consequently, determining the set of active slip systems and their associated plastic slip intensities can become an ill-posed problem. The issue of non-uniqueness in the selection of active slip systems and plastic slip intensities was first highlighted by \textcolor{black}{Taylor~\cite{taylor1938plastic}. Later Mandel~\cite{mandel1965generalisation,mandel1966contribution} extended the theory of multiple plastic potentials introduced by Koiter~\cite{koiter1960general} and investigated the problem of non-uniqueness of the set of active slip systems and of their slip intensities}. Mathematically, this problem arises in regions of the stress space where the yield surfaces of multiple slip systems intersect, specifically at the edges and corners of the global yield surface. In such cases, the flow direction may not be uniquely defined, as the normal to the yield surface forms a subdifferential set rather than a single unique normal.

Two main approaches have been adopted in the literature to address this issue. The first, and most commonly used due to its simplicity, is the viscoplastic formulation of the flow rule~\citep{peirce1983material,meric1991single,busso2005selection,eisenlohr2013spectral}. This approach introduces a rate dependence in the plastic slip rates, effectively smoothing the edges and corners of the yield surface and thereby regularizing the flow direction. The second approach employs a rate-independent formulation of the flow rule with additional constraints to uniquely determine the active slip systems and compute the associated plastic slip rates. Several studies have proposed different constraints to resolve this ambiguity. For instance, Franciosi~\citep{franciosi1991crystal} formulated a non-convex potential function whose global minimum selects the active slip systems and determines the slip intensities. Alternatively, Forest and Rubin~\cite{forest2016rate} introduced a rate-independent overstress model that replaces the standard plastic multiplier with the norm of the total rate of distortional deformation.

Algorithmic methods have also been developed to tackle this ill-posed problem of rate-independent crystal plasticity. Strategies based on generalized inverses were proposed to determine the active slip systems~\citep{anand1996computational,schroder1997aspects}, while Schmidt-Baldassari~\citep{schmidt2003numerical} introduced an algorithm leveraging Lagrange multipliers and penalization techniques. Other approaches developed in~\citep{fohrmeister2019classic,akpama2016numerical} rely on the nonlinear complementarity problem (NCP) introduced by Fischer and Burmeister~\citep{fischer1992special}. More recently, Petryk and Kursa~\cite{petryk2015incremental,rys2024spontaneous} employed a trust region method for incremental work minimization, and in~\cite{perdahciouglu2024rate,scheunemann2020novel,scheunemann2024comparison} interior point methods were utilized to frame the problem as a non-convex minimization task. A review of these and other techniques can be found in~\citep{manik2014review,pruger2020comparative}. Despite their robustness, these techniques are less widely used than the viscoplastic approach due to the complexity of their numerical implementation. More importantly, the additional constraints introduced in these methods often lack a clear physical justification. In this study, we adopt the widely used viscoplastic approach, tuning the viscosity parameters to achieve a quasi-rate-independent response, thereby preserving nearly sharp edges and corners in the yield surface.

\begin{figure}
	\centering
	\subfloat[]{
		\includegraphics[height=.24\textheight]{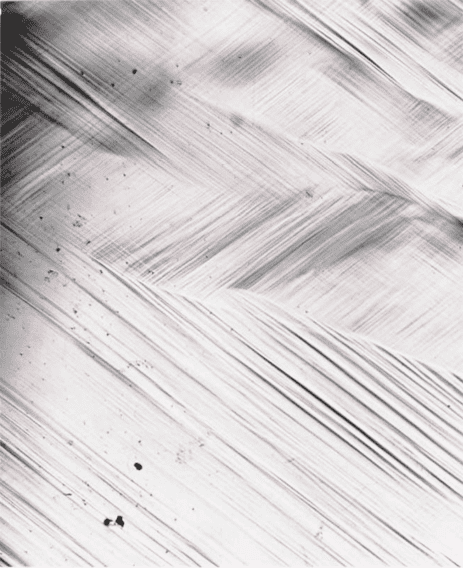}
		\includegraphics[height=.24\textheight]{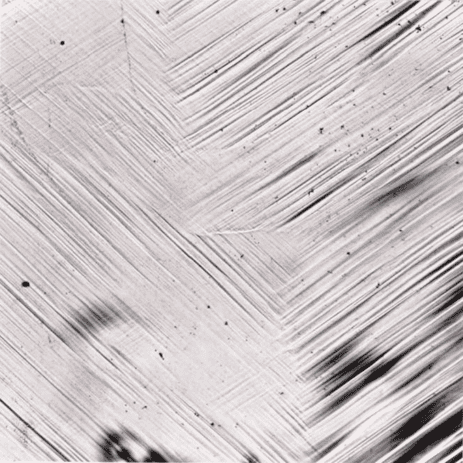}
		\label{subfig:patterns_experimental_saimoto}
	}
	\subfloat[]{	
		\includegraphics[height=.24\textheight, trim=0cm 1.cm 0cm 0cm, clip]{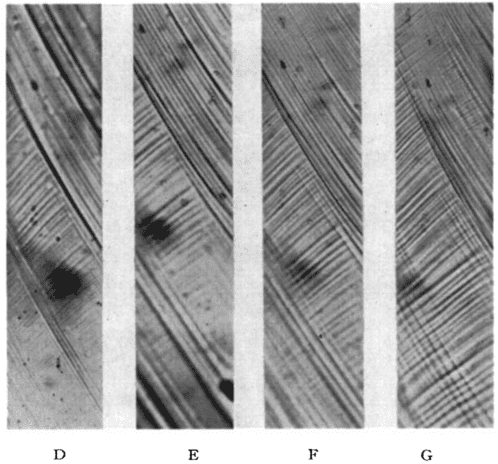}
		\label{subfig:patterns_experimental_piercy}
	}
	\caption{
		\textcolor{black}{Slip traces in (a) single crystal copper strained at \SI{78}{\kelvin} to 15\% along $[011]$ (vertical) viewed on $(0\bar{1}1)$ surface (x100 magnification)~\citep{saimoto1964} and (b) single crystal alpha-brass strained to 50\% along an unkown crystal direction (vertical)~\citep{piercy1955study}.}
	}
	\label{fig:patterns_experimental}
\end{figure}

The emergence of heterogeneous deformation fields comprising single-slip patches and deformation bands has been observed in both rate-independent~\citep{rys2024spontaneous} and rate-dependent~\citep{dequiedt2015heterogeneous,wang2018role,dequiedt2018incidence,dequiedt2023slip,rys2024spontaneous} single-crystal plasticity simulations. These patterns develop when latent hardening—the hardening effect induced by one slip system on others—is sufficiently strong. Mathematically, a strong latent hardening response leads to a non-convex energy landscape~\citep{ortiz1999nonconvex}, with energy wells corresponding to single-slip modes associated with elastic lattice rotations. In this scenario, the single crystal spontaneously partitions into patches undergoing distinct single-slip modes, forming a laminate structure~\citep{kocks1960polyslip}. Such patterns, frequently observed in crystal plasticity simulations with strong latent hardening, are also commonly reported in experimental studies. The concept of "patchy slip"~\citep{asaro1983micromechanics} was introduced to describe the surface appearance of single-crystal specimens, where slip lines organize into patches. Experimental observations of such patterns in single-crystal copper~\citep{saimoto1964,jin1984cyclic,piercy1955study} (see Figure~\ref{subfig:patterns_experimental_saimoto}), alpha-brass~\citep{piercy1955study} (see Figure~\ref{subfig:patterns_experimental_piercy}), and aluminum~\citep{wert2003deformation} reveal the tendency of single crystals to "break up"~\citep{saimoto1964} into regions with different active slip systems. As discussed, this segregation of plastic deformation into single-slip regions can be attributed to strong latent hardening. This phenomenon, that can be described as \textit{multiple-slip shyness}\footnote{The term "multiple-slip shyness" is proposed in analogy to "crown shyness," a phenomenon observed in certain tree species where the foliar crowns of individual trees do not touch, forming a network of channel-like gaps in the canopy~\citep{van2021understanding}.}, is characterized by sharp deformation boundaries separating single-slip regions. Across such boundaries, discontinuous lattice rotations occur due to the distinct active slip systems on either side. Ortiz and Repetto~\cite{ortiz1999nonconvex} linked the occurrence of these deformation patterns to the formation of dislocation structures such as cells, veins, labyrinth, mosaic, fence and carpet structures. Dislocations gliding within neighboring single-slip patches accumulate at their interfaces, forming strong junctions that, over time, lead to the development of dislocation walls.

\textcolor{black}{
Recent advances in dislocation mechanics have also shown that the emergence of heterogeneous dislocations microstructures does not necessarily require strong latent hardening. Notably, phenomenological and mesoscale field dislocation mechanics (PMFDM and MFDM) frameworks were developped to incorporate dislocation transport and explicit evolution of dislocation density fields. Roy \& Acharya~\citep{roy2006size} demonstrated the spontaneous emergence of dislocation patterns via linearized weak hyperbolicity in PMFDM. Later, Arora \& Acharya~\citep{arora2020dislocation} showed that dislocation patterning in MFDM under finite deformations can arise from the coupling between dislocation transport and strength evolution, particularly when the corresponding coupling parameter $k_0 \neq 0$. Their simulations also indicated that boundary conditions enforcing zero plastic flow can induce inhomogeneous dislocation distributions early in plastic deformation~\citep{arora2020dislocation,arora2022mechanics}. These studies suggest that multiple mechanisms—including those independent of latent hardening—can drive dislocation patterning.}

In this work, we investigate the formation of deformation patterns in single and polycrystals using a finite strain crystal viscoplasticity model based on dislocation density evolution. We provide demonstration of complex dislocation structures emerging in crystal plasticity simulations, revealing their dependence on the interaction matrix coefficients. In the conventional crystal plasticity model, which lacks an intrinsic material length scale, the characteristic size of the patterns is governed by the mesh size and boundary conditions. To address this limitation, we employ a strain gradient plasticity extension, enabling the regularization of dislocation structure size. We show that the dislocation cell size $\phi$ follows a power law with respect to the material length scale $\ell$. Furthermore, we explore deformation patterns and dislocation structures in both 2D and 3D single and polycrystals.

This article is organized as follows. Section~\ref{sec:model} presents the finite strain crystal viscoplasticity model based on dislocation density evolution and its strain gradient plasticity extension. Section~\ref{sec:plane_strain_compression} investigates the emergence of deformation patterns and dislocation structures in 2D single crystals under plane strain compression and simple shear. In Section~\ref{sec:single_crystal_3D}, the study is extended to 3D single-crystal cylinders subjected to tensile loading. The regularization of the characteristic size of dislocation structures using strain gradient plasticity is discussed in Section~\ref{sec:strain_gradient}. Section~\ref{sec:polycrystal} examines the occurrence of deformation patterns in 2D and 3D polycrystals. Finally, conclusions are drawn in Section~\ref{sec:conclusion}.



\iftrue

\section{Finite strain crystal plasticity formulation}
\label{sec:model}

\subsection{Finite strain kinematics}
\label{subsec:kinematics}

The deformation gradient $\mathbf{F}$ is assumed to undergo a multiplicative decomposition into an elastic part $\mathbf{F}^e$ and a plastic part $\mathbf{F}^p$
\begin{equation}
	\mathbf{F} = \pdv{\mathbf{x}}{\mathbf{X}} = \mathbf{F}^e \cdot \mathbf{F}^p
\end{equation}
The total velocity gradient $\mathbf{L}$, and the elastic and plastic velocity gradients $\mathbf{L}^e$ and $\mathbf{L}^p$ are given by
\begin{align}
	\mathbf{L} &= \dot{\mathbf{F}}\cdot \mathbf{F}^{-1} = \mathbf{L}^e + \mathbf{F}^e\cdot\mathbf{L}^p\cdot\mathbf{F}^{e^{-1}} \\
	\mathrm{with}\quad \mathbf{L}^e &= \dot{\mathbf{F}}^e \cdot \mathbf{F}^{e^{-1}} \quad \mathrm{and} \quad \mathbf{L}^p = \dot{\mathbf{F}}^p\cdot \mathbf{F}^{p^{-1}} = \sum_{s = 1}^N \dot{\gamma}^s \mathbf{m}^s \otimes \mathbf{n}^s
\end{align}
Here, $\mathbf{m}^s$ and $\mathbf{n}^s$ represent the slip direction and slip plane normal of the $s^{th}$ slip system, respectively. The material has $N$ slip systems, and the plastic slip rate on the $s^{th}$ slip system is denoted as $\dot{\gamma}^s$, which can be either positive or negative. The slip system definition considers $(\mathbf{m}^s, \mathbf{n}^s)$ and $(- \mathbf{m}^s, \mathbf{n}^s)$ as equivalent, with the direction of plastic slip determined by the sign of $\dot{\gamma}^s$.

The second Piola-Kirchhoff stress tensor, $\pmb{\Pi}^e$, is defined as
\begin{align}
    \pmb{\Pi}^e = \mathbb{C} : \mathbf{E}^e \quad
    \mathrm{with}\quad \mathbf{E}^e = \frac{1}{2} \left ( \mathbf{F}^{e^T} \cdot \mathbf{F}^{e} - \mathbf{I} \right )
\end{align}
where $\mathbb{C}$ is the fourth-order elasticity tensor. The Mandel stress tensor, $\pmb{\Pi}^M$, is then given by
\begin{equation}
    \pmb{\Pi}^M = \mathbf{F}^{e^T}\cdot \mathbf{F}^e\cdot \pmb{\Pi}^e 
\end{equation}

\subsection{Dislocation density based crystal plasticity}
\label{subsec:dislocation_density}

The yield function for each slip system is defined as
\begin{align}
	f^s &= |\tau^s| - \tau_c^s \quad \text{where} \quad \tau^s = \pmb{\Pi}^M : (\mathbf{m}^s \otimes \mathbf{n}^s) \label{eq:yield_function}
\end{align}
Here, $\tau_c^s$ represents the critical resolved shear stress (CRSS) for the $s^{th}$ slip system. A dislocation density-based hardening model~\citep{franciosi1980latent} is used
\begin{equation}
    \tau_c^s = \tau_0^s + \mu b \sqrt{\sum_{s=1}^N a^{su}\rho^u}
	\label{eq:franciosi}
\end{equation}
where $\tau_0^s$ is the thermal component of the CRSS, $b$ is the Burgers vector magnitude, and $a^{su}$ characterizes the interaction strength between dislocations on slip systems $s$ and $u$. These coefficients capture hardening effects of dislocation junctions and dipoles~\citep{madec2003role, devincre2006physical}. For this study, we consider an FCC lattice with $\frac{a}{2} \langle 110 \rangle \{111\}$ slip systems. Based on the work of Madec and Kubin~\cite{madec2017dislocation}, we account for seven interaction types: self-interaction, coplanar interaction, Hirth lock, Lomer junction, collinear interaction, and \SI{0}{\degree} and \SI{60}{\degree} glissile junctions. The distinction between \SI{0}{\degree} and \SI{60}{\degree} junctions results in a non-symmetric interaction matrix (Table~\ref{tab:interaction_coefficients}). Since the finite element implementation differs from standard ordering, we provide a mapping between Schmid-Boas and simulation numbering conventions. For two active slip systems with slip rates $\gamma_s$ and $\gamma_u$, the interaction coefficient $a^{su}$ is obtained from Table~\ref{tab:interaction_coefficients}.

\begin{table}
    \renewcommand{\arraystretch}{1.5}
    \centering
    \caption{Interaction coefficients $a^{su}$ for different types of dislocation interactions. The $s$ and $u$ superscript of $a^{su}$ refer to the subscript of $\gamma_s$ and $\gamma_u$ in the last column and row of the table. Schmid and Boas slip system nomenclature for FCC crystals is used. Slip planes: A $= (\bar{1}11)$, B $= (111)$, C $= (\bar{1}\bar{1}1)$, D $= (1\bar{1}1)$. Slip directions: 1 $= [011]$, 2 $= [0\bar{1}1]$, 3 =$ [101]$, 4 $= [\bar{1}01]$, 5 $= [\bar{1}10]$, 6 $= [110]$.}
    \label{tab:interaction_coefficients}
    \fontsize{7}{9}\selectfont 
    \begin{tabular}{|c||cccccccccccc||c|}
        \hline
        & A2 & A3 & A6 & B2 & B4 & B5 & C1 & C3 & C5 & D1 & D4 & D6 & $\gamma_s$ \\
        \hline \hline
        A2 & $a_{self}$ & $a_{cop}$ & $a_{cop}$ & $a_{col}$ & $a_{gli}^{0\si{\degree}}$ & $a_{gli}^{0\si{\degree}}$ & $a_{Hirth}$ & $a_{gli}^{60\si{\degree}}$ & $a_{Lomer}$ & $a_{Hirth}$ & $a_{Lomer}$ & $a_{gli}^{60\si{\degree}}$ & $\gamma_7$ \\
        A3 & $a_{cop}$ & $a_{self}$ & $a_{cop}$ & $a_{gli}^{60\si{\degree}}$ & $a_{Hirth}$ & $a_{Lomer}$ & $a_{gli}^{0\si{\degree}}$ & $a_{col}$ & $a_{gli}^{0\si{\degree}}$ & $a_{Lomer}$ & $a_{Hirth}$ & $a_{gli}^{60\si{\degree}}$ & $\gamma_9$ \\
        A6 & $a_{cop}$ & $a_{cop}$ & $a_{self}$ & $a_{gli}^{60\si{\degree}}$ & $a_{Lomer}$ & $a_{Hirth}$ & $a_{Lomer}$ & $a_{gli}^{60\si{\degree}}$ & $a_{Hirth}$ & $a_{gli}^{0\si{\degree}}$ & $a_{gli}^{0\si{\degree}}$ & $a_{col}$ & $\gamma_8$ \\
        B2 & $a_{col}$ & $a_{gli}^{0\si{\degree}}$ & $a_{gli}^{0\si{\degree}}$ & $a_{self}$ & $a_{cop}$ & $a_{cop}$ & $a_{Hirth}$ & $a_{Lomer}$ & $a_{gli}^{60\si{\degree}}$ & $a_{Hirth}$ & $a_{gli}^{60\si{\degree}}$ & $a_{Lomer}$ & $\gamma_2$ \\
        B4 & $a_{gli}^{60\si{\degree}}$ & $a_{Hirth}$ & $a_{Lomer}$ & $a_{cop}$ & $a_{self}$ & $a_{cop}$ & $a_{Lomer}$ & $a_{Hirth}$ & $a_{gli}^{60\si{\degree}}$ & $a_{gli}^{0\si{\degree}}$ & $a_{col}$ & $a_{gli}^{0\si{\degree}}$ & $\gamma_1$ \\
        B5 & $a_{gli}^{60\si{\degree}}$ & $a_{Lomer}$ & $a_{Hirth}$ & $a_{cop}$ & $a_{cop}$ & $a_{self}$ & $a_{gli}^{0\si{\degree}}$ & $a_{gli}^{0\si{\degree}}$ & $a_{col}$ & $a_{Lomer}$ & $a_{gli}^{60\si{\degree}}$ & $a_{Hirth}$ & $\gamma_3$ \\
        C1 & $a_{Hirth}$ & $a_{gli}^{60\si{\degree}}$ & $a_{Lomer}$ & $a_{Hirth}$ & $a_{Lomer}$ & $a_{gli}^{60\si{\degree}}$ & $a_{self}$ & $a_{cop}$ & $a_{cop}$ & $a_{col}$ & $a_{gli}^{0\si{\degree}}$ & $a_{gli}^{0\si{\degree}}$ & $\gamma_{12}$ \\
        C3 & $a_{gli}^{0\si{\degree}}$ & $a_{col}$ & $a_{gli}^{0\si{\degree}}$ & $a_{Lomer}$ & $a_{Hirth}$ & $a_{gli}^{60\si{\degree}}$ & $a_{cop}$ & $a_{self}$ & $a_{cop}$ & $a_{gli}^{60\si{\degree}}$ & $a_{Hirth}$ & $a_{Lomer}$ & $\gamma_{11}$ \\
        C5 & $a_{Lomer}$ & $a_{gli}^{60\si{\degree}}$ & $a_{Hirth}$ & $a_{gli}^{0\si{\degree}}$ & $a_{gli}^{0\si{\degree}}$ & $a_{col}$ & $a_{cop}$ & $a_{cop}$ & $a_{self}$ & $a_{gli}^{60\si{\degree}}$ & $a_{Lomer}$ & $a_{Hirth}$ & $\gamma_{10}$ \\
        D1 & $a_{Hirth}$ & $a_{Lomer}$ & $a_{gli}^{60\si{\degree}}$ & $a_{Hirth}$ & $a_{gli}^{60\si{\degree}}$ & $a_{Lomer}$ & $a_{col}$ & $a_{gli}^{0\si{\degree}}$ & $a_{gli}^{0\si{\degree}}$ & $a_{self}$ & $a_{cop}$ & $a_{cop}$ & $\gamma_5$ \\
        D4 & $a_{Lomer}$ & $a_{Hirth}$ & $a_{gli}^{60\si{\degree}}$ & $a_{gli}^{0\si{\degree}}$ & $a_{col}$ & $a_{gli}^{0\si{\degree}}$ & $a_{gli}^{60\si{\degree}}$ & $a_{Hirth}$ & $a_{Lomer}$ & $a_{cop}$ & $a_{self}$ & $a_{cop}$ & $\gamma_4$ \\
        D6 & $a_{gli}^{0\si{\degree}}$ & $a_{gli}^{0\si{\degree}}$ & $a_{col}$ & $a_{Lomer}$ & $a_{gli}^{60\si{\degree}}$ & $a_{Hirth}$ & $a_{gli}^{60\si{\degree}}$ & $a_{Lomer}$ & $a_{Hirth}$ & $a_{cop}$ & $a_{cop}$ & $a_{self}$ & $\gamma_6$ \\
        \hline \hline
        $\gamma_u$ & $\gamma_7$ & $\gamma_9$ & $\gamma_8$ & $\gamma_2$ & $\gamma_1$ & $\gamma_3$ & $\gamma_{12}$ & $\gamma_{11}$ & $\gamma_{10}$ & $\gamma_5$ & $\gamma_4$ & $\gamma_6$ & \\
        \hline
    \end{tabular}
\end{table}

A rate-dependent flow rule is used
\begin{align}
	\dot{\gamma}^s = \mathrm{sign}(\tau^s)\left \langle \frac{f^s}{K} \right \rangle^n \label{eq:norton}
\end{align}
\textcolor{black}{where the Macaulay brackets are defined as $\langle \bullet \rangle = \max(\bullet, 0)$, and $n$ and $K$ are viscosity-related parameters governing the rate sensitivity. A threshold-type viscoplastic flow rule is employed here, reflecting the existence of a finite stress required to initiate slip. However, the widely used threshold-free formulation could alternatively be adopted with qualitatively similar results, as discussed in~\cite{busso2005selection}.}
The evolution of dislocation densities follows
\begin{align}
    \dot{\rho}^s = \frac{|\dot{\gamma}^s|}{b} \left( \frac{\sqrt{\sum_{u\neq s} \rho^u}}{\kappa} - yb\rho^s \right)
\end{align}
\textcolor{black}{Details of the finite element numerical implementation of the dislocation density based crystal plasticity model can be found in~\cite{scherer2022implementation}.}

\subsection{Material parameters}

Unless otherwise specified, the parameters are given in Table~\ref{tab:material_parameters}. Initially, the dislocation densities are uniformly distributed, $\rho_0^s = \rho_0^{tot} / 12$. Values of $\tau_0^s$, $\mu$, $b$, $\rho_0^{tot}$, and $a^{ij}$ are selected such that, according to Eq.~\eqref{eq:franciosi}, $\tau_c^s = \SI{10}{\mega\pascal}$ initially.

The viscosity parameters $n$ and $K$ influence the deformation patterns. We choose values ensuring a quasi-rate-independent response with nearly sharp yield surface edges and corners. Large $n$ ($n \gg 1$) and small $K$ ($K \ll \tau_c^s$) lead to an extremely stiff system of equation~\citep{wulfinghoff2013equivalent}. To mitigate ill-condtionning and be able to integrate the constitutive equations, \textcolor{black}{we use small strain increments of 0.001\% for increments of time $\Delta t = \SI{0.01}{\second}$}. 
While alternative stabilization methods exist~\citep{wulfinghoff2013equivalent,rys2024spontaneous}, we opt for this approach for simplicity.
\begin{table}
    \renewcommand{\arraystretch}{1.5}
    \centering
    \caption{Material parameters for the dislocation density based crystal plasticity model.}
    \label{tab:material_parameters}
    \small
    \begin{tabular}{cccccccccc}
        \hline
        $C_{11}$ (\si{\giga\pascal}) & $C_{12}$ (\si{\giga\pascal}) & $C_{44}$ (\si{\giga\pascal}) & $\tau_0^s$ (\si{\mega\pascal}) & $\mu$ (\si{\giga\pascal}) & $b$ (\si{\angstrom}) \\ \hline \hline
        $218$ & $144$ & $125$ & $0$ & $100$ & $2.54$ \\
        \hline \\ \hline
        $K$ (\si{\mega\pascal}.\si{\second}$^{1/n}$) & $n$ (-) &  $\kappa$ (-) & $y$ (-) & $\rho_0^{tot}$ (\si{\meter}$^{-2}$) & $a_{self}$ (-) \\ \hline \hline 
        $0.1$ & $50$ & $20$ & $10$ & $10^{12}$ & $0.1$  \\ 
        \hline \\ \hline
        $a_{cop}$ (-) & $a_{Hirth}$ (-) & $a_{Lomer}$ (-) & $a_{col}$ (-) & $a_{gli}^{0\si{\degree}}$ (-) & $a_{gli}^{60\si{\degree}}$ (-) \\ \hline \hline
        $0.16$ & $0.16$ & $0.16$ & $0.16$ & $0.16$ & $0.16$ \\
        \hline
    \end{tabular}
\end{table}

\subsection{Strain gradient extension}
\label{subsec:strain_gradient}

In the following sections, we will demonstrate that emerging deformation patterns are sensitive to mesh size and the type of finite elements employed. To regularize this mesh-dependency, a characteristic material length scale must be incorporated into the model. Following the work developed in~\cite{scherer2020lagrange,phalke2022modeling,scherer2019strain}, we utilize a Lagrange multiplier relaxation of a scalar strain gradient plasticity model. An additional degree of freedom, $\gamma_\chi$, which describes the accumulated plastic slip, is introduced. The extended principle of virtual power, in the absence of body forces and under static conditions, is formulated as follows
\begin{align}
	\forall D_0 \subset \Omega,\, \forall \dot{\pmb{u}},\, \forall \dot{\gamma}_\chi\quad
	\int_{D_0} \left(
		\pmb{S} : \dot{\pmb{F}}
		+ S \dot{\gamma}_\chi
		+ \pmb{M}\cdot \nabla_X \dot{\gamma}_\chi
	\right)\, \mathrm{d}V_0
	= \int_{\partial D_0} \left(
		\pmb{T} \cdot \pmb{\dot{u}}
		+ M \dot{\gamma}_\chi
	\right)\, \mathrm{d}S_0
	\label{eq:virtual_power}
\end{align}
where $\pmb{S}$ represents the Boussinesq (or nominal 1$^{st}$ Piola-Kirchhoff) stress tensor, related to the Cauchy stress tensor $\pmb{\sigma}$ by $\pmb{S} = \det(\pmb{F})\, \pmb{\sigma}\cdot \pmb{F}^{-T}$. Higher-order scalar and vector stresses, $S$ and $\pmb{M}$, are introduced as power conjugates to the additional degree of freedom $\gamma_\chi$ and its gradient $\nabla_X \gamma_\chi$. The notation $\nabla_X$ denotes the gradient operator with respect to the reference configuration. The conventional traction vector is denoted by $\pmb{T}$, while $M$ represents a higher-order traction scalar. The strong form and boundary conditions associated with Eq.~\eqref{eq:virtual_power} are given by
\begin{alignat}{2}
    &\forall \pmb{X} \in \Omega &\quad \nabla_X \cdot \pmb{S} &= \pmb{0} \\
    &\forall \pmb{X} \in \partial \Omega &\quad \pmb{T} &= \pmb{S} \cdot \pmb{n}_0 \\
    &\forall \pmb{X} \in \Omega &\quad \nabla_X \cdot \pmb{M} - S &= 0 \label{eq:equilibrium_M_S} \\
    &\forall \pmb{X} \in \partial \Omega &\quad M &= \pmb{M} \cdot \pmb{n}_0
\end{alignat}
where $\pmb{n}_0$ is the outward unit normal to the surface $S_0$.

The model developed in~\cite{scherer2020lagrange} links the additional degree of freedom $\gamma_\chi$ to the accumulated plastic slip $\gamma_{cum} = \int_{t} \sum_{s=1}^N |\dot{\gamma}^s|\, \mathrm{d}t$ using a Lagrange multiplier $\lambda$. The postulated form of the free energy density is:
\begin{align}
	\psi(\pmb{E}^e, \gamma_{cum}, r^s, \gamma_\chi, \nabla_X \gamma_\chi, \lambda) &= \frac{1}{2} \pmb{E}^e:\mathbb{C}:\pmb{E}^e
	+ \frac{A}{2} \nabla_X \gamma_\chi \cdot \nabla_X \gamma_\chi
	+ \lambda \left( \gamma_\chi - \gamma_{cum} \right) \nonumber \\
	&+ \frac{\mu_\chi}{2} \left( \gamma_\chi - \gamma_{cum} \right)^2
	+ \psi_h (r^s, \gamma_{cum})
	\label{eq:free_energy_density}
\end{align}
where $A$ and $\mu_\chi$ are higher-order material parameters, and $\psi_h$ represents the hardening potential. The coefficient $A$ has the dimension of force (\si{\newton=\pascal\cdot\meter\squared}) and thus introduces a characteristic length into the formulation. \textcolor{black}{This gradient regularization based on the accumulated plastic slip follows the approach originally proposed by Wulfinghoff and Böhlke~\citep{wulfinghoff2012equivalent}. While this formulation is not directly tied to a physically measurable quantity—unlike models based on Nye’s tensor~\citep{acharya2000lattice,cermelli2001characterization,cordero2012grain}—it offers a significant computational advantage by introducing only two additional variables ($\gamma_\chi$ and a Lagrange multiplier $\lambda$). This simplification retains the essential regularizing effect needed to mitigate mesh dependence, while enabling the simulation of complex microstructural patterns at the crystal scale.}

From the free energy potential Eq.~\eqref{eq:free_energy_density} and the application of the first and second principles of thermodynamics, the following state laws were postulated in~\cite{scherer2020lagrange}
\begin{align}
	\pmb{\Pi}^e &= \pdv{\psi}{\pmb{E}^e} = \mathbb{C} : \pmb{E}^e \label{eq:state_elasticity}\\
	S &= \pdv{\psi}{\gamma_\chi} = \lambda + \mu_\chi (\gamma_\chi - \gamma_{cum}) \label{eq:state_S}\\
	\pmb{M} &= \pdv{\psi}{\nabla_X \gamma_\chi} = A \nabla_X \gamma_\chi \label{eq:state_M}
\end{align}
The Lagrange multiplier stress $\lambda$ weakly enforces the constraint $\gamma_\chi = \gamma_{cum}$, while $\mu_\chi$ can be interpreted as a penalty parameter. Combining the equilibrium equation~\eqref{eq:equilibrium_M_S} with the state laws Eqs.~\eqref{eq:state_S} and \eqref{eq:state_M}, we obtain the following second-order differential equation for $\gamma_\chi$:
\begin{align}
	0 = \nabla_X \cdot \pmb{M} - S &= \nabla_X \cdot \left( A \nabla_X \gamma_\chi \right) - (\lambda + \mu_\chi (\gamma_\chi - \gamma_{cum}) ) \\
	&= A\Delta_X \gamma_\chi - (\lambda + \mu_\chi (\gamma_\chi - \gamma_{cum}) ) \label{eq:equilibrium_gamma_chi}
\end{align}
Finally, the expression of the residual mechanical dissipation leads to the following definition of extended yield functions for the relaxed strain gradient crystal plasticity model~\cite{scherer2020lagrange}
\begin{align}
	f^s &= |\tau^s| - (\tau_c^s - S ) \nonumber \\
		&= |\tau^s| - (\tau_c^s - \lambda - \mu_\chi (\gamma_\chi - \gamma_{cum})) \label{eq:yield_function_extended}
\end{align}
Eq.~\eqref{eq:yield_function_extended} replaces Eq.~\eqref{eq:yield_function} in the rate-dependent flow rule Eq.~\eqref{eq:norton}.

\section{Deformation patterns under plane strain loadings}
\label{sec:plane_strain_compression}

\textcolor{black}{
In this section, we study the emergence of deformation patterns and dislocation structures in a single crystal under plane strain loadings. The strain gradient crystal plasticity extension of the model presented in Section~\ref{subsec:strain_gradient} is not used in this section, but will be analyzed later under the same loading conditions in Section~\ref{sec:strain_gradient}. 
}


\subsection{Plane strain compression}
\label{subsec:plane_strain_compression}

Following Ry{\'s} \textit{et al.}~\cite{rys2024spontaneous}, we analyze the plane strain compression of an FCC single crystal oriented along the $[00\bar{1}]$ direction within the $(110)$ plane. The finite element mesh for the square specimen, with dimensions $L \times L \times h$, \textcolor{black}{is such that $x_1 = [1\bar{1}0]$, $x_2 = [00\bar{1}]$, and $x_3 = [110]$}. The mesh consists of $200 \times 200 \times 1$ quadratic hexahedral elements of size $h$. Unless otherwise specified, fully integrated elements (C3D20) are utilized. The bottom boundary of the specimen is constrained in the vertical direction while remaining free to move horizontally. To prevent rigid body motion, a single node on the bottom boundary is fixed in the horizontal direction. Plane strain conditions are enforced by restricting the normal displacement of the front and back faces to zero. The top boundary is subjected to a uniform downward displacement, inducing an applied strain rate of $\dot{u}_{2}(x_2=L)/L = -10^{-3}\, \si{\second}^{-1}$. Unless otherwise stated, the specimen undergoes compression up to a macroscopic strain of $1\%$, which is achieved at $t_f = \SI{10}{\second}$. As previously noted, small time increments of $\Delta t = \SI{0.01}{\second}$ are employed to integrate the constitutive equations, which are inherently ill-conditioned due to the chosen viscosity parameters.

As discussed in~\cite{rys2024spontaneous}, the slip systems exhibiting an equal maximum Schmid factor of $\frac{1}{\sqrt{6}}$, consistent with plane strain compression along $[00\bar{1}]$ in the $(110)$ plane, are
\begin{itemize}
	\item D4: $[\bar{1}01](1\bar{1}1)$ $\longrightarrow$ $\gamma_4$
	\item D1: $[011](1\bar{1}1)$ $\longrightarrow$ $\gamma_5$
	\item A2: $[0\bar{1}1](\bar{1}11)$ $\longrightarrow$ $\gamma_7$
	\item A3: $[101](\bar{1}11)$ $\longrightarrow$ $\gamma_9$
\end{itemize}
The interactions between dislocations within these slip systems, as detailed in Table~\ref{tab:interaction_coefficients}, include:
\begin{itemize}
	\item Self-interaction: D4 -- D4, D1 -- D1, A2 -- A2, A3 -- A3 $\longrightarrow$ $a_{self}=a^{44}=a^{55}=a^{77}=a^{99}$
	\item Coplanar interaction: D4 -- D1, A2 -- A3 $\longrightarrow$ $a_{cop}=a^{45}=a^{54}=a^{79}=a^{97}$
	\item Hirth lock: D4 -- A3, D1 -- A2 $\longrightarrow$ $a_{Hirth}=a^{49}=a^{94}=a^{57}=a^{75}$
	\item Lomer junction: D1 -- A3, D4 -- A2 $\longrightarrow$ $a_{Lomer}=a^{59}=a^{95}=a^{47}=a^{74}$
\end{itemize}

\subsubsection*{Sensitivity to material parameters} 
We first examine the influence of material parameters and crystal orientation relative to the loading direction. Figure~\ref{fig:plane_strain_compression_010} presents the plastic slips on slip systems D4, D1, A2, and A3 at $1\%$ macroscopic strain, using the material parameters specified in Table~\ref{tab:material_parameters}. In Figure~\ref{subfig:plane_strain_compression_010_grid}, the slip distributions $\gamma_4$ (D4), $\gamma_5$ (D1), $\gamma_7$ (A2), and $\gamma_9$ (A3) are plotted separately. Meanwhile, Figure~\ref{subfig:plane_strain_compression_010_superimposed} displays the maximum slip magnitude at each Gauss point, where the color corresponds to the most active slip system according to the colorscale in Figure~\ref{subfig:plane_strain_compression_010_grid}. A distinct deformation microstructure emerges, characterized by four domains, each dominated by a single slip system. Within each domain, the plastic slip is organized into smaller parallelogram-shaped regions, featuring two short, nearly horizontal edges and two inclined edges. \textcolor{black}{The pattern of deformation bands observed in Figure~\ref{fig:plane_strain_compression_010} is reminiscent of the herring-bone patterns reported in pure aluminium, copper and nickel single crystals after plane strain compression loading along the $[00\bar{1}]$ direction~\cite{basson2000deformation}. 
}

Following the framework proposed by Ortiz and Repetto~\cite{ortiz1999nonconvex}, the pattern in Figure~\ref{fig:plane_strain_compression_010} corresponds to a rank-two laminate microstructure. The first-order lamination decomposes the single crystal into successive horizontal layers normal to $[00\bar{1}]$. while the second-order lamination further subdivides these layers into inclined bands, forming the parallelogram-shaped regions where single-slip deformation occurs. The boundary normals between single-slip regions are $[00\bar{1}]$ (D1 -- A3 and D4 -- A2), and at $+54.74\si{\degree}$ (D4 -- D1) and $-54.74\si{\degree}$ (A2 -- A3) with respect to $[1\bar{1}0]$. The projections of the slip directions onto the $(110)$ plane for the four active slip systems are shown in Figure~\ref{subfig:plane_strain_compression_010_superimposed}. Notably, the slip directions are oriented perpendicular to the longer edges of the parallelograms. This observation indicates that the deformation microstructure is characterized by a pattern of short kink bands.
\begin{figure}
	\centering
	\subfloat[]{
		\hspace{-1.3cm}
		\includegraphics[width=0.55\textwidth]{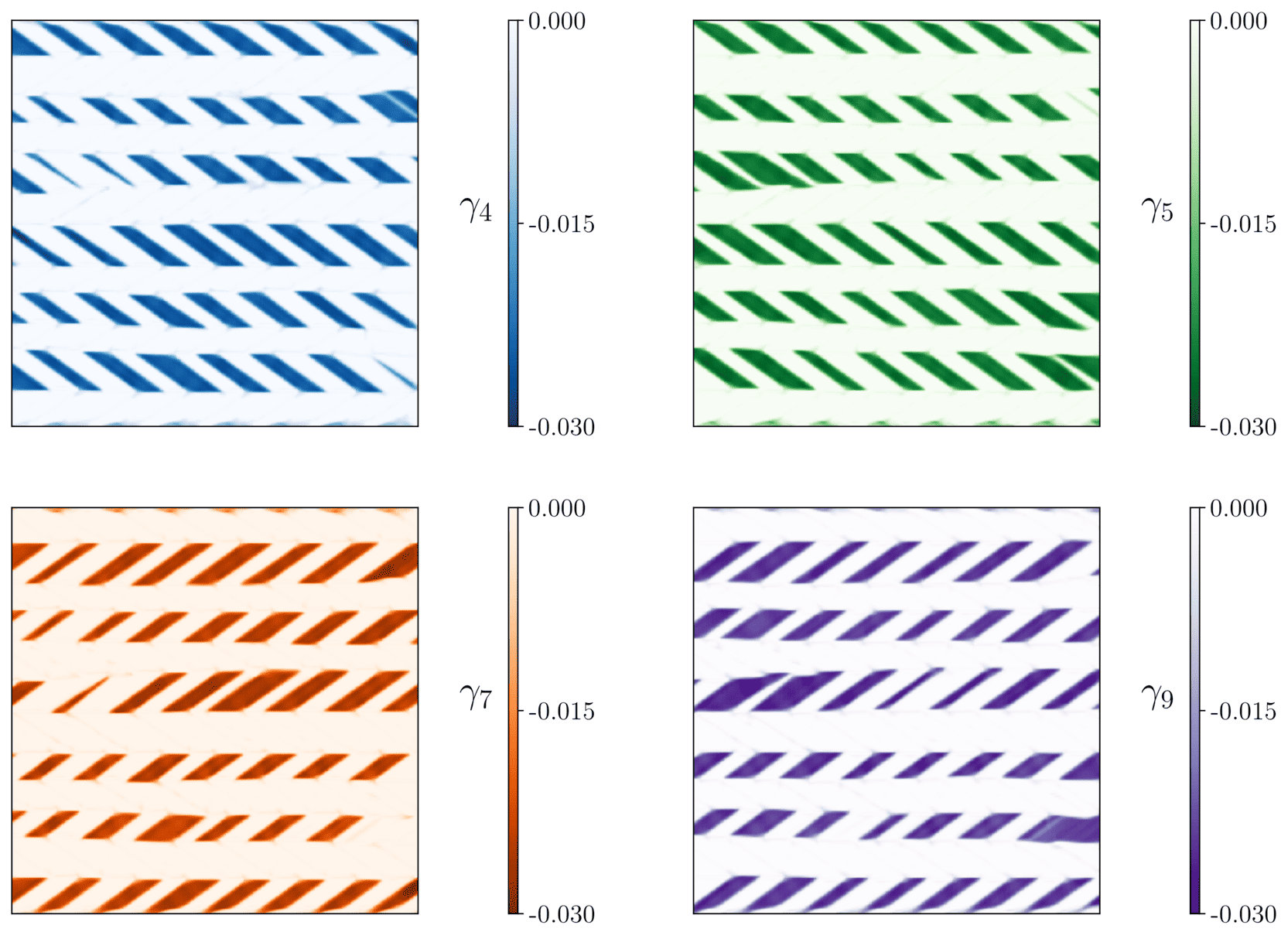}
		\label{subfig:plane_strain_compression_010_grid}
	}
	\subfloat[]{
        \raisebox{0cm}{\includegraphics[width=0.13\textwidth]{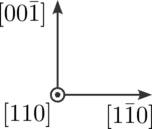}}\hspace{-1.3cm}
        \raisebox{.65cm}{\includegraphics[width=0.3\textwidth, trim=0.115cm 0.15cm 8.5cm 0cm, clip]{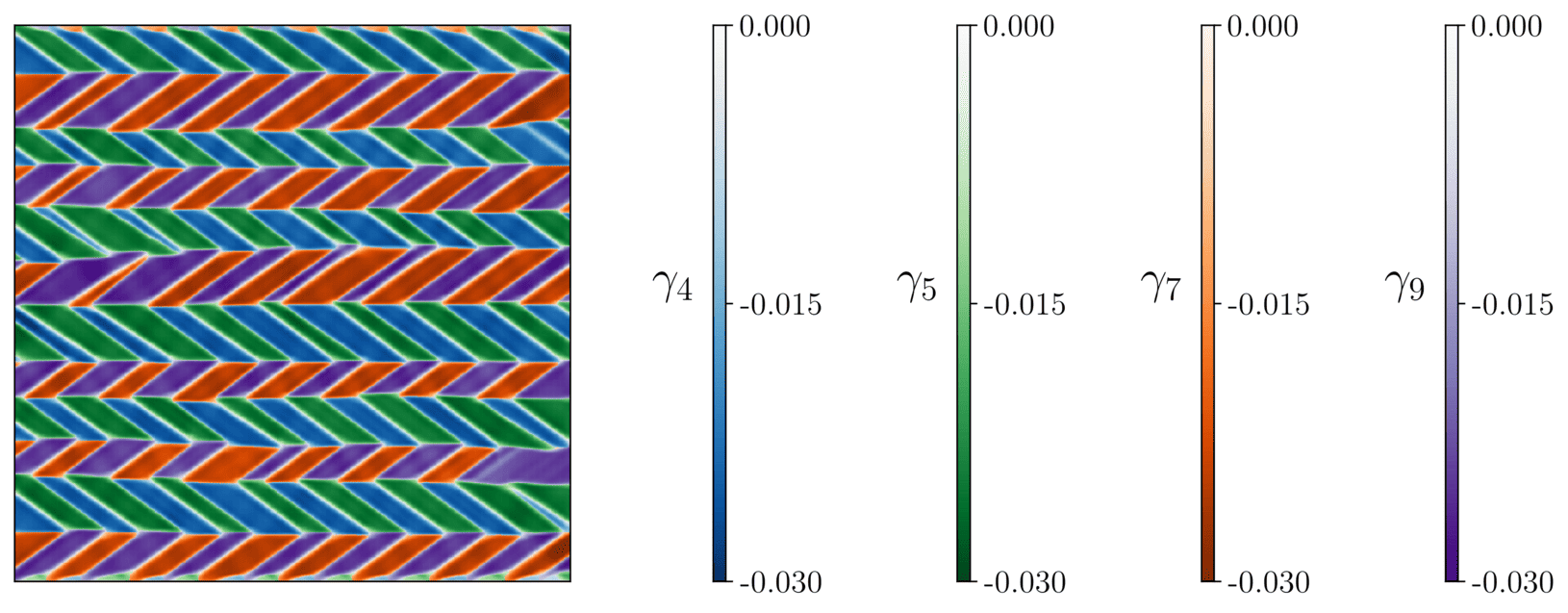}}\hspace{-4.2cm}
		\raisebox{5.9cm}{\includegraphics[width=0.15\textwidth]{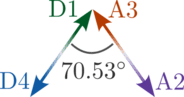}}
		\label{subfig:plane_strain_compression_010_superimposed}
	}
	\caption{Plane strain compression along $[00\bar{1}]$ in the $(110)$ plane at 1$\%$ macroscopic strain for $a^{ii} = 0.10$ and $a^{ij} = 0.16$. (a) Plastic slips on slip systems D4 ($\gamma_4$), D1 ($\gamma_5$), A2 ($\gamma_7$) and A3 ($\gamma_9$). (b) Superimposed plastic slips, where the color is selected according to the slip system with maximum slip activity. Arrows indicate the projections of slip directions of the active slip systems on the $(110)$ plane.}
	\label{fig:plane_strain_compression_010}
\end{figure}

The formation of this deformation pattern is attributed to strong latent hardening effects. Specifically, the off-diagonal elements of the hardening matrix are set to $0.16$, whereas the diagonal elements are $0.10$ (see Table~\ref{tab:material_parameters}). Although all four slip systems experience identical loading conditions at the onset of plasticity, the structure of the hardening matrix ensures that multiple slip results in greater hardening compared to single slip. Consequently, even minor numerical perturbations—such as round-off errors or convergence tolerances—can induce local imbalances in plastic slip activity, ultimately leading to the segregation of plastic slip into distinct single-slip domains. Notably, the emergence of this instability exhibits sensitivity to several computational parameters, including the distribution of workload among processing threads in the multithreaded finite element computations. To illustrate this effect, Figure~\ref{fig:plane_strain_compression_010_redo} presents an alternative solution obtained under the same initial and boundary conditions as those used in Figure~\ref{fig:plane_strain_compression_010}, but with a different thread allocation strategy. While the characteristic parallelogram-shaped regions persist, the overall deformation microstructure differs significantly. By constraining the computation to a single processing thread, we verified that the solutions remain deterministic—identical results are consistently obtained for the same initial and boundary conditions.
\begin{figure}
	\centering
	\subfloat[]{
		\hspace{-1.3cm}
		\includegraphics[width=0.55\textwidth]{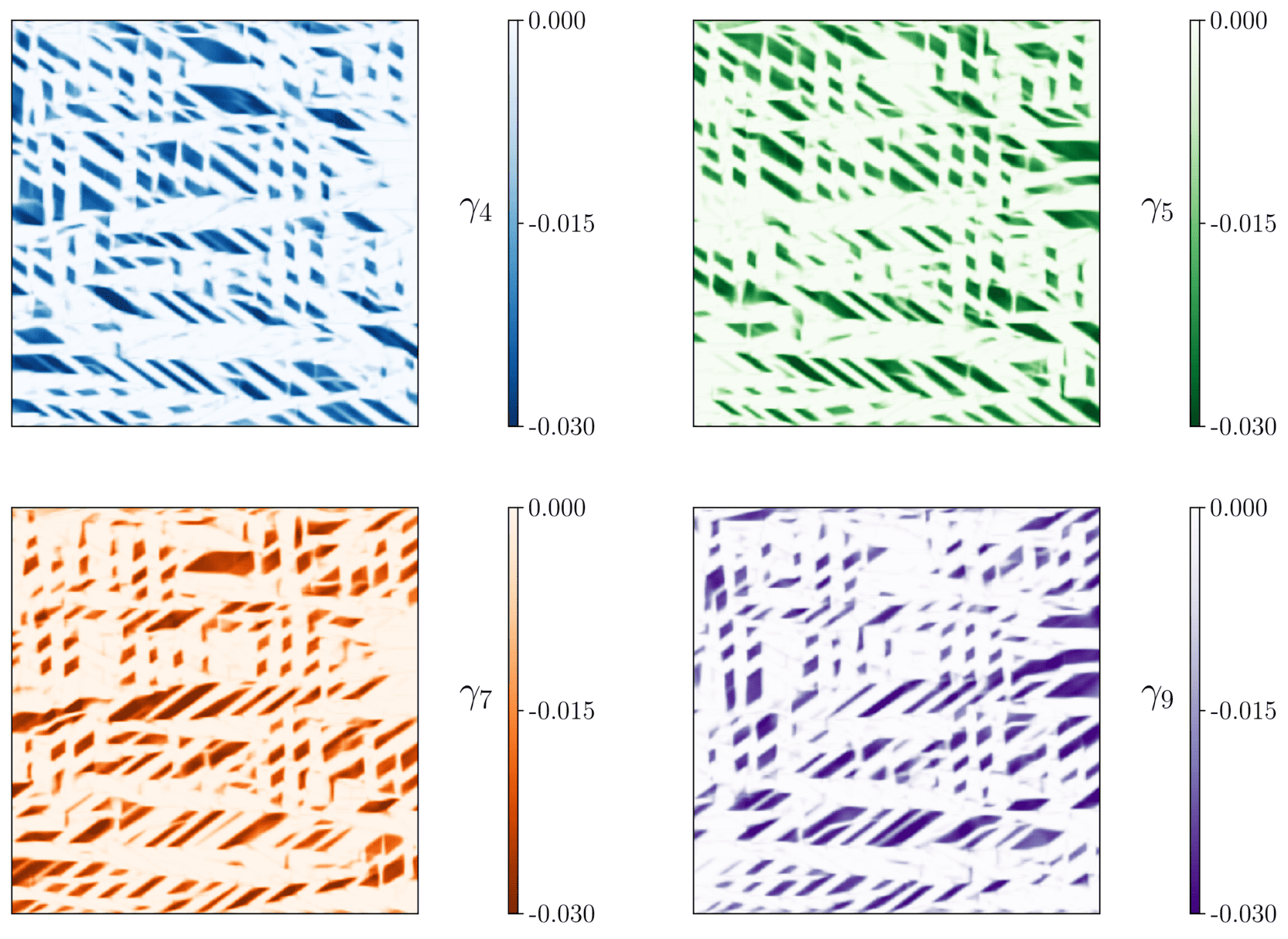}
		\label{subfig:plane_strain_compression_010_redo_grid}
	}
	\subfloat[]{
        \raisebox{0cm}{\includegraphics[width=0.13\textwidth]{FiguresReduced_axes.png}}\hspace{-1.3cm}
        \raisebox{.65cm}{\includegraphics[width=0.3\textwidth, trim=0.115cm 0.15cm 8.5cm 0cm, clip]{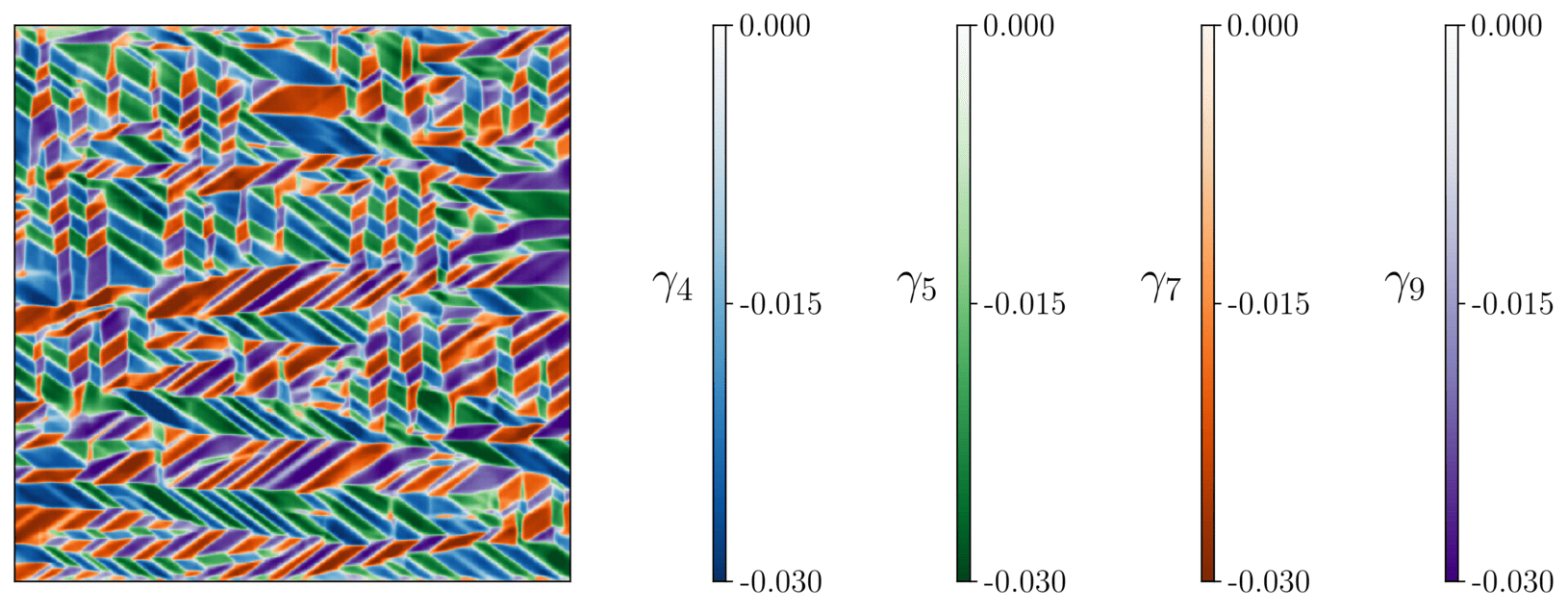}}\hspace{-4.2cm}
		\raisebox{5.9cm}{\includegraphics[width=0.15\textwidth]{FiguresReduced_slip_systems.png}}
		\label{subfig:plane_strain_compression_010_redo_superimposed}
	}
	\caption{Plane strain compression along $[00\bar{1}]$ in the $(110)$ plane at 1$\%$ macroscopic strain for $a^{ii} = 0.10$ and $a^{ij} = 0.16$. (a) Plastic slips on slip systems D4 ($\gamma_4$), D1 ($\gamma_5$), A2 ($\gamma_7$) and A3 ($\gamma_9$). (b) Superimposed plastic slips, where the color is selected according to the slip system with maximum slip activity. Arrows indicate the projections of slip directions of the active slip systems on the $(110)$ plane.}
	\label{fig:plane_strain_compression_010_redo}
\end{figure}

The components of the deformation gradient $\pmb{F}$ are presented in Figure~\ref{fig:deformation_gradient_010_redo}. The fragmentation of the single crystal within single-slip regions results in heterogeneous deformation fields. Sharp deformation bands are observed in the in-plane diagonal components of the deformation gradient, $F_{11}$ and $F_{22}$, coinciding with the interfaces between single-slip regions. The shear components of the deformation gradient also exhibit the signature of single crystal fragmentation. The first-order lamination microstructure is distinctly visible in the $F_{12}$ component, where alternating layers of positive and negative shear extend along the vertical direction. In contrast, the second-order lamination manifests in the $F_{23}$ component, with shear strain alternating in sign along the horizontal direction. Due to the imposed plane strain conditions, the $F_{3\bullet}$ components remain close to zero. These simulated deformation gradient components align with experimental observations of patchy slip, as illustrated in Figure~\ref{fig:patterns_experimental}. Furthermore, the heterogeneity of plastic deformation at the single crystal scale has been demonstrated using high-resolution digital image correlation~\citep{di2013plastic,di2015experimental,charpagne2020automated,sperry2021comparison,black2023high,texier2024strain} and topotomography~\citep{proudhon2018incipient,charpagne2021multi,stinville2022observation,stinville2023insights}. While, here, individual slip lines or slip line bundles are not resolved as in high-resolution DIC, the formation of deformation patterns with sharp interfaces between distinct regions remains consistent with experimental findings. \textcolor{black}{The components of Nye's dislocation density tensor $\pmb{\alpha}$ were also computed and are analyzed in~\ref{sec:nye_tensor}.
}
\begin{figure}
	\centering
	\subfloat{\includegraphics[width=0.32\textwidth]{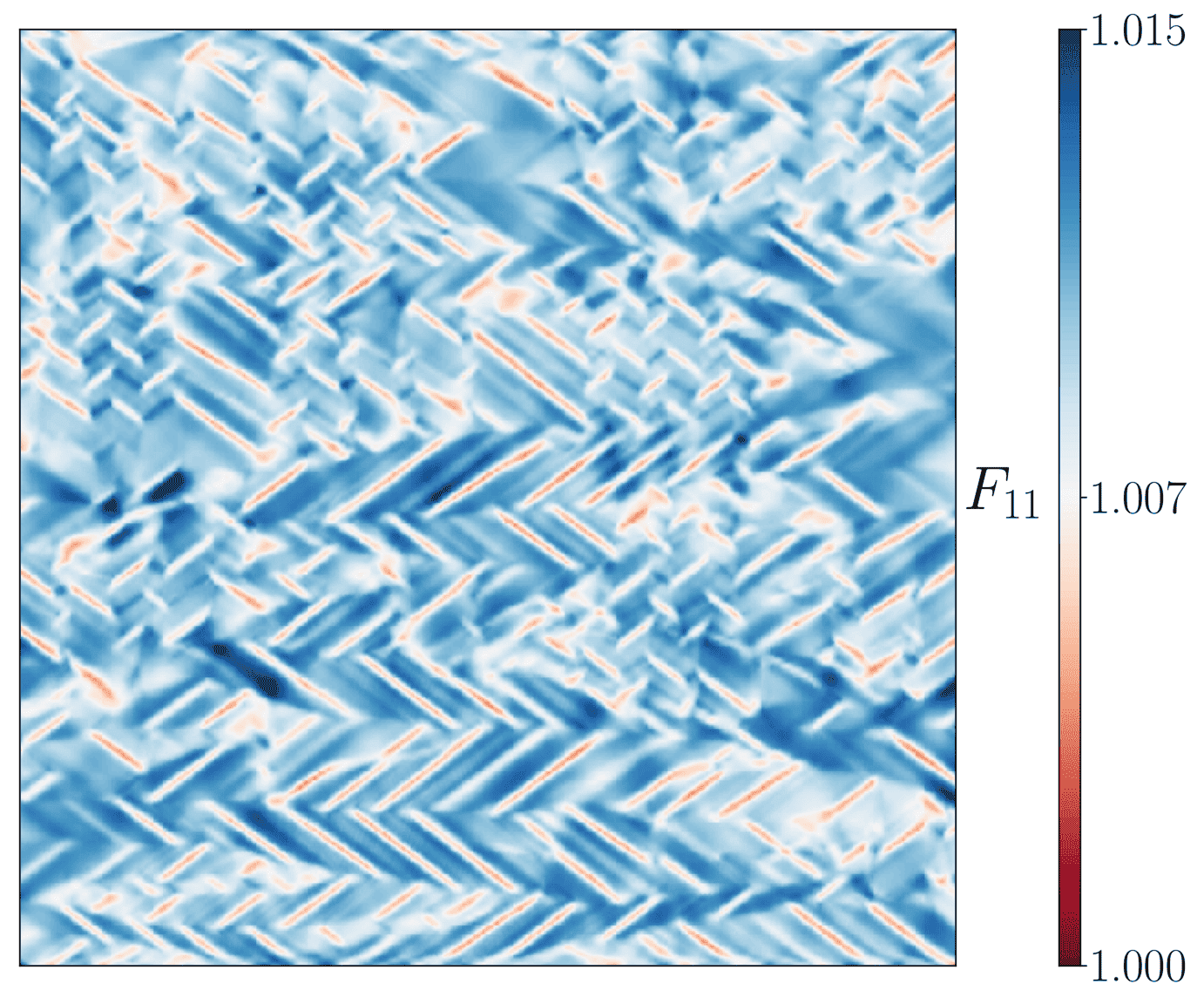}}
	\subfloat{\includegraphics[width=0.32\textwidth]{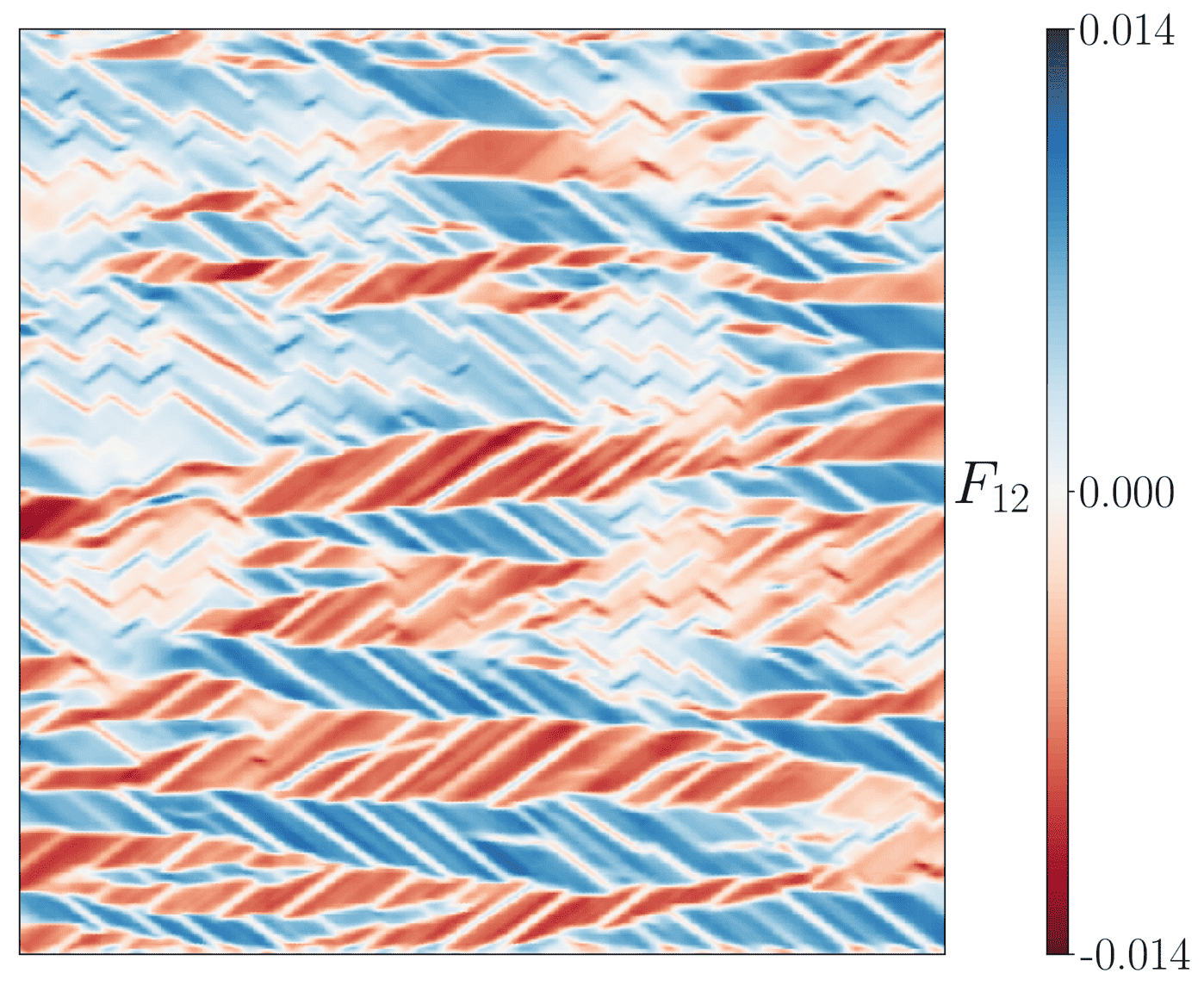}}
	\subfloat{\includegraphics[width=0.32\textwidth]{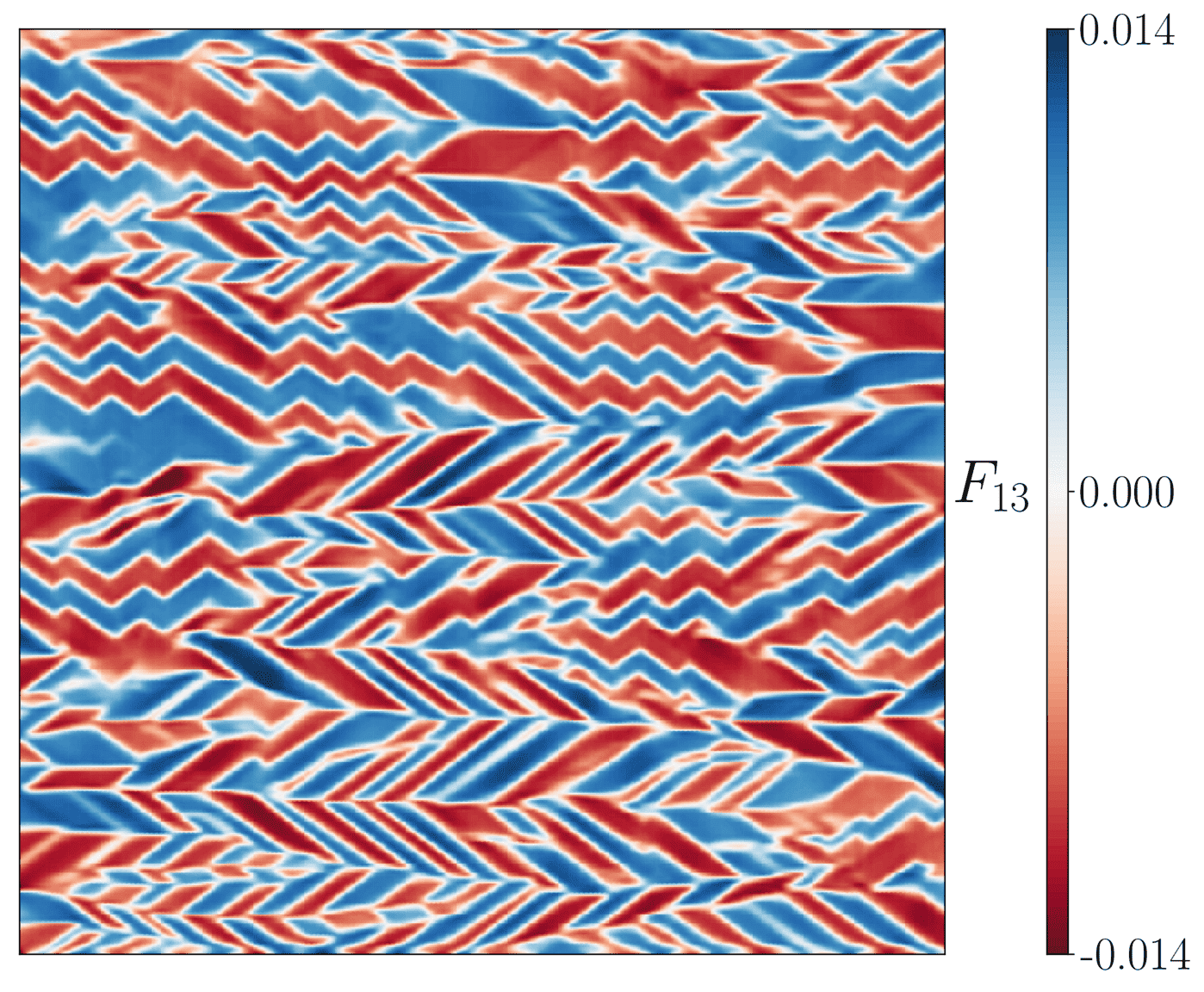}}\\
	\subfloat{\includegraphics[width=0.32\textwidth]{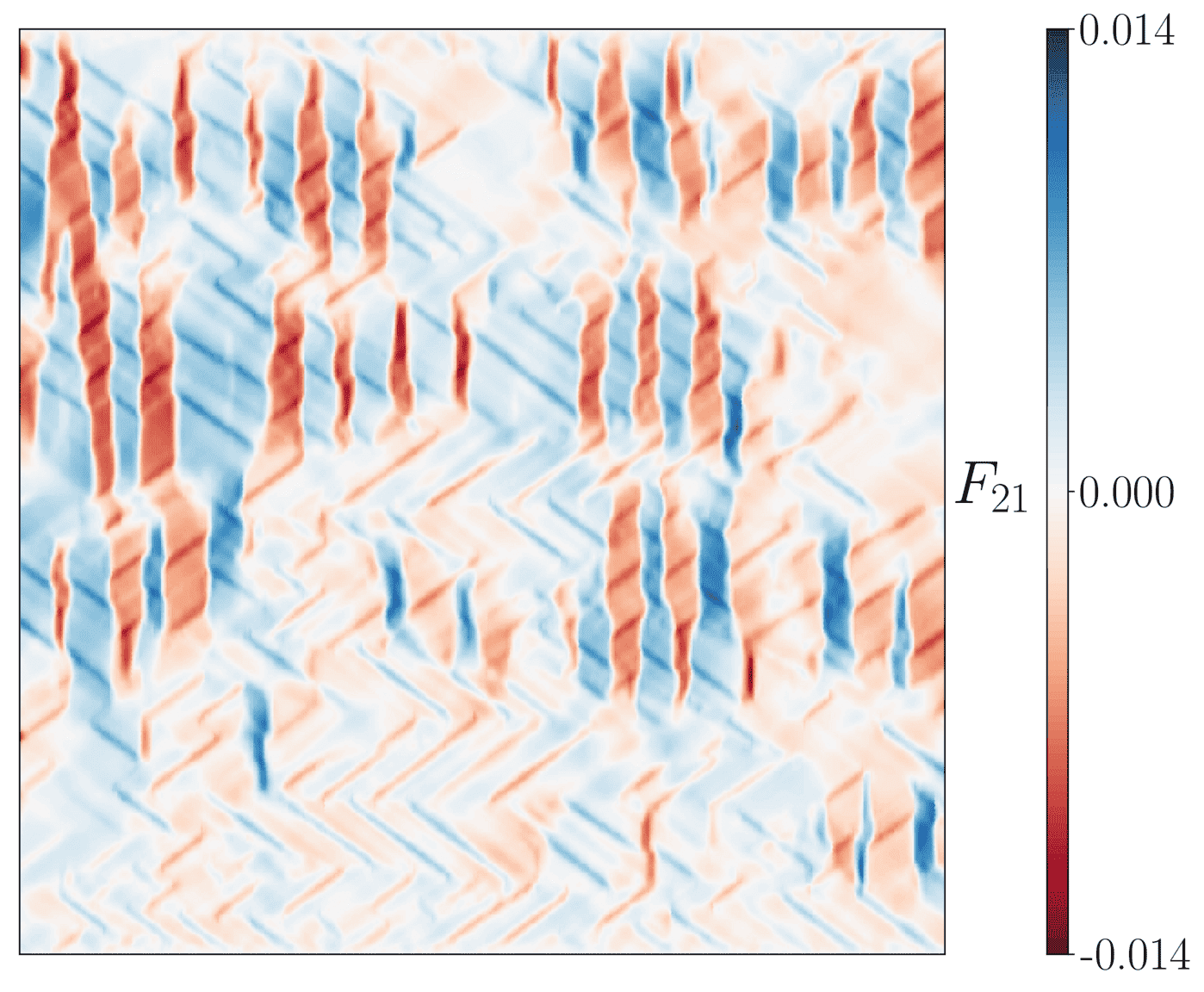}}
	\subfloat{\includegraphics[width=0.32\textwidth]{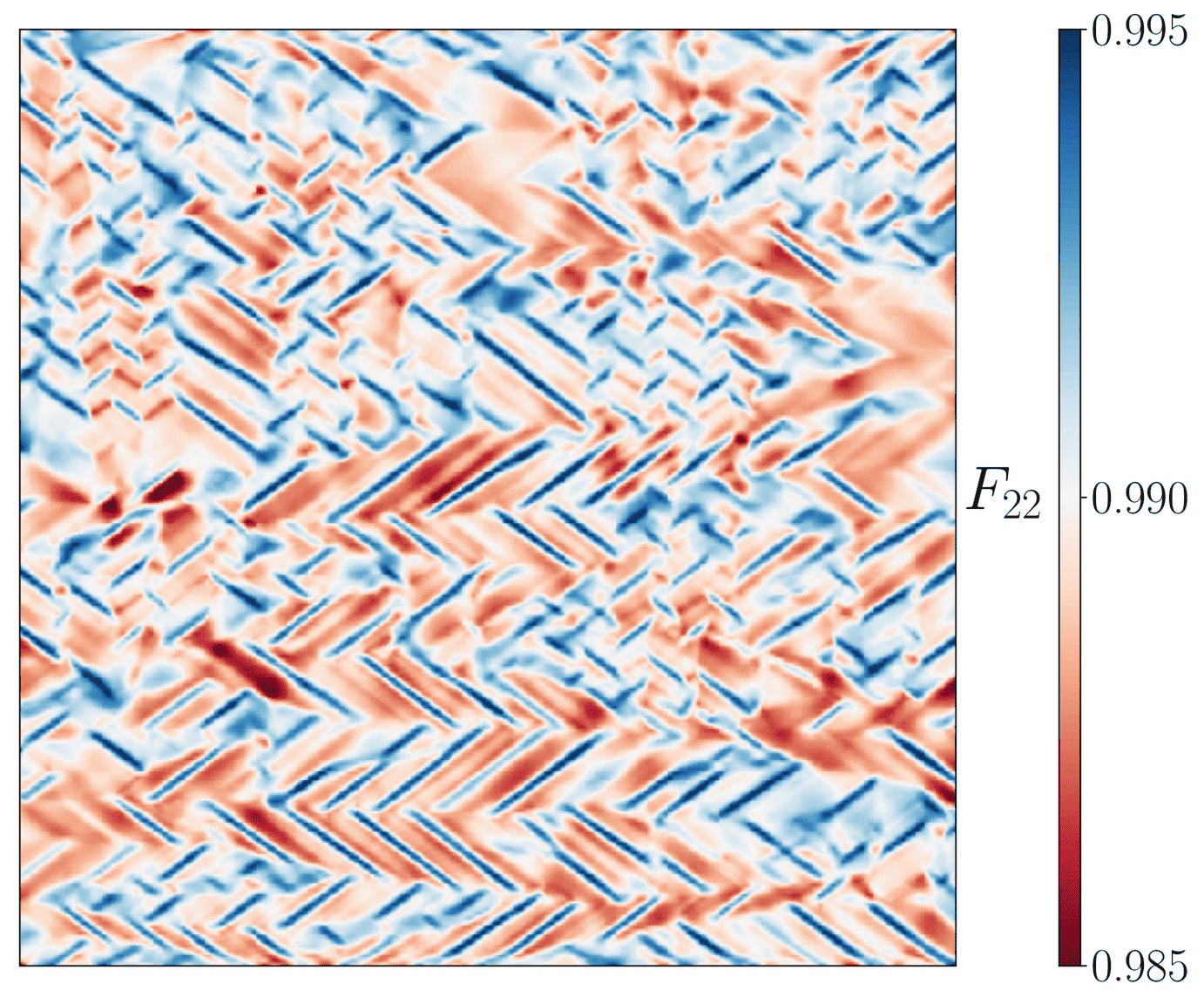}}
	\subfloat{\includegraphics[width=0.32\textwidth]{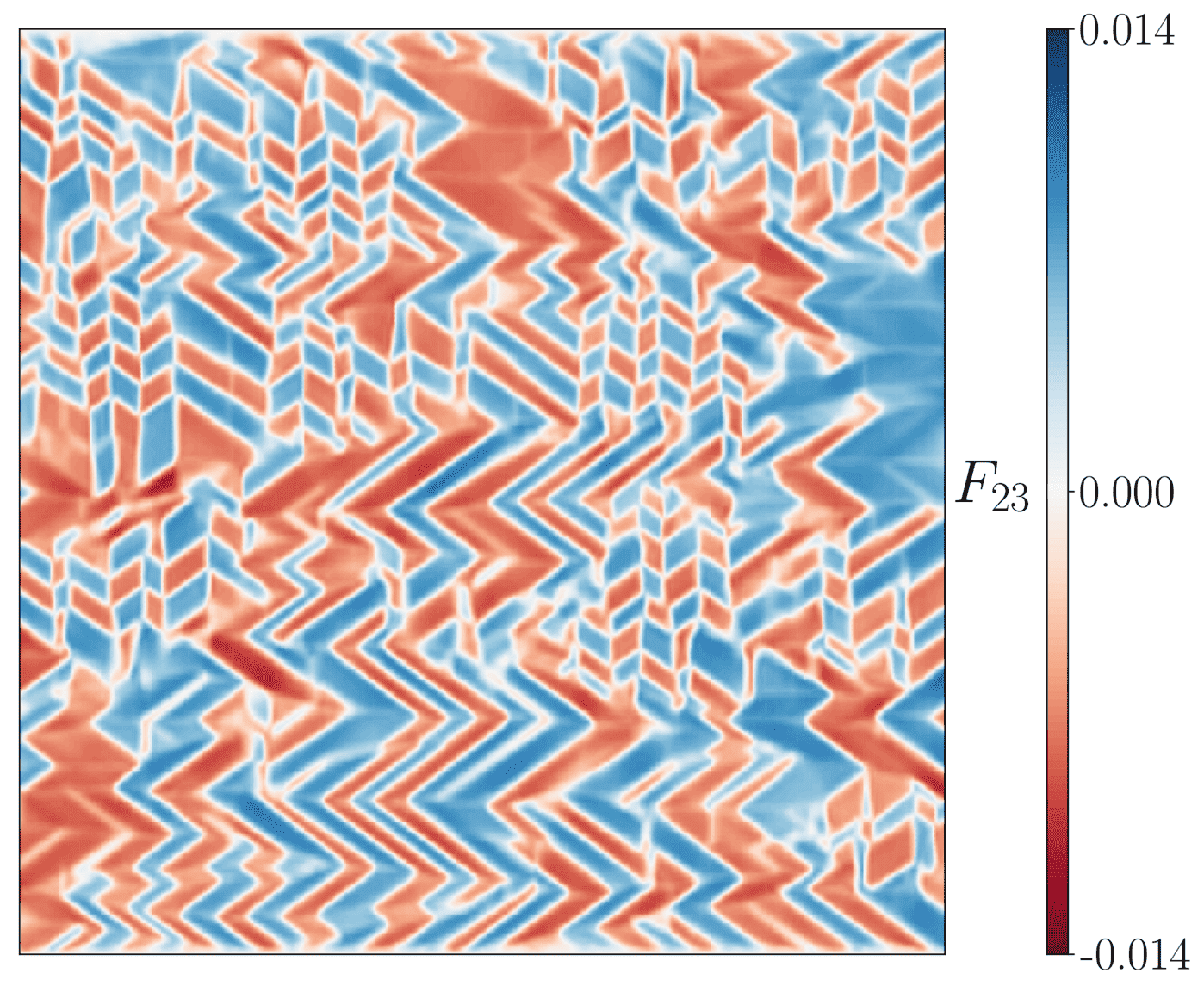}}\\
	\subfloat{\includegraphics[width=0.32\textwidth]{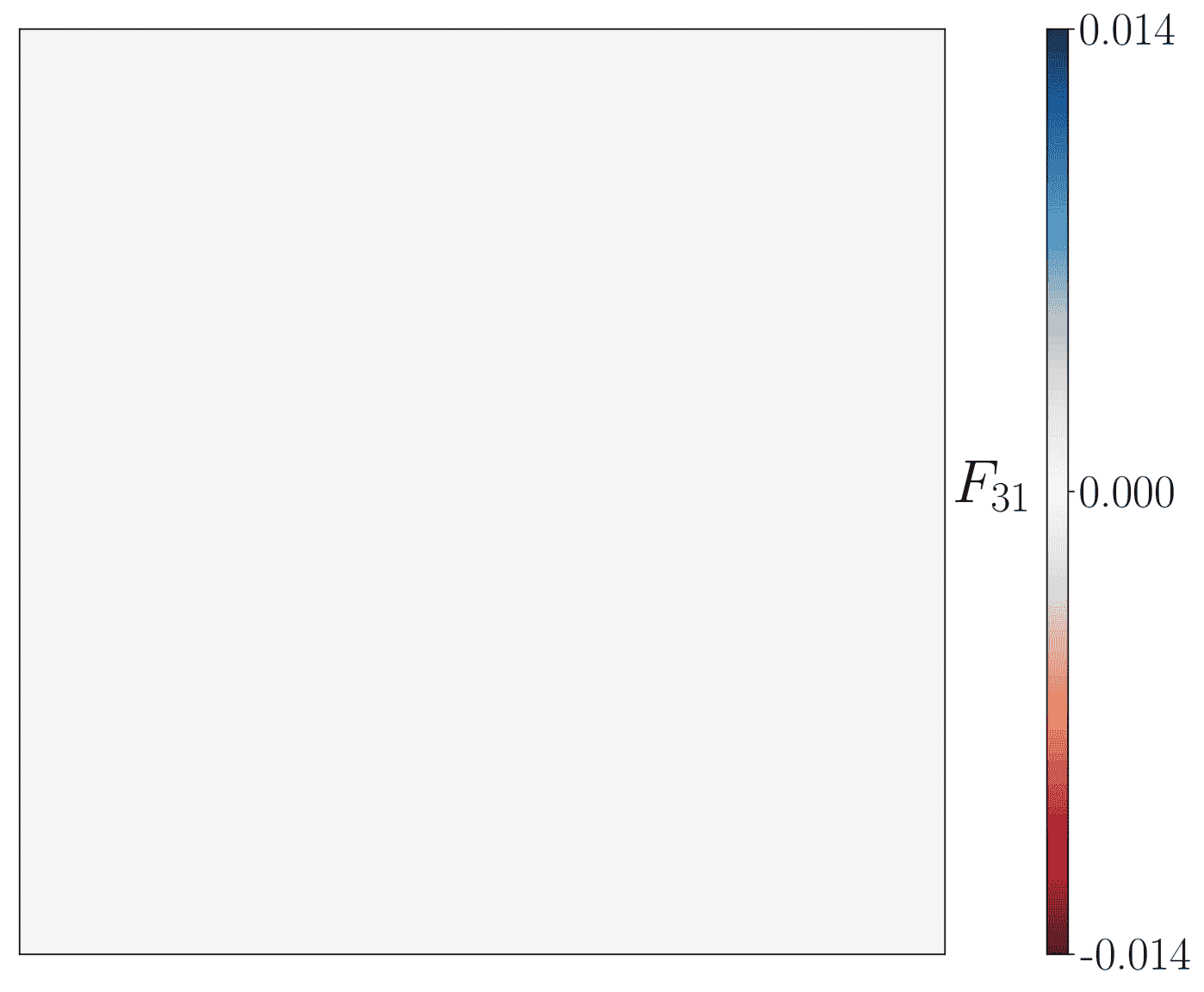}}
	\subfloat{\includegraphics[width=0.32\textwidth]{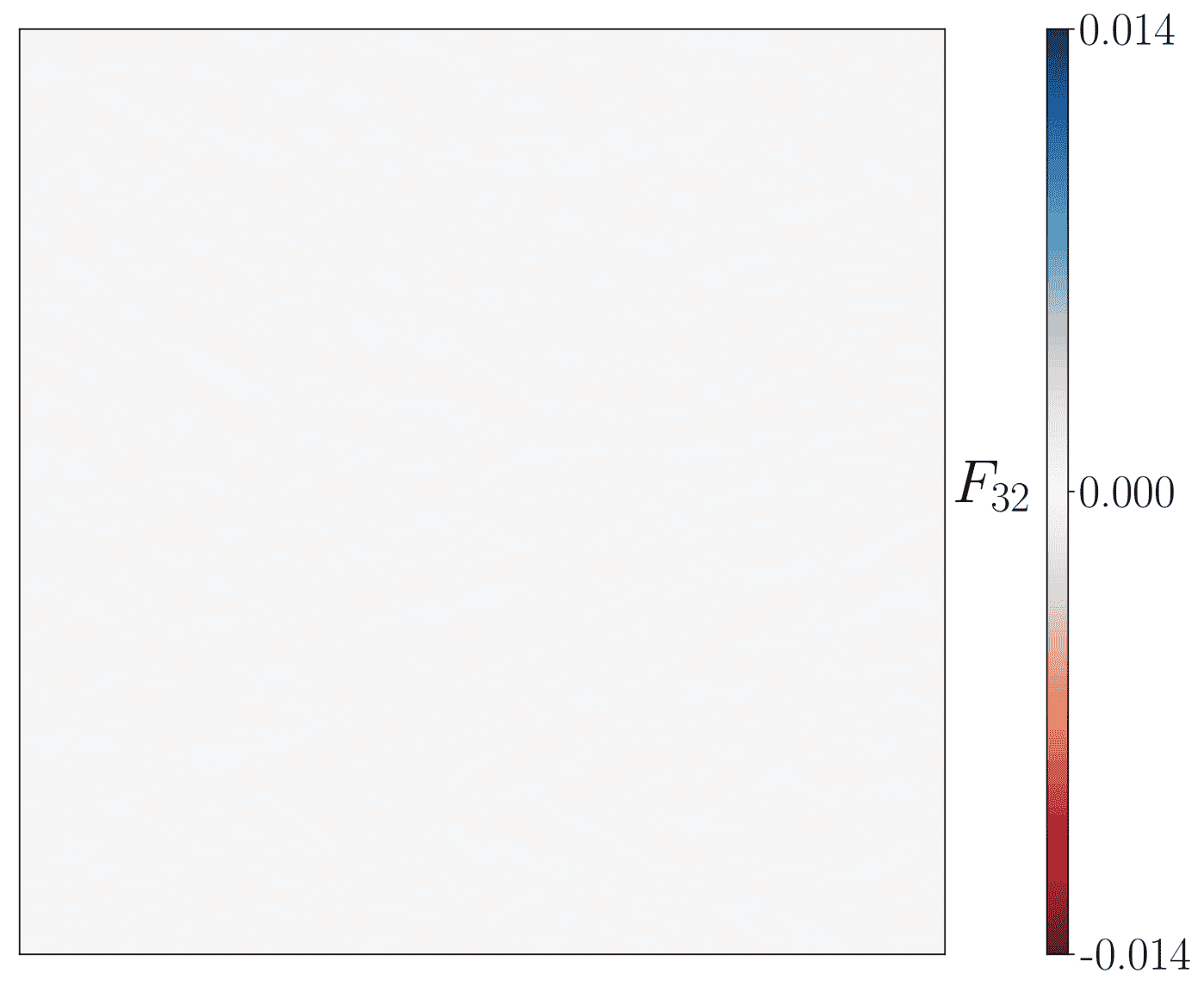}}
	\subfloat{\includegraphics[width=0.32\textwidth]{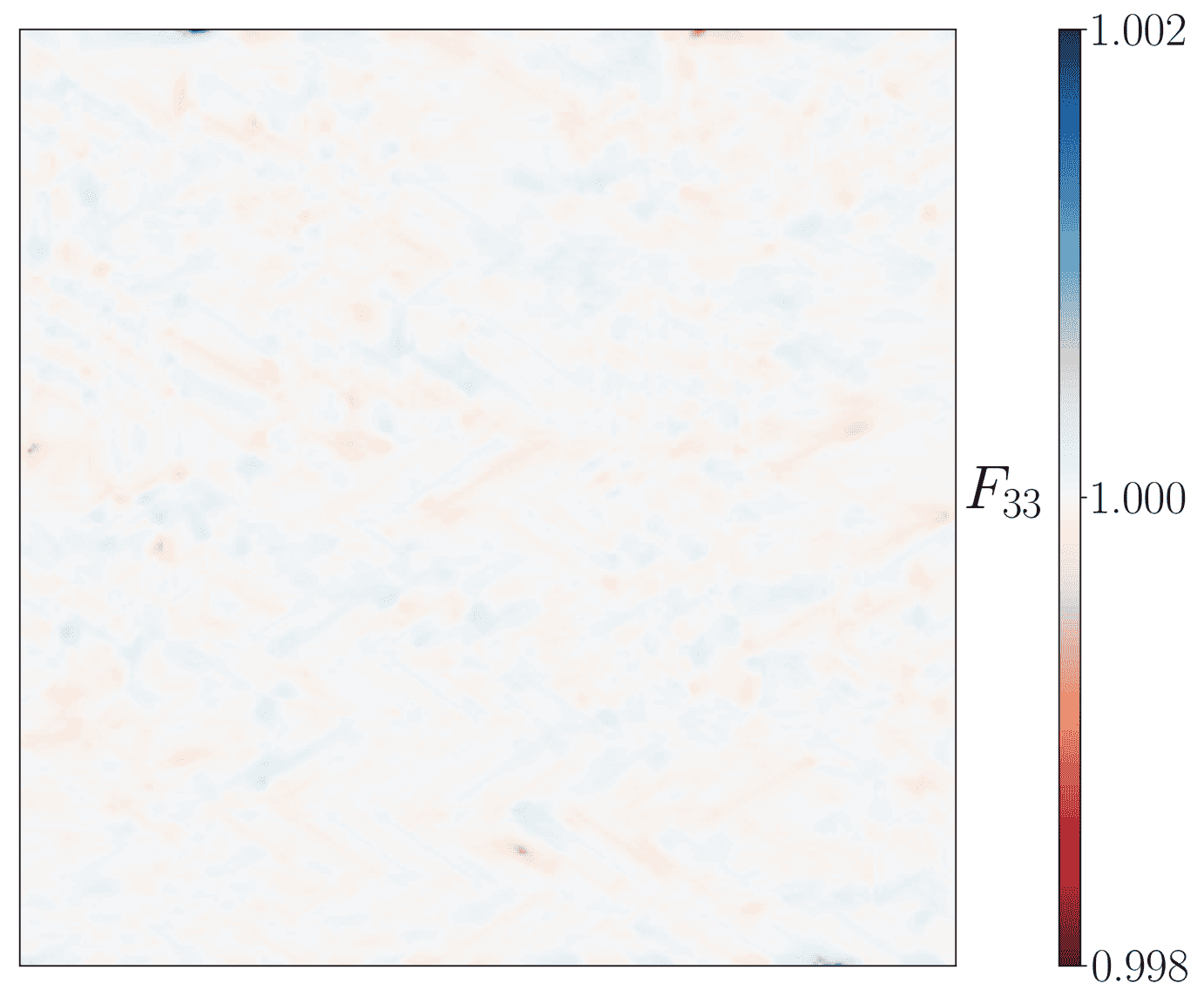}}
	\caption{Components of the deformation gradient $\pmb{F}$ at 1\% macroscopic strain under plane strain compression along $[00\bar{1}]$ in the $(110)$ plane.}
	\label{fig:deformation_gradient_010_redo}
\end{figure}

To analyze the crystal lattice rotation induced by the deformation pattern, the simulation depicted in Figure~\ref{fig:plane_strain_compression_010} was extended to $40\%$ macroscopic strain. The pole figures at $0\%$, $10\%$, $20\%$, and $40\%$ macroscopic strain are presented in Figure~\ref{fig:pole_figures}, corresponding to the $(110)$ and $(00\bar{1})$ planes. Initially (see Figure~\ref{subfig:pf_time_0}), the pole figures \textcolor{black}{are that of a single crystal with $x_1 = [1\bar{1}0]$, $x_2 = [00\bar{1}]$, and $x_3 = [110]$}. By $10\%$ macroscopic strain (Figure~\ref{subfig:pf_time_50}), a symmetric fragmentation of the orientation distribution into four distinct modes emerges, each corresponding to a single-slip region shown in Figure~\ref{fig:plane_strain_compression_010_redo}. As deformation progresses, the fragmentation intensifies, and the orientation distribution becomes less localized, signifying the development of orientation gradients. At $20\%$ and $40\%$ macroscopic strain (Figures~\ref{subfig:pf_time_100} and~\ref{subfig:pf_time_200}), the single crystal undergoes complete fragmentation into a mosaic of four distinct sub-grains. Supplementary videos illustrate the evolution of plastic slip fields and pole figures from $0\%$ to $40\%$ macroscopic strain, revealing that at higher strains, a rank-three laminate microstructure develops. This observation aligns with the trend of the model proposed in~\cite{ortiz1999nonconvex}, which relates the characteristic size of the deformation pattern $\phi$ to the applied strain $\gamma$ through an inverse power law: $\phi \propto \gamma^{-1/2}$. 
\begin{figure}
	\centering
	\subfloat[$0\%$ macroscopic strain]{
	    \includegraphics[width=0.24\textwidth]{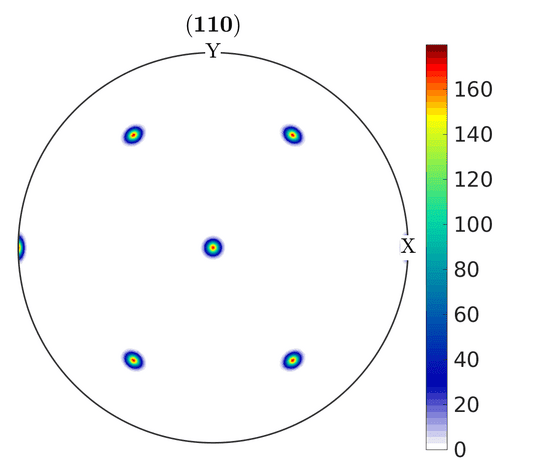}
	    \includegraphics[width=0.24\textwidth]{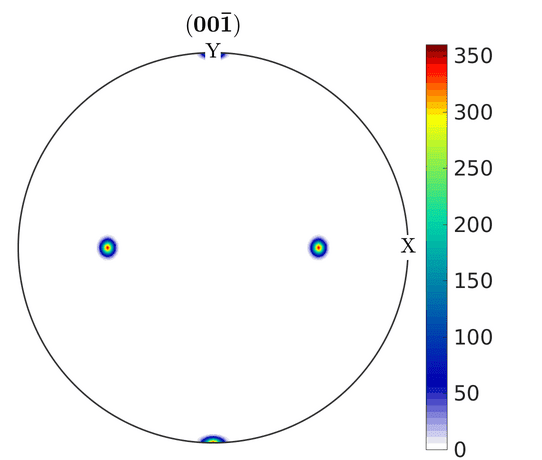}
	    \label{subfig:pf_time_0}}
	\subfloat[$10\%$ macroscopic strain]{
	    \includegraphics[width=0.24\textwidth]{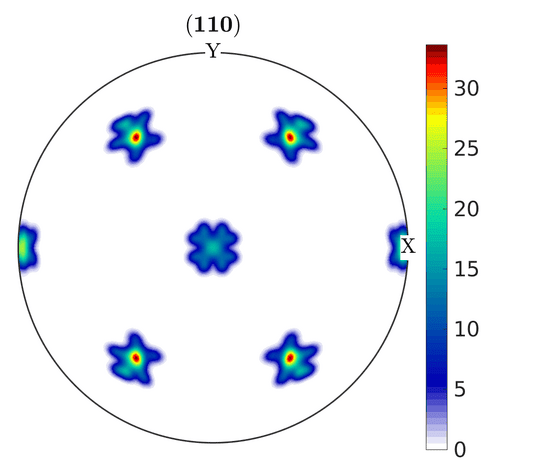}
	    \includegraphics[width=0.24\textwidth]{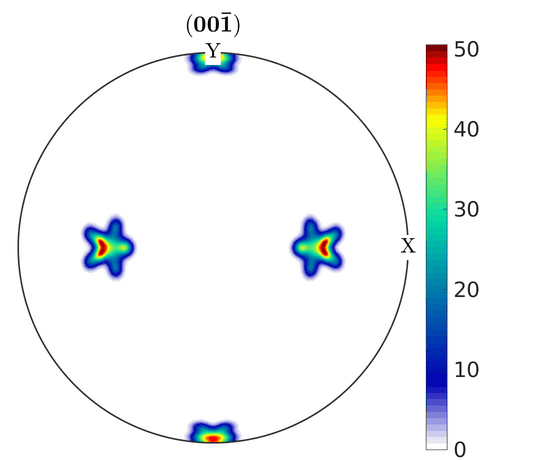}
	    \label{subfig:pf_time_50}}\\
	\subfloat[$20\%$ macroscopic strain]{
	    \includegraphics[width=0.24\textwidth]{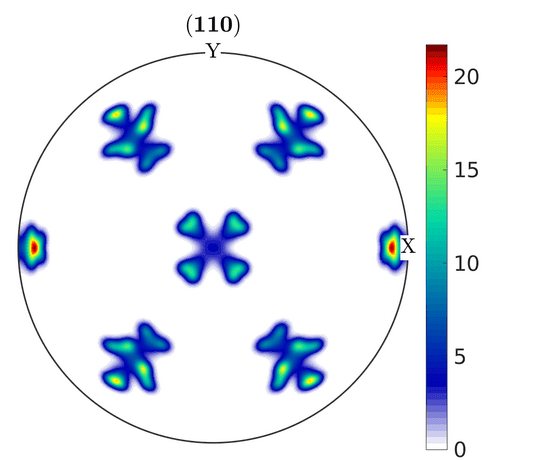}
	    \includegraphics[width=0.24\textwidth]{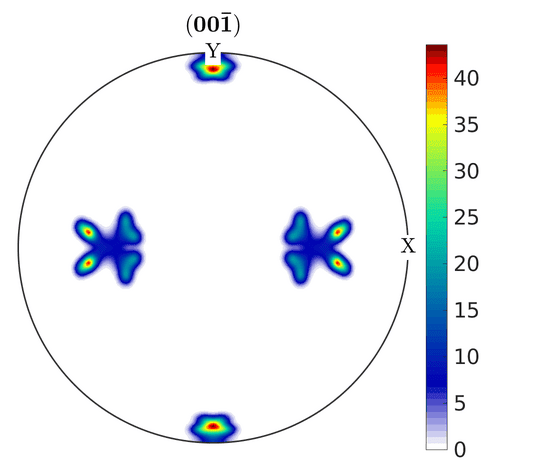}
	    \label{subfig:pf_time_100}}
	\subfloat[$40\%$ macroscopic strain]{
	    \includegraphics[width=0.24\textwidth]{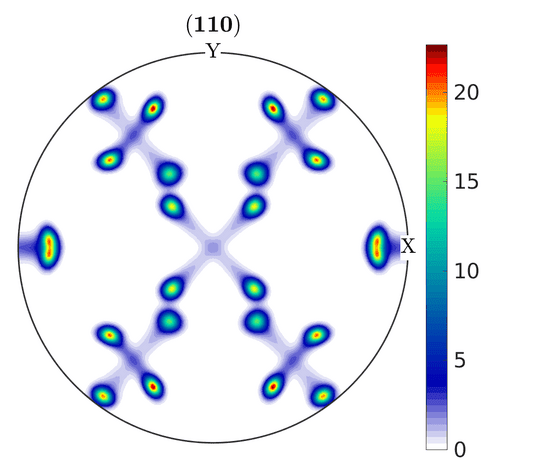}
	    \includegraphics[width=0.24\textwidth]{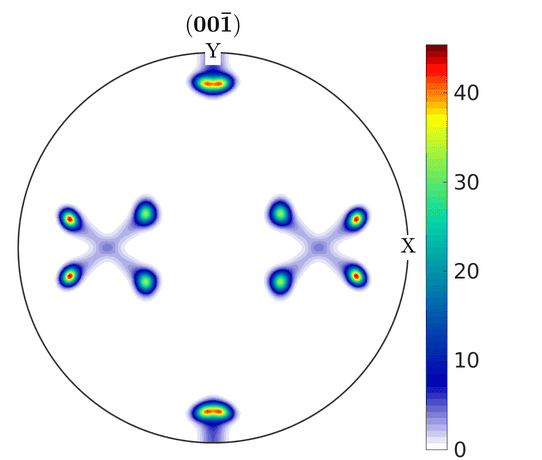}
	    \label{subfig:pf_time_200}}
	\caption{Pole figures at (a) $0\%$, (b) $10\%$, (c) $20\%$ and (d) $40\%$ macroscopic strain for plane strain compression along $[00\bar{1}]$ in the $(110)$ plane. In each subplot, the $(110)$ pole figure is shown on the left and the $(00\bar{1})$ pole figure is shown on the right.}
	\label{fig:pole_figures}
\end{figure}

The role of the interaction matrix is further investigated by modifying the value of one of the interaction coefficients, specifically the coplanar interaction coefficient $a_{cop}$. As discussed earlier, coplanar interactions occur between dislocations gliding on slip systems D4 and D1 or A2 and A3. We reduce the coplanar interaction coefficient to $0.05$ while keeping the other interaction coefficients unchanged. Figure~\ref{fig:plane_strain_compression_010_coplanar} presents the deformation pattern obtained at 1\% macroscopic strain with this modified interaction matrix. The plastic slip is now divided into only two domains, forming a \textit{labyrinth}-like microstructure. The width of the labyrinth channels is on the order of the mesh size. These channel boundaries are oriented along $[00\bar{1}]$ (vertical) and $[1\bar{1}0]$ (horizontal). Within each of the two domains, two coplanar slip systems are simultaneously active with nearly equal intensity. Multiple slip is now feasible because the coplanar interactions are weaker than in the previous case and, in particular, weaker than the self-interaction. This finding highlights the significant sensitivity of deformation patterning to the interaction matrix, demonstrating that varying the relative strength of dislocation interactions results in substantially different patterns. Previous studies have examined this phenomenon~\cite{dequiedt2015heterogeneous,wang2018role}, particularly the impact of collinear interaction and its relatively higher strength compared to other interactions~\citep{madec2003role} on deformation pattern formation.
\begin{figure}
	\centering
	\subfloat[]{
	\hspace{-1.3cm}
	\includegraphics[width=0.55\textwidth]{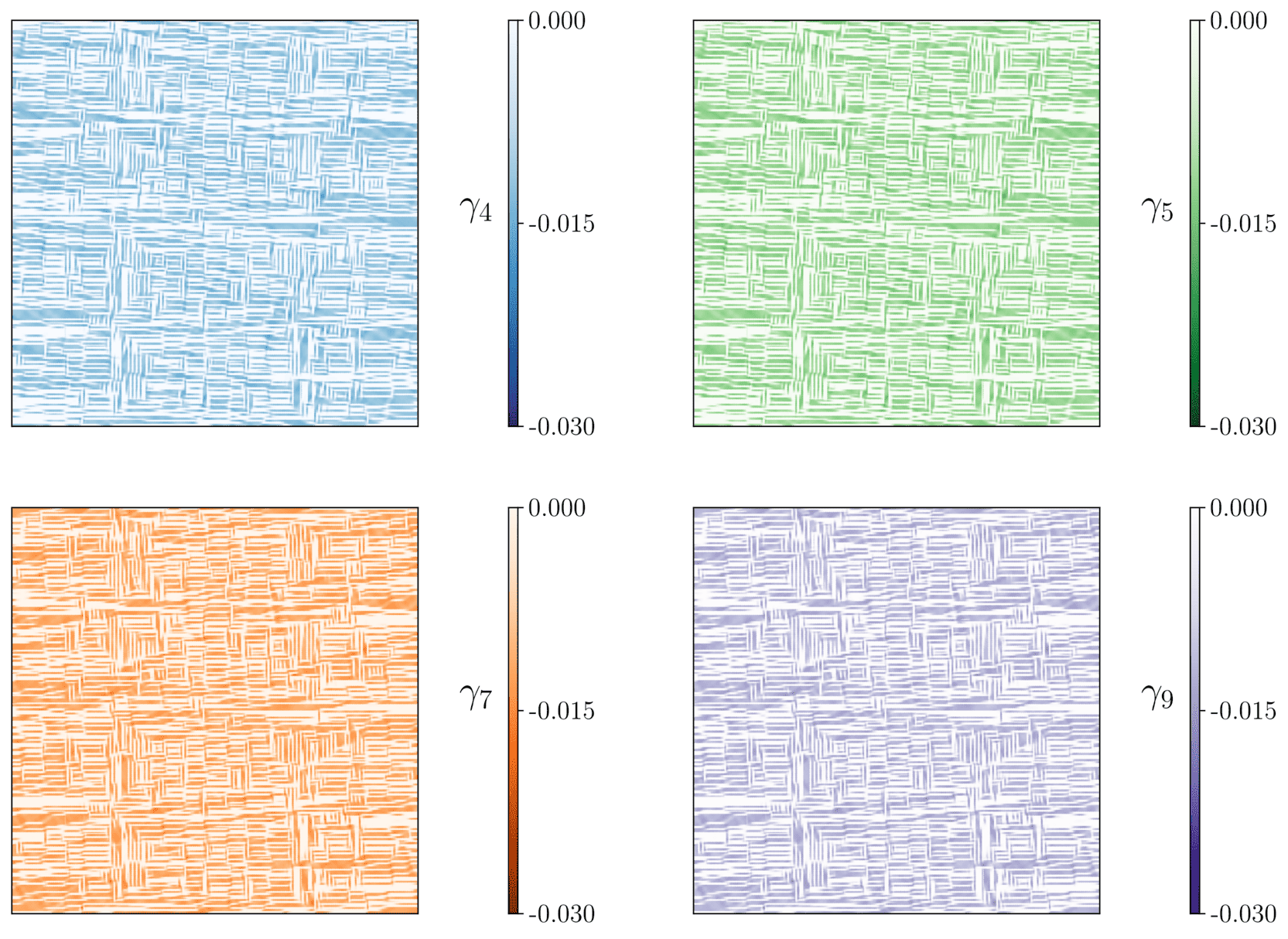}
	\label{subfig:plane_strain_compression_010_coplanar_grid}
	}
	\subfloat[]{
	\raisebox{0cm}{\includegraphics[width=0.13\textwidth]{FiguresReduced_axes.png}}\hspace{-1.3cm}
	\raisebox{.65cm}{\includegraphics[width=0.3\textwidth, trim=0.115cm 0.15cm 8.5cm 0cm, clip]{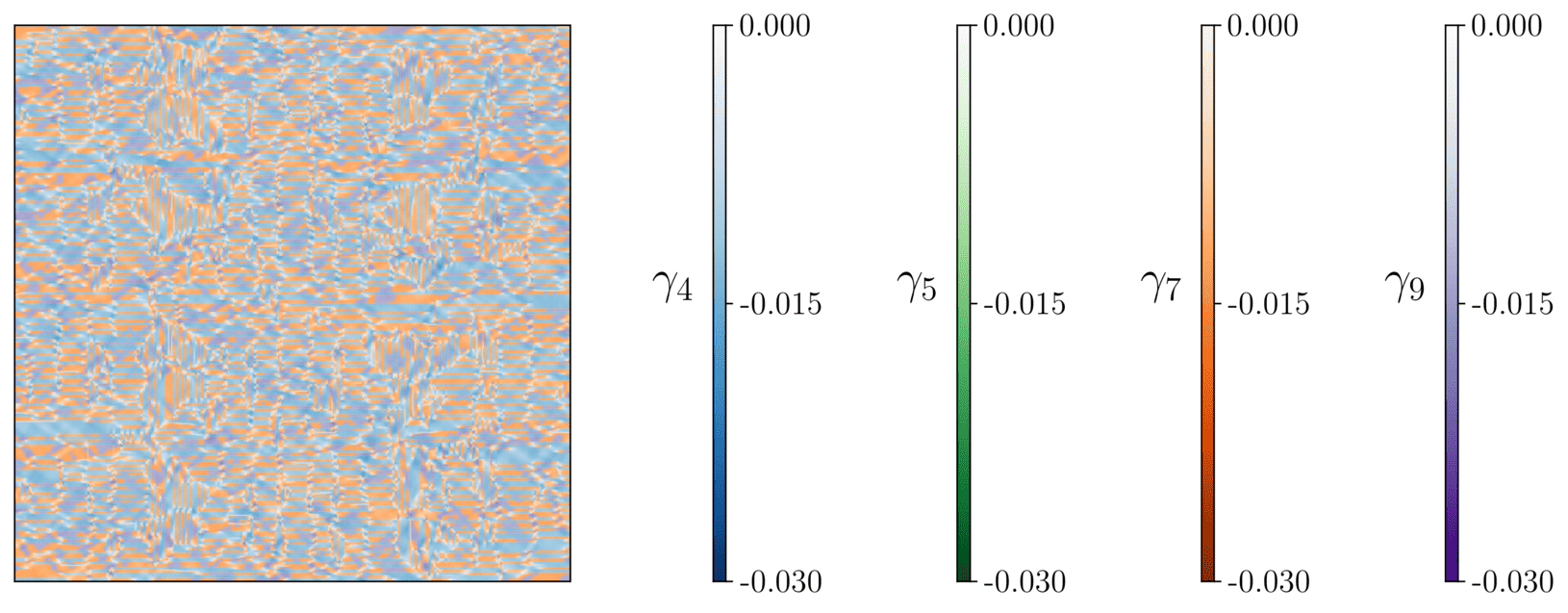}}\hspace{-4.2cm}
	\raisebox{5.9cm}{\includegraphics[width=0.15\textwidth]{FiguresReduced_slip_systems.png}}
	\label{subfig:plane_strain_compression_010_coplanar_superimposed}
	}
	\caption{Plane strain compression along $[00\bar{1}]$ in the $(110)$ plane at 1$\%$ macroscopic strain for $a_{coplanar}=0.05$. (a) Plastic slips on slip systems D4 ($\gamma_4$), D1 ($\gamma_5$), A2 ($\gamma_7$) and A3 ($\gamma_9$). (b) Superimposed plastic slips, where the color is selected according to the slip system with maximum slip activity. Arrows indicate the projections of slip directions of active slip systems on the $(110)$ plane.}
	\label{fig:plane_strain_compression_010_coplanar}
\end{figure}

We use again the interaction matrix such that $a^{ii} = 0.10$ and $a^{ij} = 0.16$ for $i \neq j$, but we now slightly rotate the crystal around the $[110]$ axis by $-\SI{4}{\degree}$ as shown in Figure~\ref{fig:plane_strain_compression_010_misorientation}. This slight misorientation with respect to the loading direction reduces the Schmid factor on slip systems A2 and A3 ($\sim 0.384 < 1/\sqrt{6}$) and increases it slightly on slip systems D4 and D1 ($\sim 0.424 > 1/\sqrt{6}$). As a result, plastic slip activity on A2 and A3 is nearly absent, except in the vicinity of the bottom and top edges. A deformation pattern with two single-slip domains is obtained. These two domains exhibit a layered pattern of kink bands forming a rank-one microstructure. The deformation bands are indeed inclined such that the slip directions (oriented at $\SI{54.74}{\degree}-\SI{4}{\degree} = \SI{50.74}{\degree}$) are orthogonal to the bands. The width of these bands varies but remains overall comparable to what was observed previously in Figures~\ref{fig:plane_strain_compression_010} and~\ref{fig:plane_strain_compression_010_redo}. 
\begin{figure}
	\centering
	\subfloat[]{
		\hspace{-1.3cm}
		\includegraphics[width=0.55\textwidth]{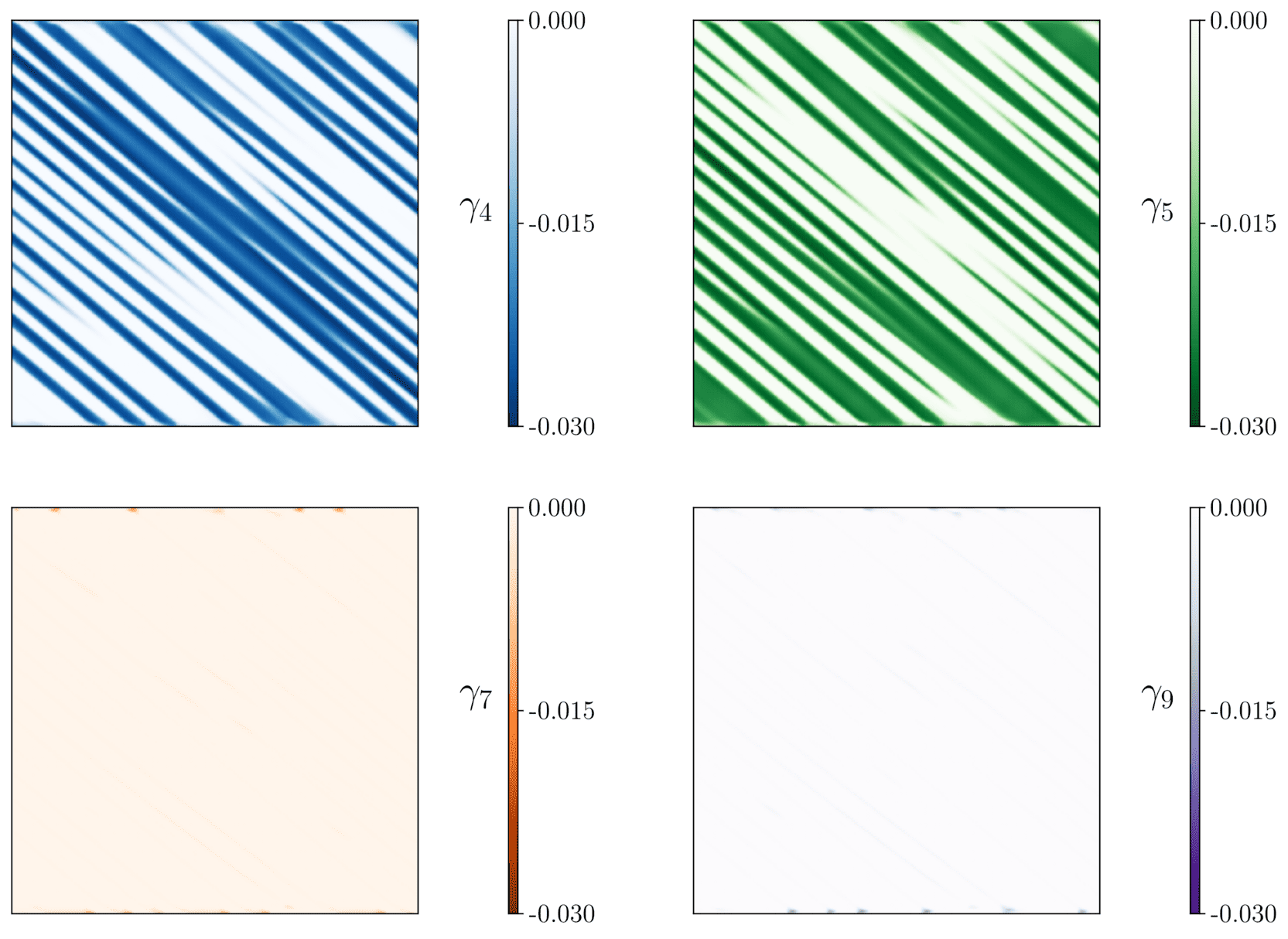}
		\label{subfig:plane_strain_compression_010_misorientation_grid}
	}
	\subfloat[]{
        \raisebox{0cm}{\includegraphics[width=0.13\textwidth]{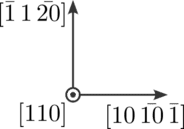}}\hspace{-1.3cm}
        \raisebox{.65cm}{\includegraphics[width=0.3\textwidth, trim=0.115cm 0.15cm 8.5cm 0cm, clip]{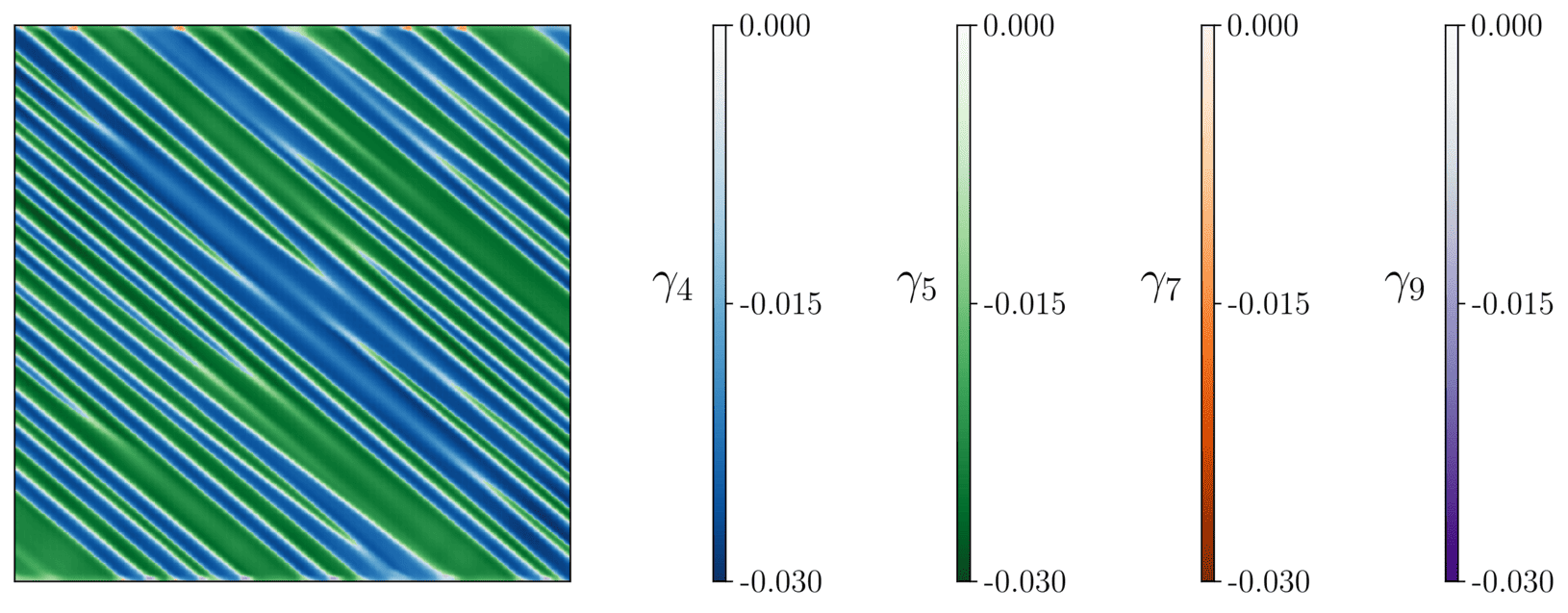}}\hspace{-4.2cm}
		\raisebox{5.9cm}{\includegraphics[width=0.15\textwidth,angle=-4]{FiguresReduced_slip_systems.png}}
		\label{subfig:plane_strain_compression_010_misorientation_superimposed}
	}
	\caption{Plane strain compression along $[\bar{1}\,1\,\bar{20}]$ in the $(110)$ plane at 1$\%$ macroscopic strain for $a^{ii} = 0.10$ and $a^{ij} = 0.16$. (a) Plastic slips on slip systems D4 ($\gamma_4$), D1 ($\gamma_5$), A2 ($\gamma_7$) and A3 ($\gamma_9$). (b) Superimposed plastic slips, where the color is selected according to the slip system with maximum slip activity. Arrows indicate the projections of slip directions of active slip systems on the $(110)$ plane.}
	\label{fig:plane_strain_compression_010_misorientation}
\end{figure}

We now consider a different crystal orientation, where the directions are defined as $x_1 = [0\bar{1}0]$, $x_2 = [00\bar{1}]$, and $x_3 = [100]$, as illustrated in Figure~\ref{fig:plane_strain_compression_010_001}. In this configuration, the active slip systems are as follows
\begin{itemize}
	\item B2: $[0\bar{1}1](111)$ $\longrightarrow$ $\gamma_2$
	\item D1: $[011](1\bar{1}1)$ $\longrightarrow$ $\gamma_5$
	\item A2: $[0\bar{1}1](\bar{1}11)$ $\longrightarrow$ $\gamma_7$
	\item C1: $[011](\bar{1}\bar{1}1)$ $\longrightarrow$ $\gamma_{12}$
\end{itemize}
For these slip systems, the potential interactions between dislocations are as follows
\begin{itemize}
	\item Self interaction: B2 -- B2, D1 -- D1, A2 -- A2, C1 -- C1 $\longrightarrow$ $a_{self}=a^{22}=a^{55}=a^{77}=a^{12\,12}$
	\item Hirth lock: B2 -- D1, A2 -- C1, B2 -- C1, D1 -- A2  $\longrightarrow$ $a_{Hirth}=a^{25}=a^{52}=a^{7\, 12}=a^{12\, 7}=a^{2\, 12}=a^{57}$
	\item Collinear interaction: B2 -- A2, D1 -- C1 $\longrightarrow$ $a_{col}=a^{27}=a^{72}=a^{5\, 12}=a^{12\, 5}$
\end{itemize}
As shown in Figure~\ref{subfig:plane_strain_compression_010_001_superimposed}, a rank-two laminate microstructure is formed. The single crystal is divided into four single-slip domains. The first-order lamination divides the crystal into roughly equally sized rectangular cells, with the normals to these cells being $[0\bar{1}0]$ and $[00\bar{1}]$. The second-order lamination further subdivides each rectangular cell into arrays of parallel bands with normals at $\SI{45}{\degree}$, along $[0\bar{1}\bar{1}]$ (B2 -- A2, B2 -- C1, B2 -- D1) and at $-\SI{45}{\degree}$, along $[01\bar{1}]$ (D1 -- C1, D1 -- A2, A2 -- C1). In this case, all bands are slip bands, as the slip directions are parallel to the bands. It is noteworthy that the pattern formed exhibits an anti-periodic character. The cells and bands that terminate at one part of the domain's edge are of the opposite type to those that terminate at the corresponding part of the opposite edge. The laminate microstructure also contains mirror planes, which bisect the domain along the $[0\bar{1}0]$ and $[00\bar{1}]$ directions, as well as at $\pm 45 \si{\degree}$ along the $[0\bar{1}\bar{1}]$ and $[01\bar{1}]$ directions.
\begin{figure}
	\centering
	\subfloat[]{
		\hspace{-1.3cm}
		\includegraphics[width=0.55\textwidth]{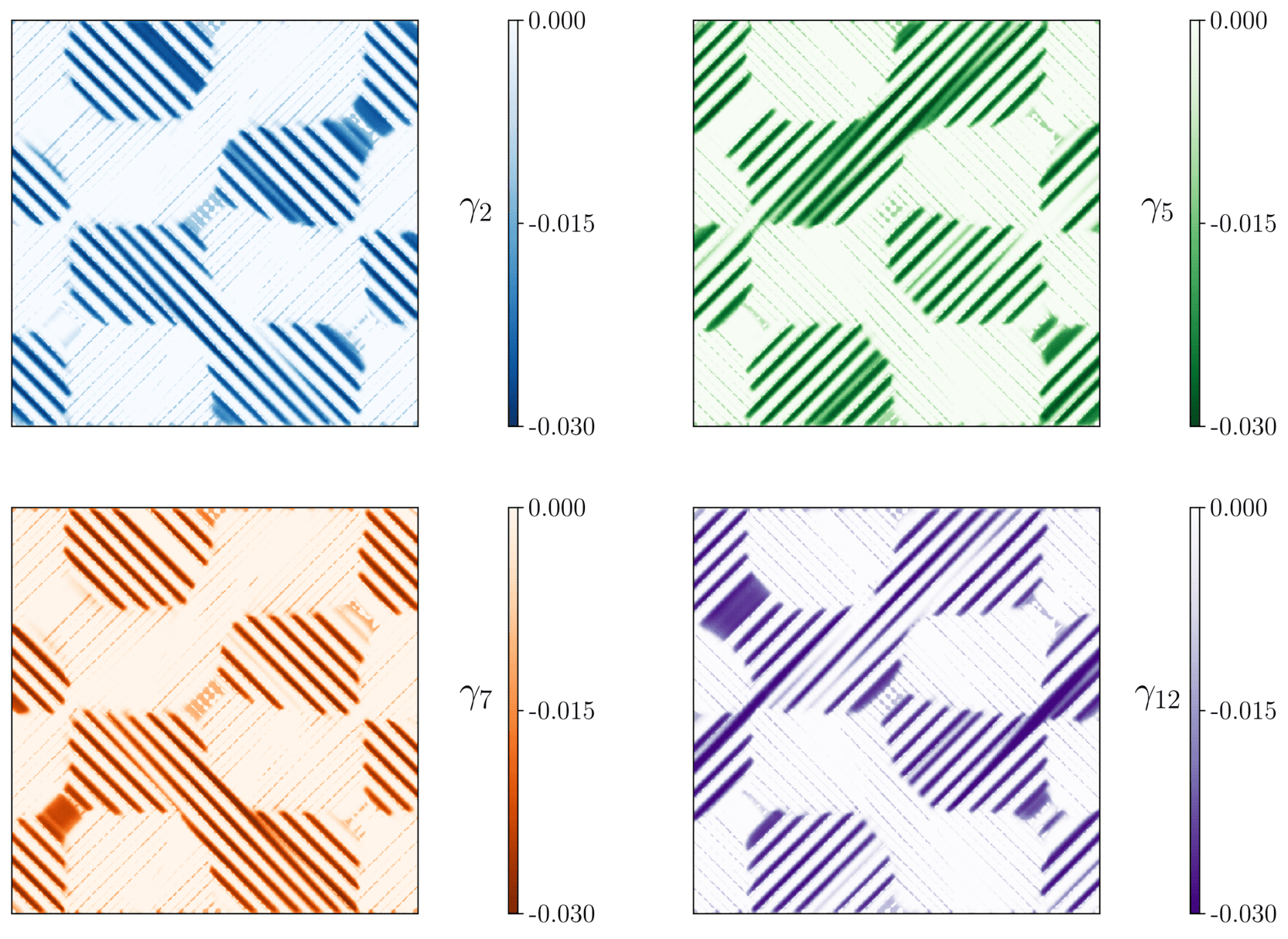}
		\label{subfig:plane_strain_compression_010_001_grid}
	}
	\subfloat[]{
        \raisebox{0cm}{\includegraphics[width=0.13\textwidth]{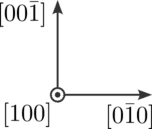}}\hspace{-1.3cm}
        \raisebox{.65cm}{\includegraphics[width=0.3\textwidth, trim=0.115cm 0.15cm 8.5cm 0cm, clip]{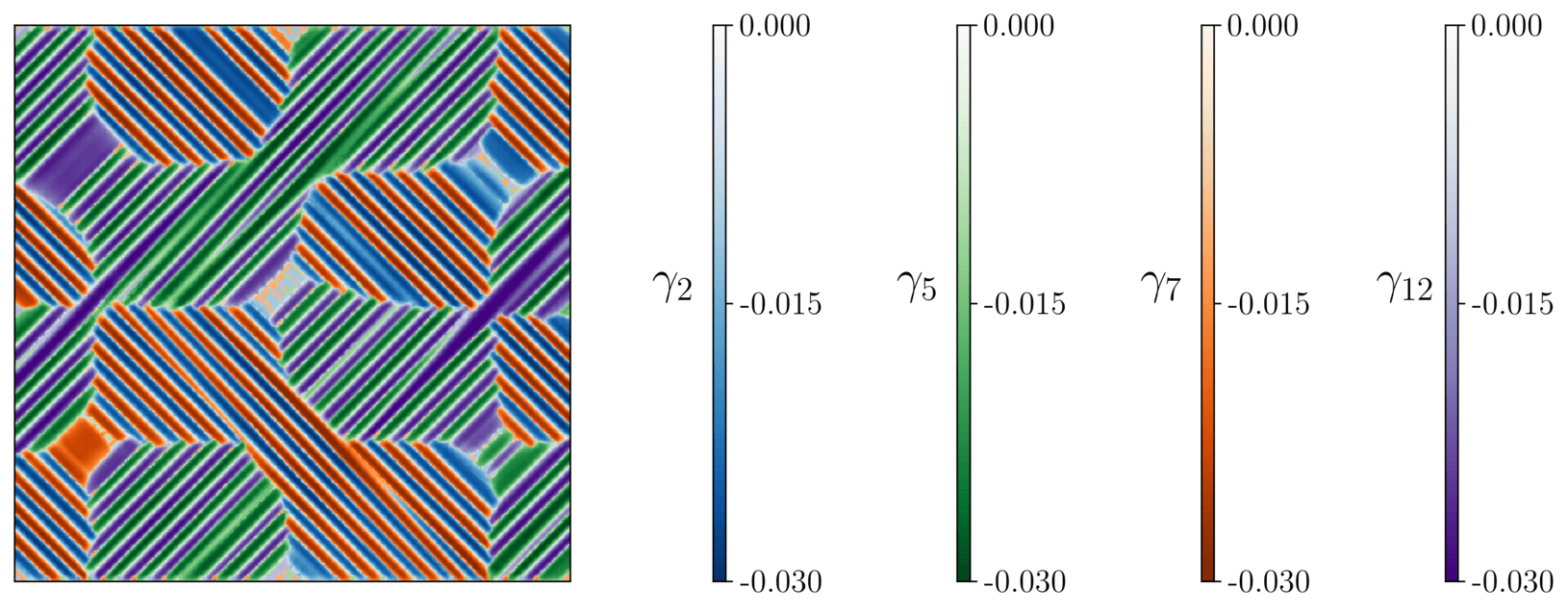}}\hspace{-4.2cm}
		\raisebox{5.9cm}{\includegraphics[width=0.15\textwidth]{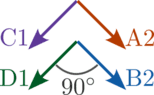}}
		\label{subfig:plane_strain_compression_010_001_superimposed}
	}
	\caption{Plane strain compression along $[00\bar{1}]$ in the $(100)$ plane at 1$\%$ macroscopic strain. (a) Plastic slips on slip systems B2 ($\gamma_2$), D1 ($\gamma_5$), A2 ($\gamma_7$) and C1 ($\gamma_{12}$). (b) Superimposed plastic slips, where the color is selected according to the slip system with maximum slip activity. Arrows indicate the projections of slip directions of active slip systems on the $(100)$ plane.}
	\label{fig:plane_strain_compression_010_001}
\end{figure}

In Figure~\ref{fig:plane_strain_compression_010_rho}, the total dislocation density fields corresponding to the deformation microstructures presented in Figures~\ref{fig:plane_strain_compression_010},~\ref{fig:plane_strain_compression_010_coplanar},~\ref{fig:plane_strain_compression_010_misorientation}, and~\ref{fig:plane_strain_compression_010_001} are shown. The dislocation density microstructures are highly varied, reflecting the complexity of the underlying deformation patterns. Even at just 1\% macroscopic strain, the dislocation density already varies by approximately one order of magnitude across the specimens. For the cases where the single crystal is divided into single-slip regions (Figures~\ref{subfig:plane_strain_compression_010_rho},~\ref{subfig:plane_strain_compression_010_misorientation_rho}, and~\ref{subfig:plane_strain_compression_010_001_rho}), the dislocation density is concentrated at the boundaries between the single-slip regions. These interfaces can thus be described as \textit{dislocation walls}, while the regions of lower dislocation density, bounded by these sub-grain boundaries, can be referred to as \textit{dislocation cells}~\citep{kocks2003physics}. Such features are frequently observed in transmission electron microscopy (TEM) experiments~\citep{jin1984cyclic,feaugas1999origin,ortiz1999nonconvex}, as well as in high-resolution electron backscatter diffraction (EBSD) studies~\citep{wang2023dislocation} and both continuum~\citep{sandfeld2015pattern} and discrete dislocation plasticity simulations~\citep{amodeo1988review,madec2002simulation}. In the case where the coplanar interaction is reduced (Figure~\ref{subfig:plane_strain_compression_010_coplanar_rho}), the dislocation density field exhibits variations with shorter wavelengths than in the other cases. This is consistent with the deformation microstructure described earlier in Figure~\ref{fig:plane_strain_compression_010_coplanar}, where the channel width is approximately the same size as the mesh size. Notably, the labyrinth deformation microstructure is not directly visible in the dislocation density field, though such dislocation patterns have been observed in TEM studies~\citep{ortiz1999nonconvex}. Our results suggest that TEM observations of dislocation structures could be utilized in the future to improve the calibration of the interaction matrix used in crystal plasticity models. While the coefficients of this interaction matrix can be estimated through discrete dislocation dynamics (DDD) simulations~\citep{madec2017dislocation}, experimental characterization of these interactions remains challenging~\citep{scherer2024tensile}. The dislocation patterns presented in this work may provide an indirect method to assess the strength of interactions between dislocations. A comprehensive analysis of the morphologies of dislocation structures as a function of the relative strength of various interactions is beyond the scope of the present study, but will be explored in future research.
\begin{figure}
	\centering
	\subfloat[{Plane strain compression along $[00\bar{1}]$ (Figure~\ref{fig:plane_strain_compression_010})}]{
		\includegraphics[width=0.48\textwidth,trim=0.075cm 0.075cm 0.cm 0cm,clip]{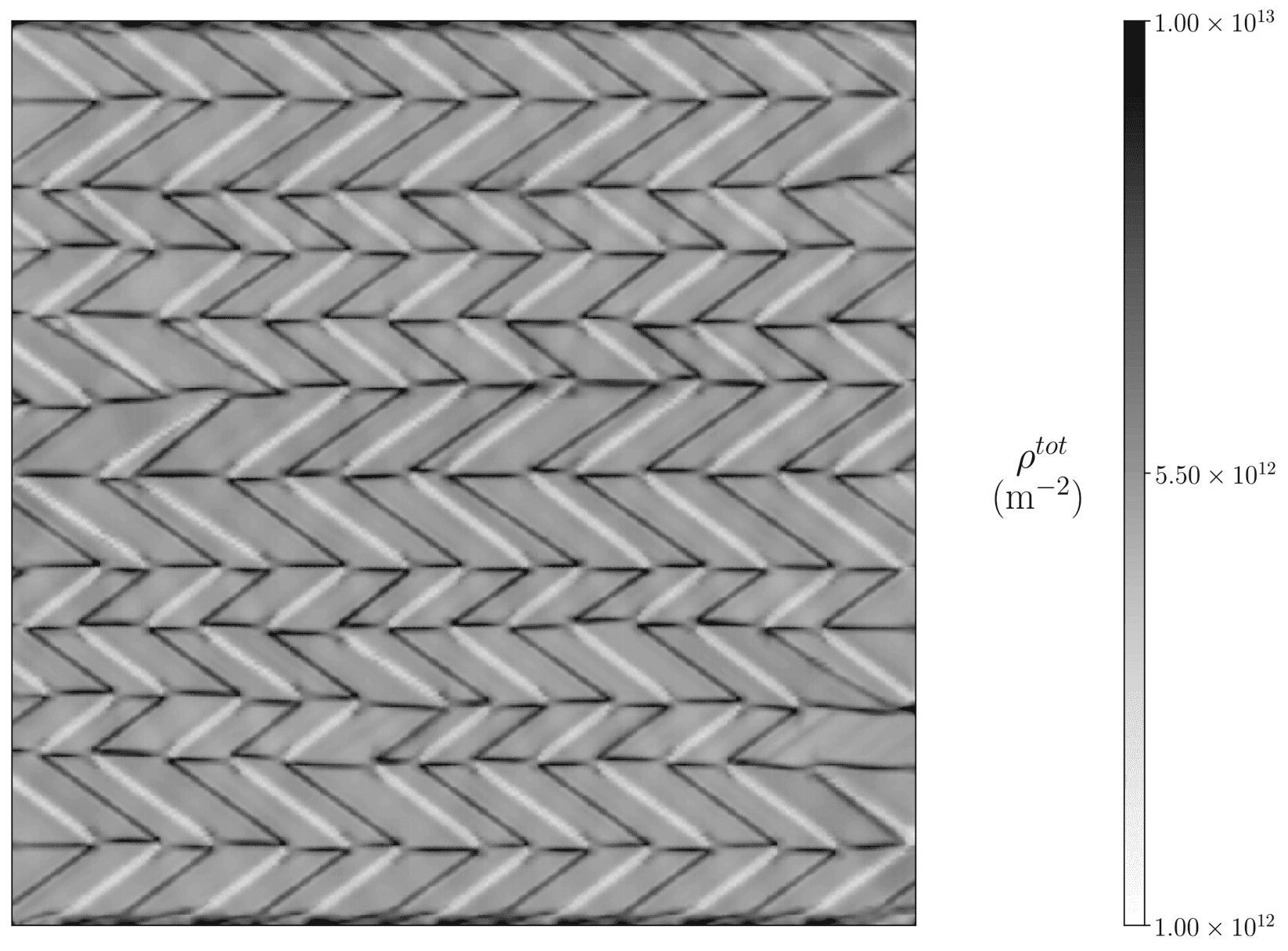}
		\label{subfig:plane_strain_compression_010_rho}
	}\hspace{.0275cm}
	\subfloat[{Lower coplanar interaction $a_{cop}=0.05$ (Figure~\ref{fig:plane_strain_compression_010_coplanar})}]{
		\includegraphics[width=0.48\textwidth,trim=0.075cm 0.075cm 0.cm 0cm,clip]{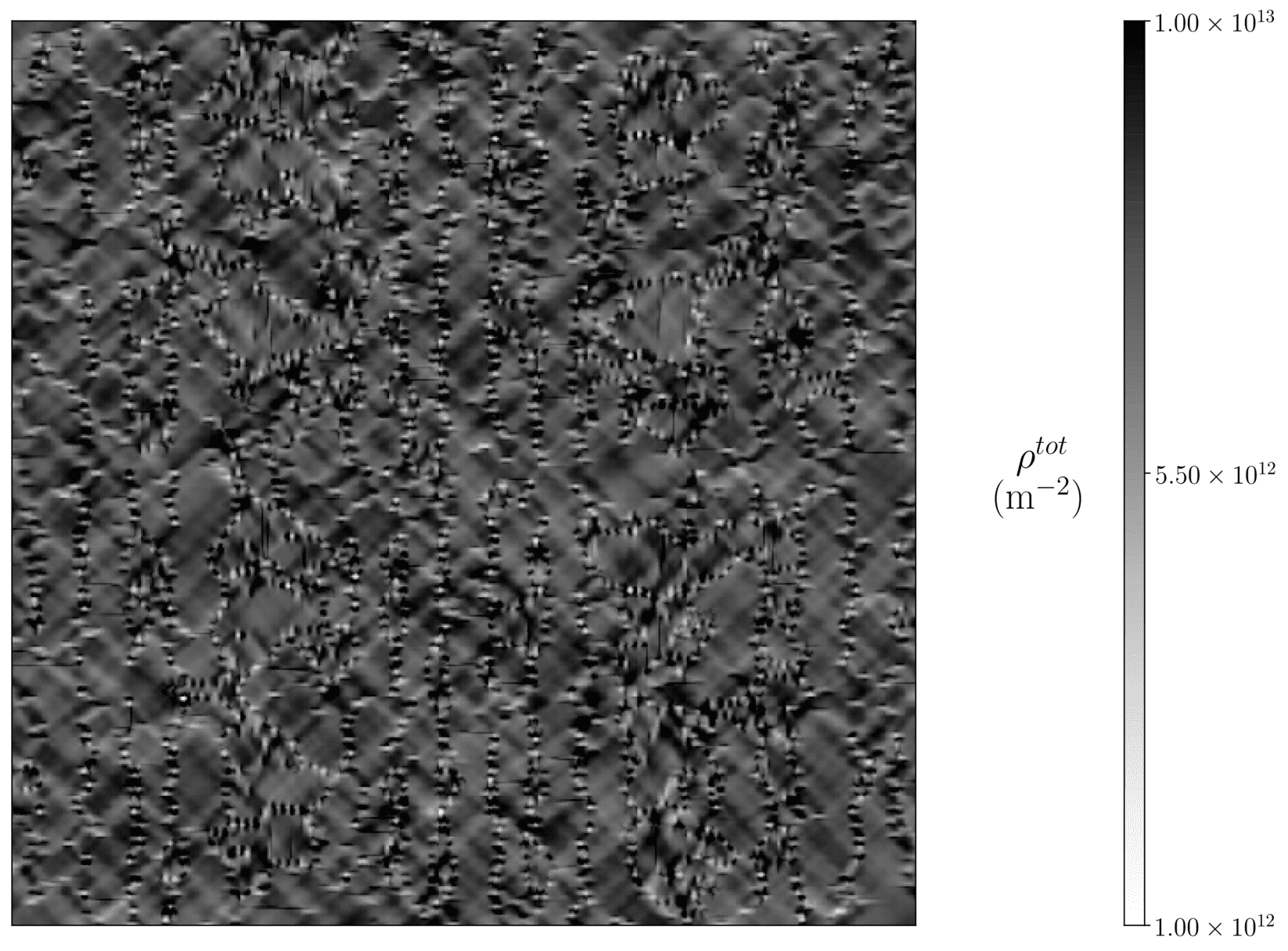}
		\label{subfig:plane_strain_compression_010_coplanar_rho}
	}\\
	\subfloat[{ \SI{4}{\degree} misorientation (Figure~\ref{fig:plane_strain_compression_010_misorientation})}]{
		\includegraphics[width=0.48\textwidth,trim=0.075cm 0.075cm 0.cm 0cm,clip]{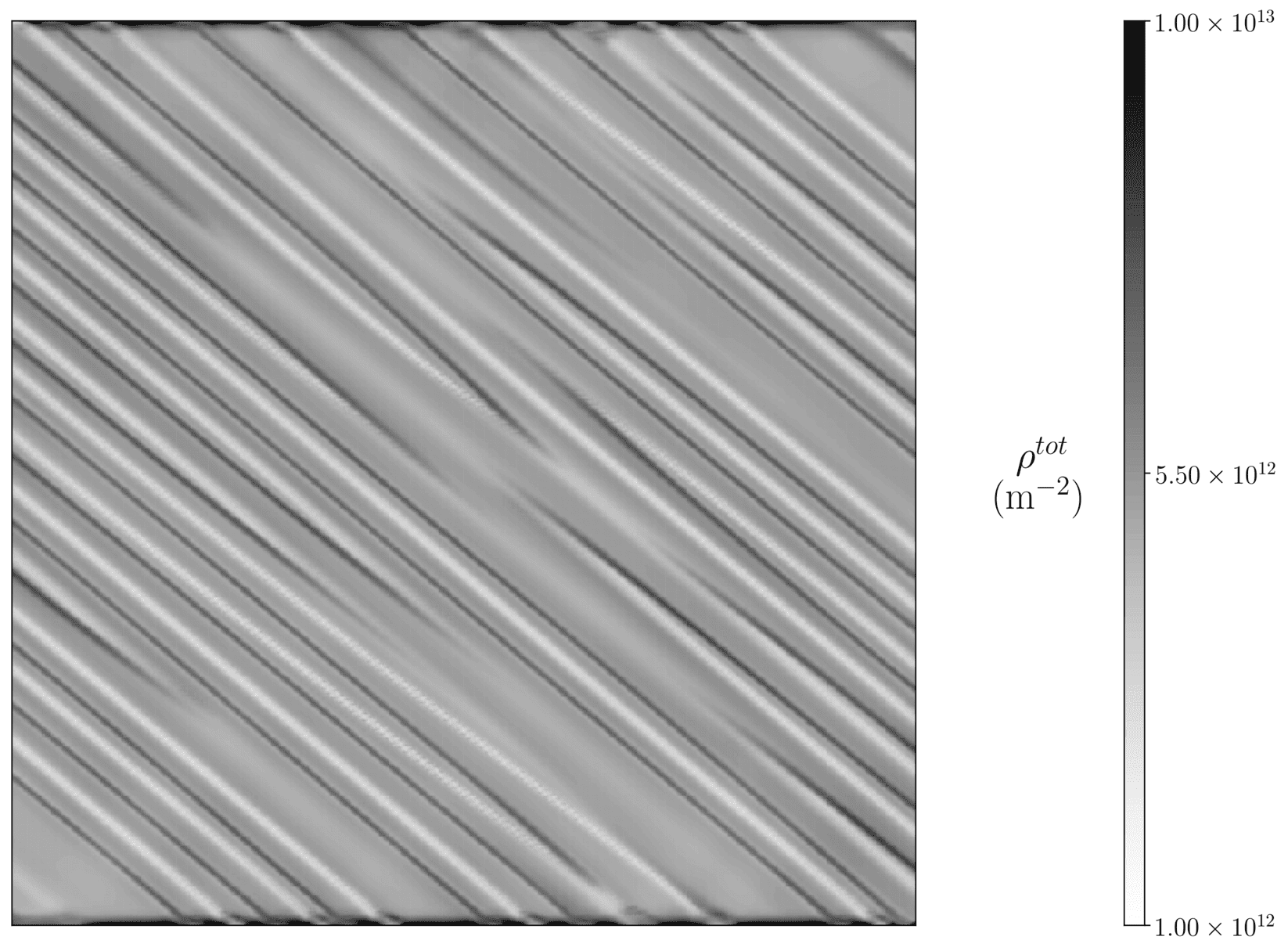}
		\label{subfig:plane_strain_compression_010_misorientation_rho}
	}\hspace{.0275cm}
	\subfloat[{Compression along $[00\bar{1}]$ in the $(100)$ plane (Figure~\ref{fig:plane_strain_compression_010_001})}]{
		\includegraphics[width=0.48\textwidth,trim=0.075cm 0.075cm 0.cm 0cm,clip]{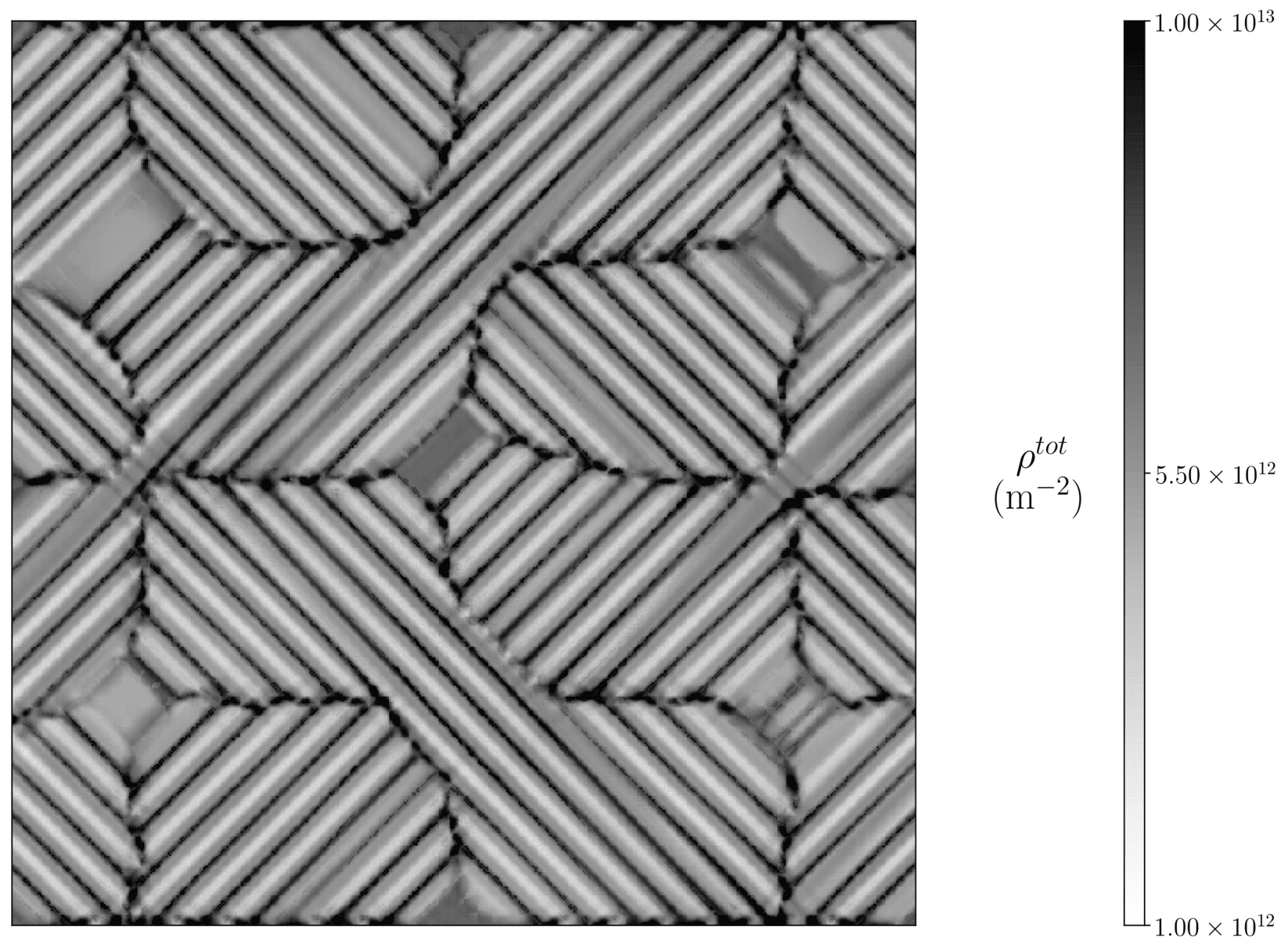}
		\label{subfig:plane_strain_compression_010_001_rho}
	}
	\caption{Total dislocation density at 1$\%$ macroscopic strain corresponding to the deformation microstructures shown in Figures~\ref{fig:plane_strain_compression_010},~\ref{fig:plane_strain_compression_010_coplanar},~\ref{fig:plane_strain_compression_010_misorientation},~\ref{fig:plane_strain_compression_010_001}.}
	\label{fig:plane_strain_compression_010_rho}
\end{figure}

\subsubsection*{Sensitivity to numerical parameters}
We now investigate the sensitivity of the deformation patterns to the type of finite elements used and the mesh size. For the following simulations, the material parameters are those listed in Table~\ref{tab:material_parameters}, and the crystal orientation is \textcolor{black}{such that} $x_1 = [1\bar{1}0]$, $x_2 = [00\bar{1}]$, and $x_3 = [110]$.

In~\cite{rys2024spontaneous}, 9-node quadrilateral finite elements with biquadratic shape functions were used, employing a full integration scheme with $3 \times 3$ Gauss quadrature points. They chose quadratic elements to avoid issues like hourglass or shear locking phenomena that can occur with lower-order elements. For the simulations in Figure~\ref{fig:plane_strain_compression_010}, we use fully integrated, quadratic, hexahedral finite elements (C3D20) with $3 \times 3 \times 3$ Gauss points. In this section, we instead employ quadratic hexahedral elements with reduced integration (C3D20r) and $2 \times 2 \times 2$ Gauss points. The deformation pattern obtained at 1\% macroscopic strain, shown in Figure~\ref{fig:plane_strain_compression_010_reduced}, reveals a substantial difference from the one obtained using full integration elements. The plastic activity is divided into four distinct domains, with a single slip system active in each domain. However, the resulting deformation microstructure exhibits pronounced oscillations, with a wavelength corresponding to the distance between the Gauss points. This laminate-like microstructure consists of vertical bands, which are further subdivided into thin horizontal layers, each of a thickness corresponding to one Gauss point. These layers exhibit single slip on slip systems D4 and D1, or A2 and A3. The deformation pattern is, therefore, highly sensitive to the mesh size when hexahedral elements with reduced integration are used. This observation strongly suggests that a material length scale should be introduced into the model to regularize the characteristic size of the deformation microstructure. To address this issue, a strain gradient crystal plasticity regularization based on the model presented in Section~\ref{subsec:strain_gradient} will be employed in Section~\ref{sec:strain_gradient}.
\begin{figure}
	\centering
	\subfloat[]{
		\hspace{-1.3cm}
		\includegraphics[width=0.55\textwidth]{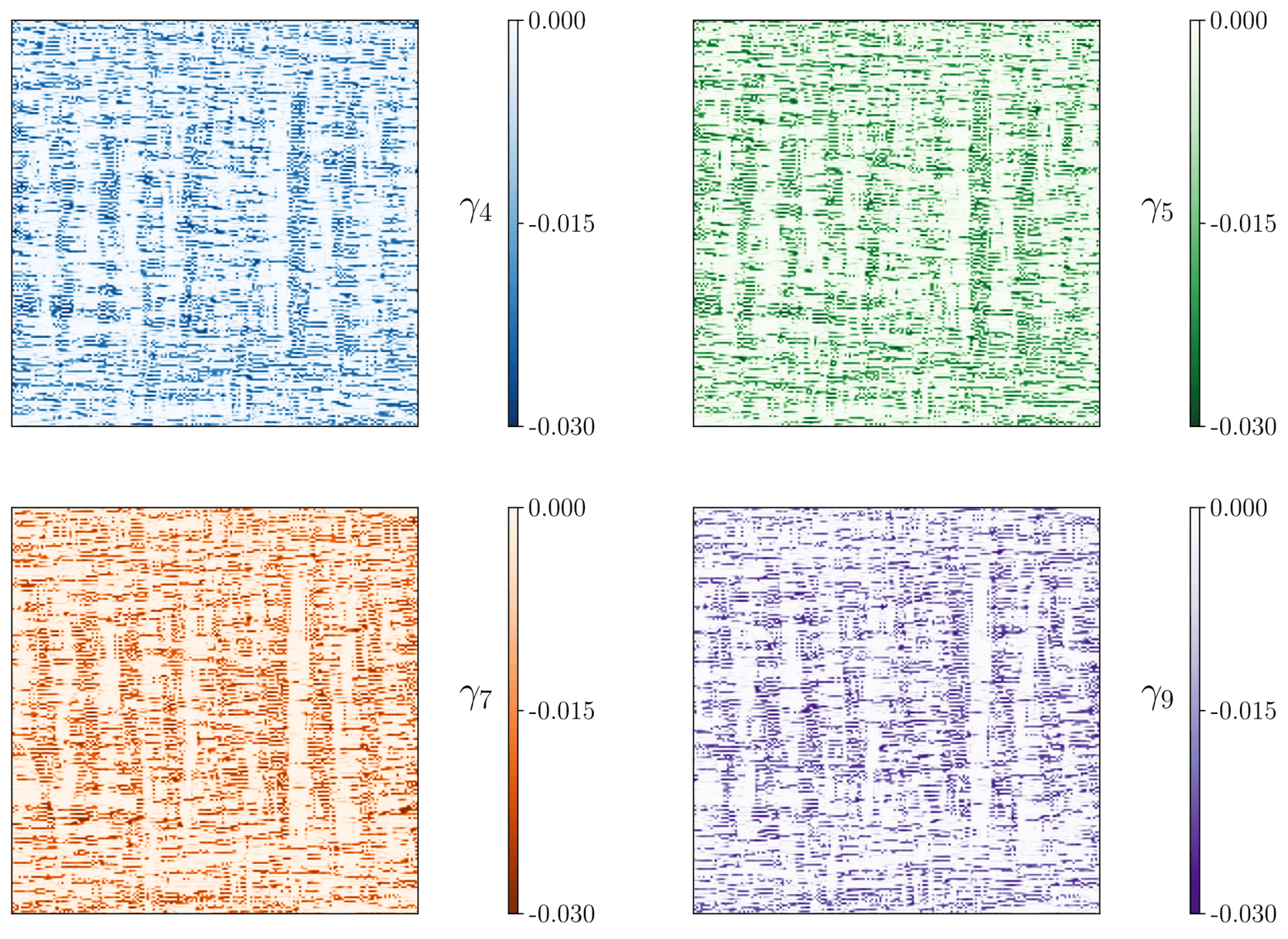}
		\label{subfig:plane_strain_compression_010_reduced_grid}
	}
	\subfloat[]{
        \raisebox{0cm}{\includegraphics[width=0.13\textwidth]{FiguresReduced_axes.png}}\hspace{-1.3cm}
        \raisebox{.65cm}{\includegraphics[width=0.3\textwidth, trim=0.115cm 0.15cm 8.5cm 0cm, clip]{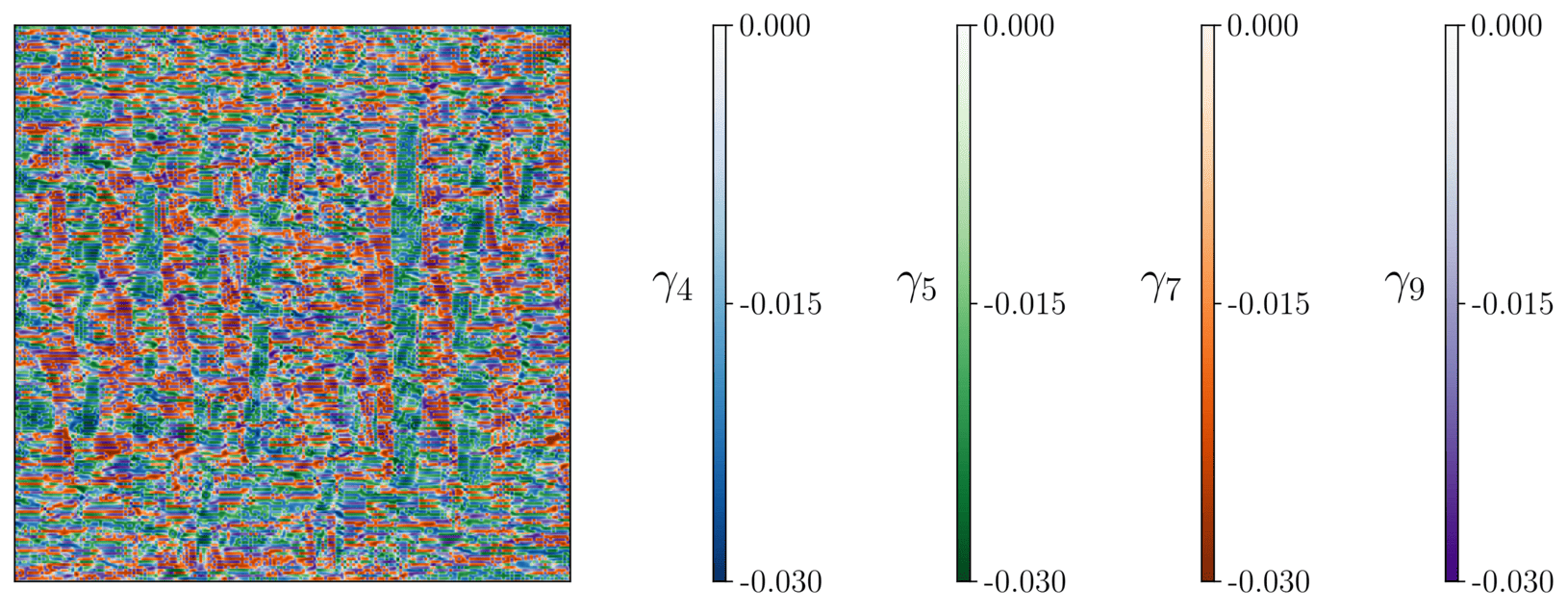}}\hspace{-4.2cm}
		\raisebox{5.9cm}{\includegraphics[width=0.15\textwidth]{FiguresReduced_slip_systems.png}}
		\label{subfig:plane_strain_compression_010_reduced_superimposed}
	}
	\caption{Plane strain compression along $[00\bar{1}]$ in the $(110)$ plane at 1$\%$ macroscopic strain with reduced quadratic hexahedral finite elements (C3D20r). (a) Plastic slips on slip systems D4 ($\gamma_4$), D1 ($\gamma_5$), A2 ($\gamma_7$) and A3 ($\gamma_9$). (b) Superimposed plastic slips, where the color is selected according to the slip system with maximum slip activity. Arrows indicate the projections of slip directions of active slip systems on the $(110)$ plane.}
	\label{fig:plane_strain_compression_010_reduced}
\end{figure}

We now consider finite elements with reduced integration, using quadratic tetrahedral elements (C3D10\_4) instead of quadratic hexahedral elements (C3D20r). The deformation pattern obtained at 1\% macroscopic strain is shown in Figure~\ref{fig:plane_strain_compression_010_tetra}. The pattern predicted with tetrahedral elements is similar to that obtained with hexahedral elements with full integration (see Figures~\ref{fig:plane_strain_compression_010} and~\ref{fig:plane_strain_compression_010_redo}). However, the sizes of the single-slip regions show greater variability with tetrahedral elements than with hexahedral elements. Some single-slip regions occupy a large fraction of the domain, while others have sizes comparable to those observed with fully integrated hexahedral elements. Importantly, the reduced integration scheme with tetrahedral elements does not lead to the strong oscillations observed with reduced hexahedral elements (see Figure~\ref{fig:plane_strain_compression_010_reduced}).
\begin{figure}
	\centering
	\subfloat[]{
		\hspace{-1.3cm}
		\includegraphics[width=0.55\textwidth]{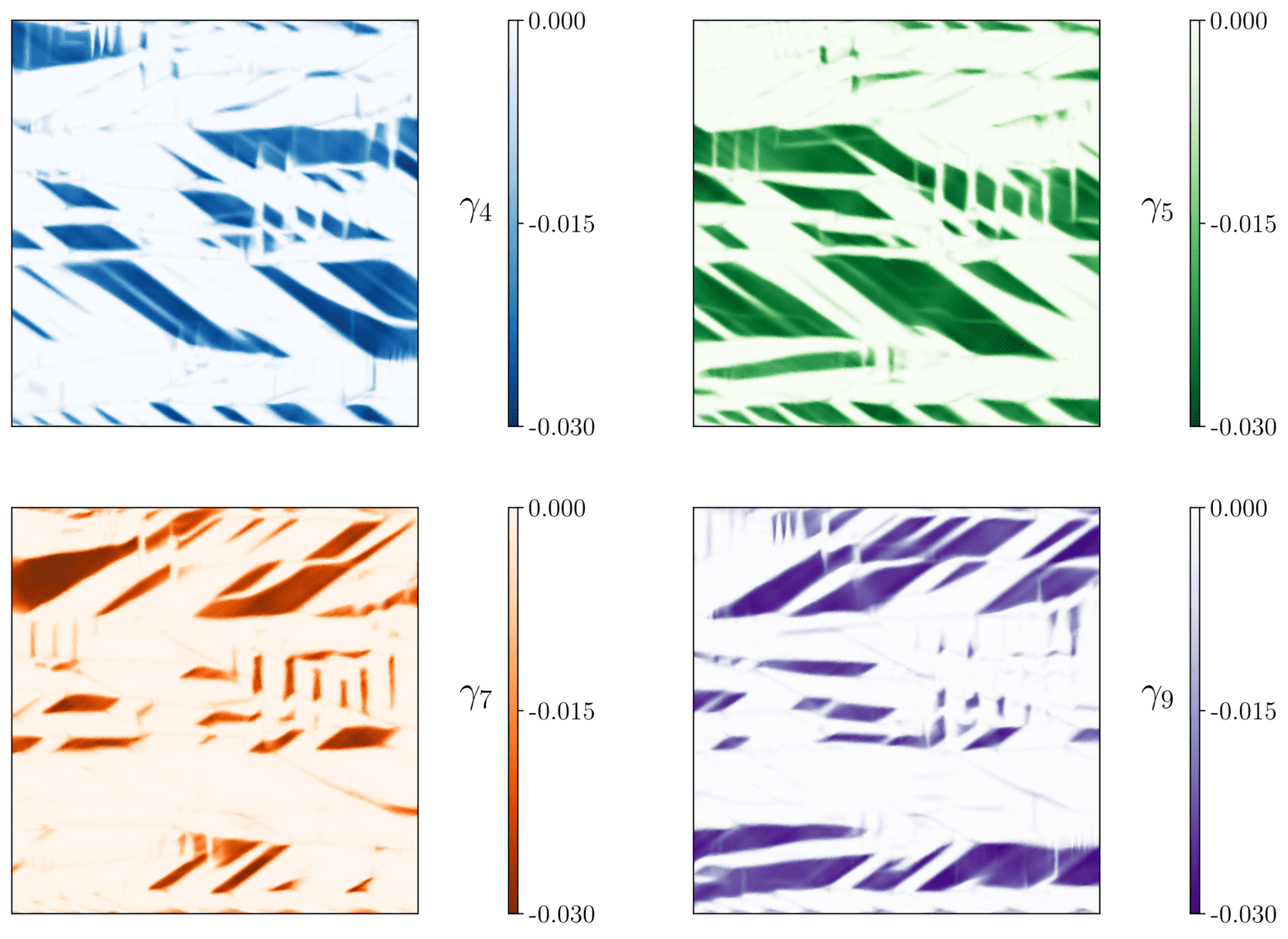}
		\label{subfig:plane_strain_compression_010_tetra_grid}
	}
	\subfloat[]{
        \raisebox{0cm}{\includegraphics[width=0.13\textwidth]{FiguresReduced_axes.png}}\hspace{-1.3cm}
        \raisebox{.65cm}{\includegraphics[width=0.3\textwidth, trim=0.115cm 0.15cm 8.5cm 0cm, clip]{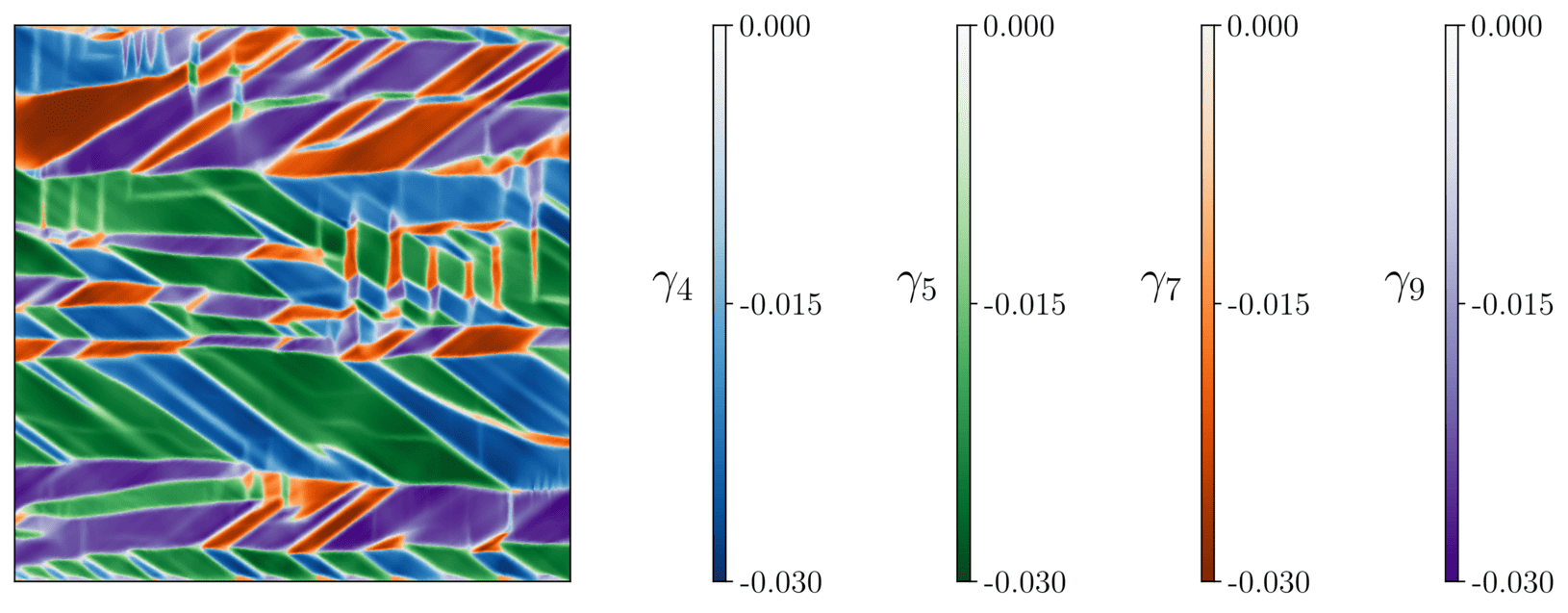}}\hspace{-4.2cm}
		\raisebox{5.9cm}{\includegraphics[width=0.15\textwidth]{FiguresReduced_slip_systems.png}}
		\label{subfig:plane_strain_compression_010_tetra_superimposed}
	}
	\caption{Plane strain compression along $[00\bar{1}]$ in the $(110)$ plane at 1$\%$ macroscopic strain with reduced quadratic tetrahedral finite elements (C3D10\_4). (a) Plastic slips on slip systems D4 ($\gamma_4$), D1 ($\gamma_5$), A2 ($\gamma_7$) and A3 ($\gamma_9$). (b) Superimposed plastic slips, where the color is selected according to the slip system with maximum slip activity. Arrows indicate the projections of slip directions of active slip systems on the $(110)$ plane.}
	\label{fig:plane_strain_compression_010_tetra}
\end{figure}

It is well known that volumetric locking can occur in finite element simulations of quasi-incompressible materials when using standard variational formulations of equilibrium~\citep{brezzi2012mixed}. The issue arises because the usual finite element spaces are not rich enough to satisfy the discrete incompressibility constraint, \textit{i.e.}, $\boldsymbol{\nabla} \cdot \boldsymbol{u}^h = 0$, where $\boldsymbol{u}^h$ is the discrete displacement field. This leads to spurious oscillations in the hydrostatic pressure field. To mitigate this, pressure-controlled integration schemes have been proposed~\citep{simo1985variational, chen2022simulation}. In crystal plasticity, quasi-incompressibility progressively develops as the plastic strain increases, since plastic slip is purely deviatoric. To prevent oscillations in the hydrostatic pressure field observed in previous simulations (not shown), we employ a pressure-controlled integration scheme here. This method treats the hydrostatic pressure and associated volume change as additional degrees of freedom. The pressure is computed as a Lagrange multiplier enforcing the incompressibility constraint. Quadratic interpolation is used for the displacement field, while linear interpolation is applied for the pressure and volume variation, as described in~\cite{chen2022simulation}. Hexahedral elements with a full integration scheme are used. The deformation pattern obtained at 1\% macroscopic strain is shown in Figure~\ref{fig:plane_strain_compression_010_gen3f}. The resulting microstructure is similar to that obtained without pressure control (see Figures~\ref{fig:plane_strain_compression_010} and~\ref{fig:plane_strain_compression_010_redo}). While the hydrostatic pressure field is smooth in this case (not shown), it does not significantly affect the pattern. Specifically, the sizes, shapes, and orientations of the single-slip regions remain the same as those observed with standard finite elements without pressure control.
\begin{figure}
	\centering
	\subfloat[]{
		\hspace{-1.3cm}
		\includegraphics[width=0.55\textwidth]{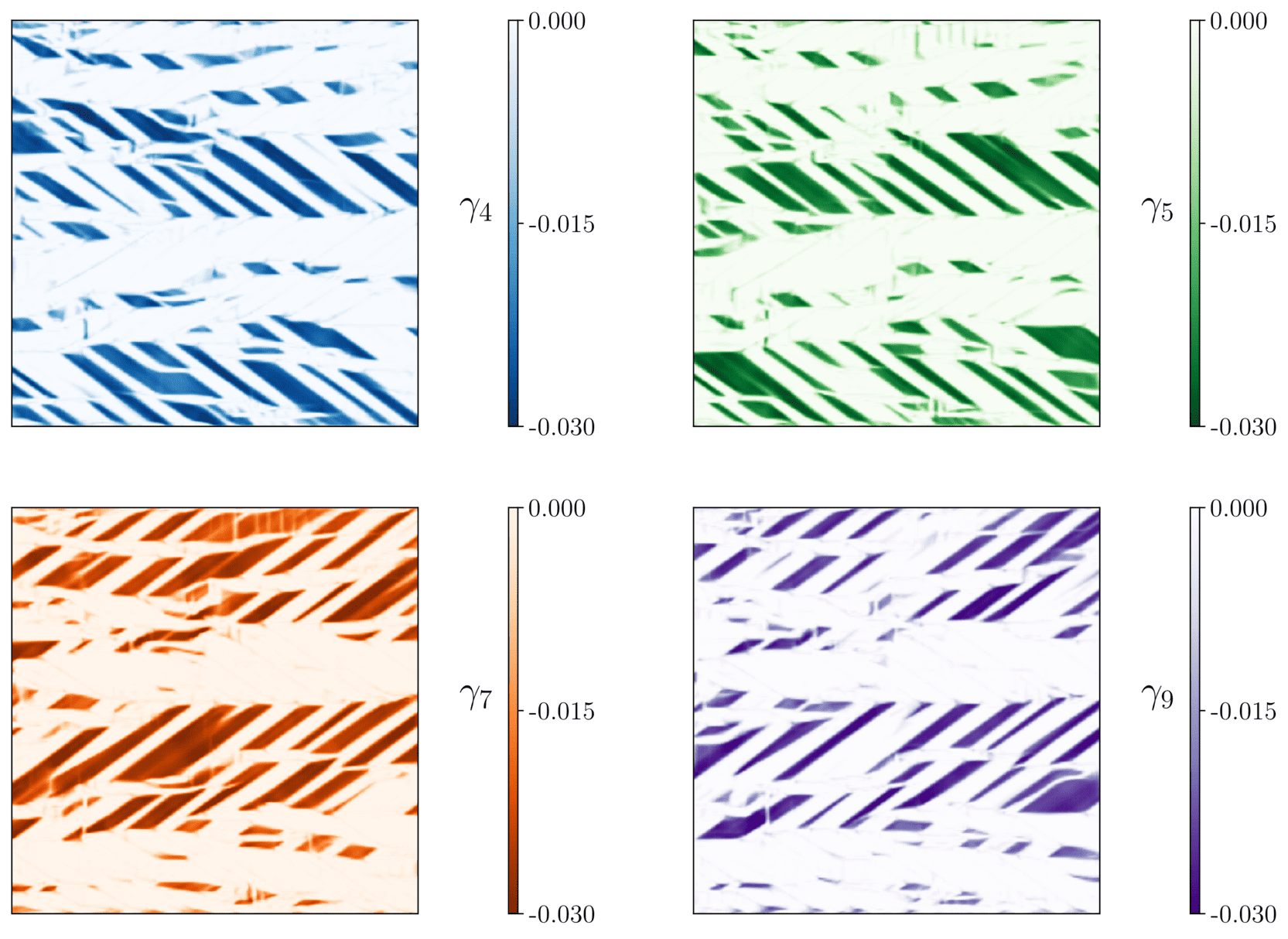}
		\label{subfig:plane_strain_compression_010_gen3f_grid}
	}
	\subfloat[]{
        \raisebox{0cm}{\includegraphics[width=0.13\textwidth]{FiguresReduced_axes.png}}\hspace{-1.3cm}
        \raisebox{.65cm}{\includegraphics[width=0.3\textwidth, trim=0.115cm 0.15cm 8.5cm 0cm, clip]{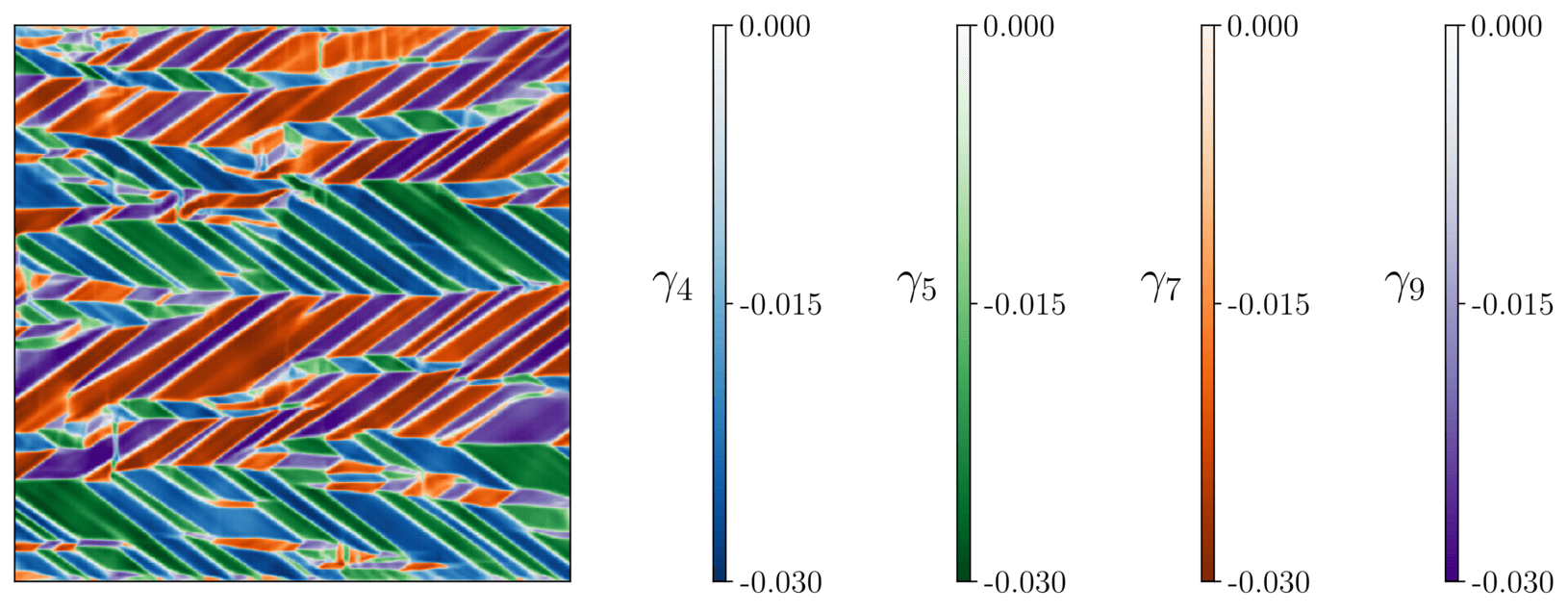}}\hspace{-4.2cm}
		\raisebox{5.9cm}{\includegraphics[width=0.15\textwidth]{FiguresReduced_slip_systems.png}}
		\label{subfig:plane_strain_compression_010_gen3f_superimposed}
	}
	\caption{Plane strain compression along $[00\bar{1}]$ in the $(110)$ plane at 1$\%$ macroscopic strain with pressure-controlled integration. 
	(a) Plastic slips on slip systems D4 ($\gamma_4$), D1 ($\gamma_5$), A2 ($\gamma_7$) and A3 ($\gamma_9$). (b) Superimposed plastic slips, where the color is selected according to the slip system with maximum slip activity. Arrows indicate the projections of slip directions of active slip systems on the $(110)$ plane.}
	\label{fig:plane_strain_compression_010_gen3f}
\end{figure}

Instead of the $200 \times 200 \times 1$ hexahedral mesh \textcolor{black}{used previously}, we now use a coarser mesh with $100 \times 100 \times 1$ full integration elements (C3D20). The deformation pattern obtained at 1\% macroscopic strain is shown in Figure~\ref{fig:plane_strain_compression_010_100x100}. The average size of the single-slip regions is significantly larger compared to the $200 \times 200 \times 1$ mesh. This is consistent with the observations made by Ry{\'s} \textit{et al.}~\cite{rys2024spontaneous} for discretizations ranging from $10 \times 10$ to $160 \times 160$ elements. This mesh size dependence arises because the model does not incorporate a material length scale, so the characteristic size of the deformation microstructure is dictated by extrinsic parameters such as mesh size, domain size, or boundary conditions. While Ry{\'s} \textit{et al.}~\cite{rys2024spontaneous} observed only vertical and horizontal deformation bands in their simulations, independent of mesh size, we show here that the orientation of the bands changes as the mesh becomes coarser. In Figure~\ref{fig:plane_strain_compression_010_100x100}, some deformation bands are vertical or horizontal, while others are inclined. In contrast, with the $200 \times 200 \times 1$ mesh, all deformation bands were inclined perpendicular to their corresponding slip direction (see Figure~\ref{fig:plane_strain_compression_010}). Thus, the mesh size not only affects the size of the single-slip deformation bands but also their orientation.
\begin{figure}
	\centering
	\subfloat[]{
		\hspace{-1.3cm}
		\includegraphics[width=0.55\textwidth]{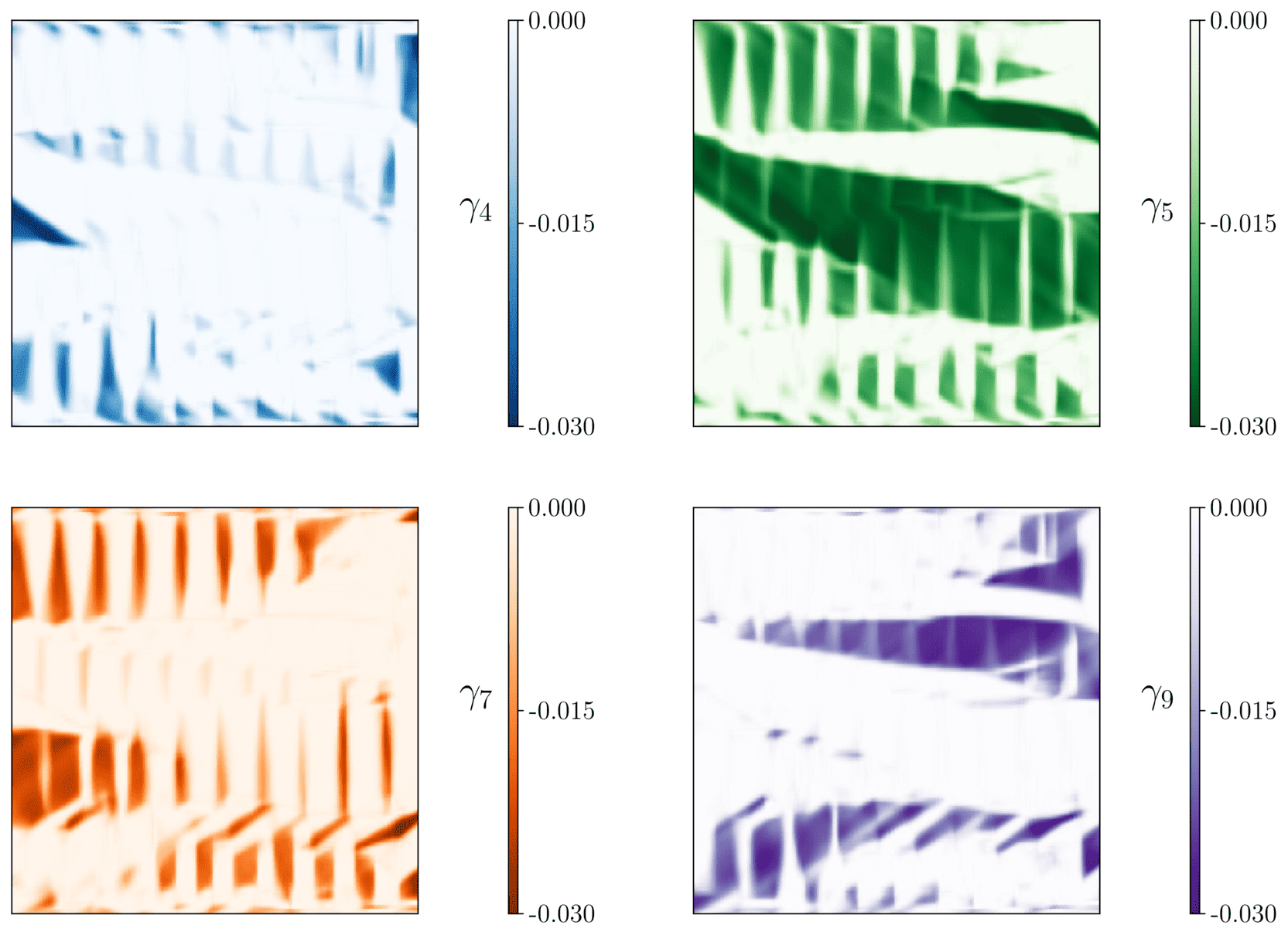}
		\label{subfig:plane_strain_compression_010_100x100_grid}
	}
	\subfloat[]{
        \raisebox{0cm}{\includegraphics[width=0.13\textwidth]{FiguresReduced_axes.png}}\hspace{-1.3cm}
        \raisebox{.65cm}{\includegraphics[width=0.3\textwidth, trim=0.115cm 0.15cm 8.5cm 0cm, clip]{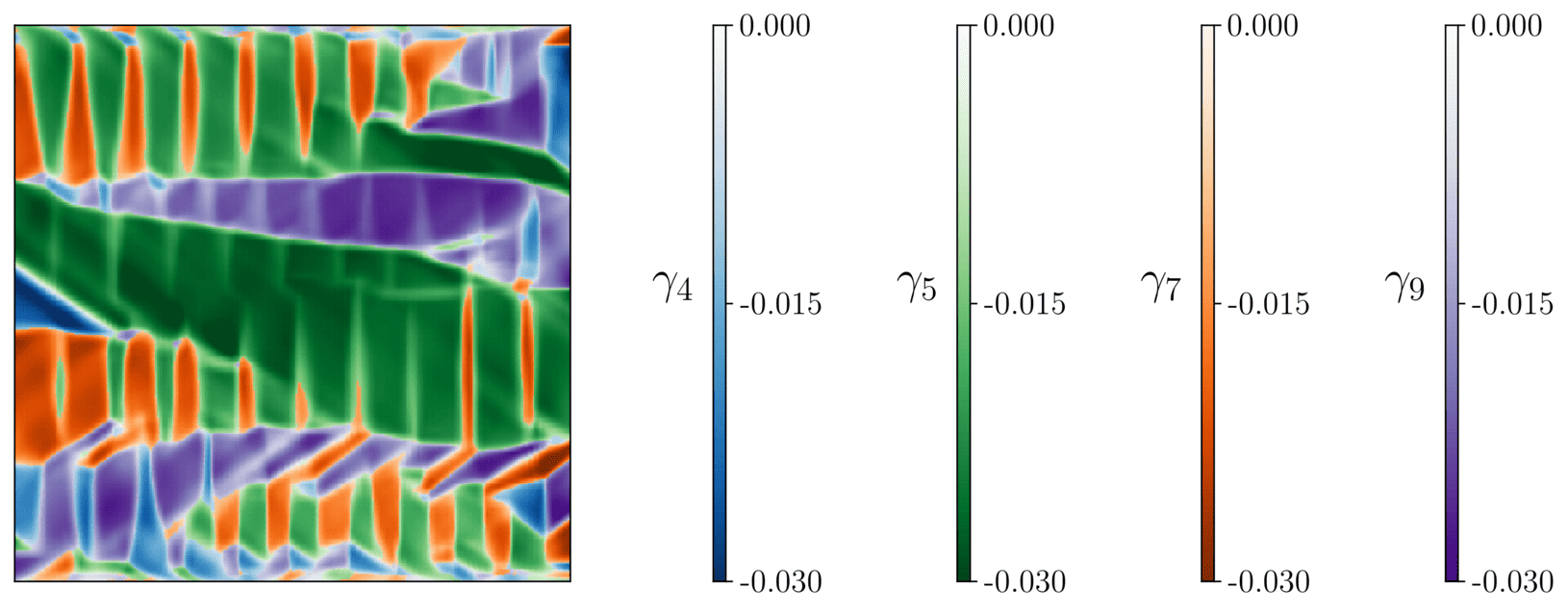}}\hspace{-4.2cm}
		\raisebox{5.9cm}{\includegraphics[width=0.15\textwidth]{FiguresReduced_slip_systems.png}}
		\label{subfig:plane_strain_compression_010_100x100_superimposed}
	}
	\caption{Plane strain compression along $[00\bar{1}]$ in the $(110)$ plane at 1$\%$ macroscopic strain with a $100\times 100\times 1$ finite element mesh. (a) Plastic slips on slip systems D4 ($\gamma_4$), D1 ($\gamma_5$), A2 ($\gamma_7$) and A3 ($\gamma_9$). (b) Superimposed plastic slips, where the color is selected according to the slip system with maximum slip activity. Arrows indicate the projections of slip directions of active slip systems on the $(110)$ plane.}
	\label{fig:plane_strain_compression_010_100x100}
\end{figure}

The total dislocation density fields corresponding to the deformation microstructures shown in Figures~\ref{fig:plane_strain_compression_010_reduced},~\ref{fig:plane_strain_compression_010_tetra},~\ref{fig:plane_strain_compression_010_gen3f}, and~\ref{fig:plane_strain_compression_010_100x100} are presented in Figure~\ref{fig:plane_strain_compression_010_rho_num}. The dislocation density field obtained with reduced integration elements (Figure~\ref{subfig:plane_strain_compression_010_reduced_rho}) exhibits oscillations with a wavelength corresponding to the distance between the Gauss points. The dislocation density pattern obtained with tetrahedral or pressure-controlled finite elements, shown in Figures~\ref{subfig:plane_strain_compression_010_tetra_rho} and~\ref{subfig:plane_strain_compression_010_gen3f_rho}, is similar to that obtained with full integration quadratic hexahedral elements (Figure~\ref{subfig:plane_strain_compression_010_rho}). Specifically, the dislocation cell walls are straight, and their shape follows the boundaries of the single-slip regions. In contrast, the dislocation cell walls observed with a coarser mesh (Figure~\ref{subfig:plane_strain_compression_010_100x100_rho}) are much thicker, though their shape still follows the boundary of the single-slip deformation bands.
\begin{figure}
	\centering
	\subfloat[{Reduced integration quadratic finite\\ elements (C3D20r) (Figure~\ref{fig:plane_strain_compression_010_reduced})}]{
		\includegraphics[width=0.48\textwidth,trim=0.075cm 0.075cm 0.cm 0cm,clip]{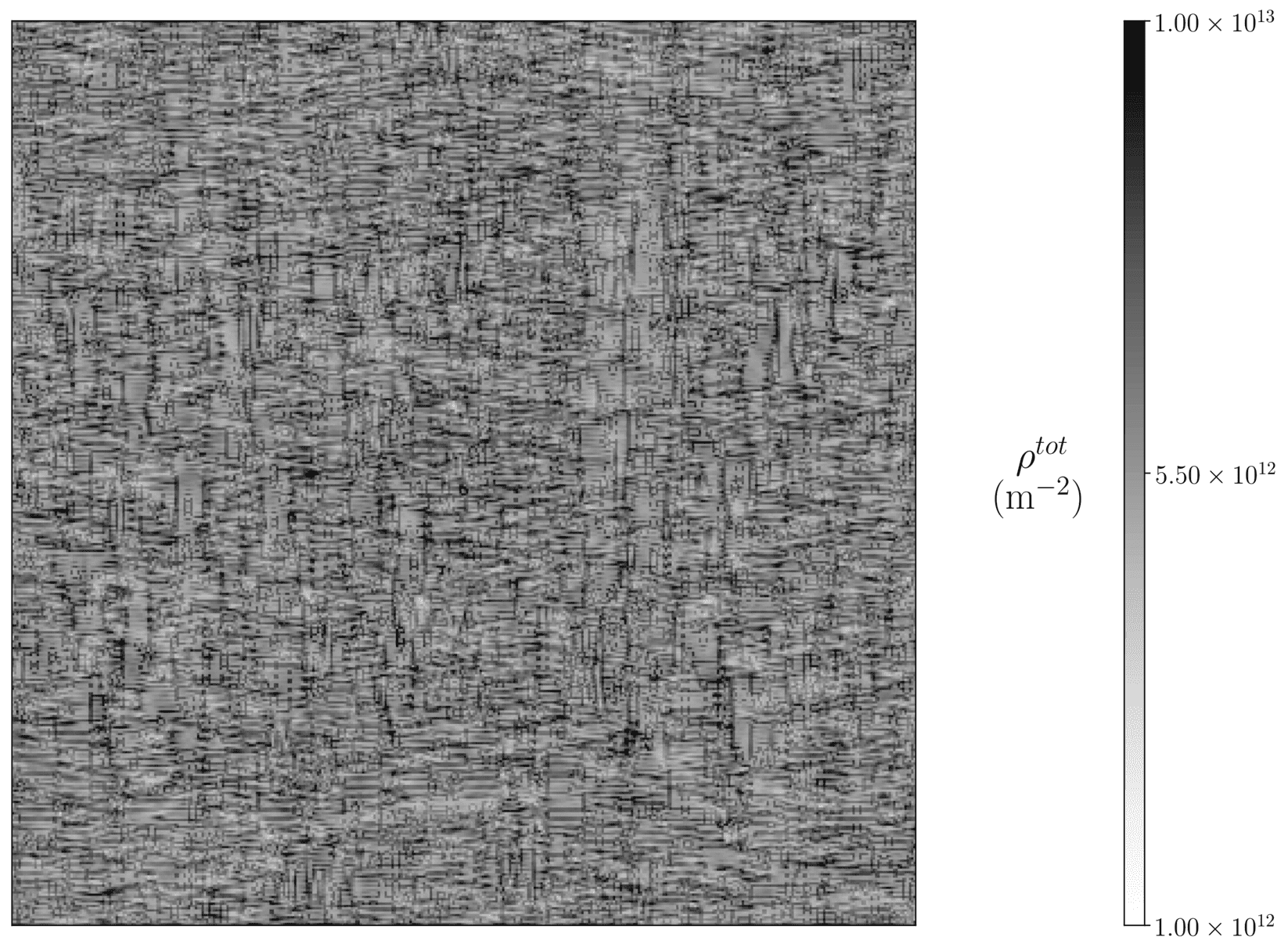}
		\label{subfig:plane_strain_compression_010_reduced_rho}
	}
	\subfloat[{Quadratic tetrahedral finite\\ elements (C3D10\_4) (Figure~\ref{fig:plane_strain_compression_010_tetra})}]{
		\includegraphics[width=0.48\textwidth,trim=0.075cm 0.075cm 0.cm 0cm,clip]{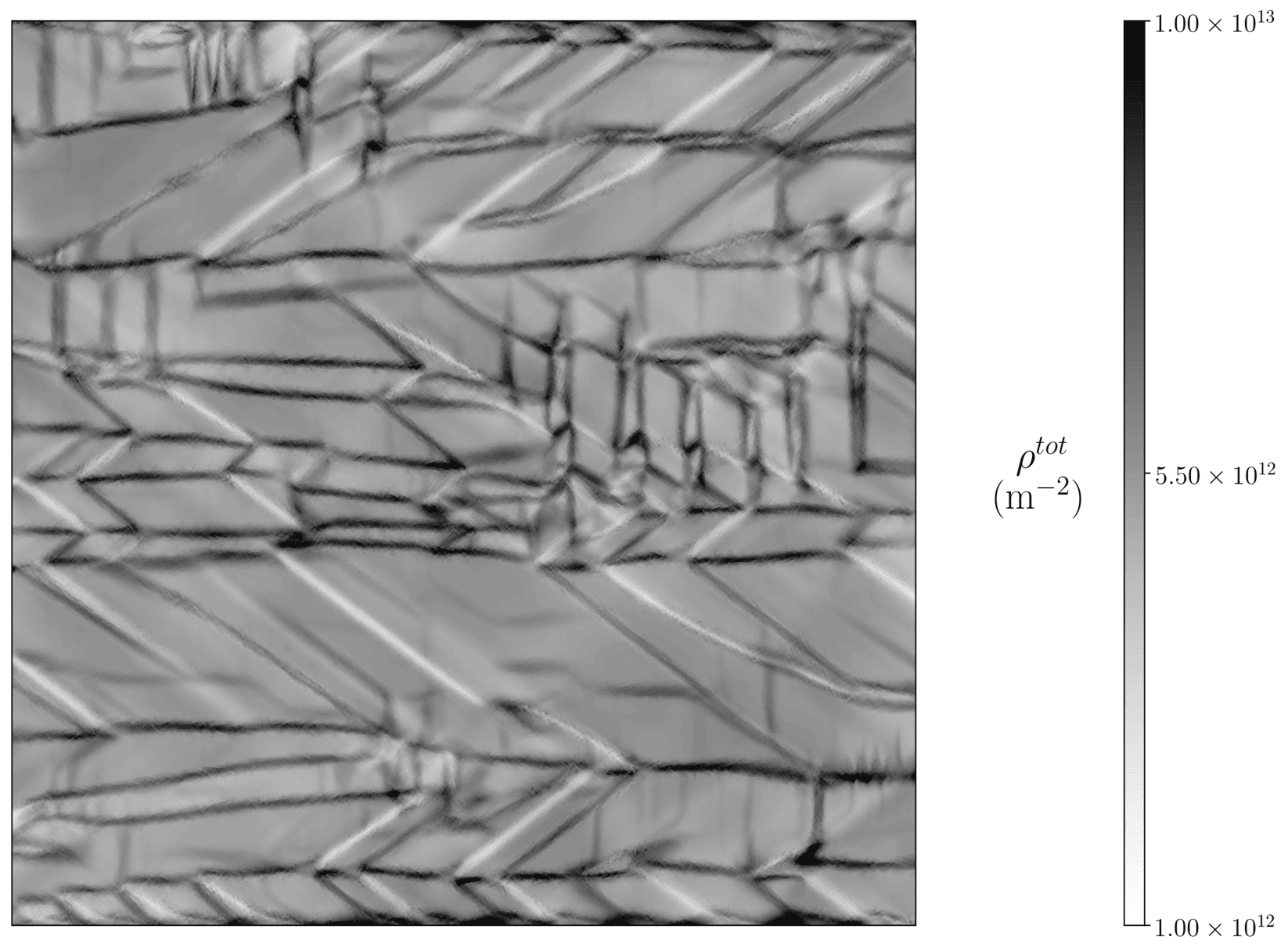}
		\label{subfig:plane_strain_compression_010_tetra_rho}
	}\\
	\subfloat[{Pressure-controlled integration\\ (Figure~\ref{fig:plane_strain_compression_010_gen3f})}]{
		\includegraphics[width=0.48\textwidth,trim=0.075cm 0.075cm 0.cm 0cm,clip]{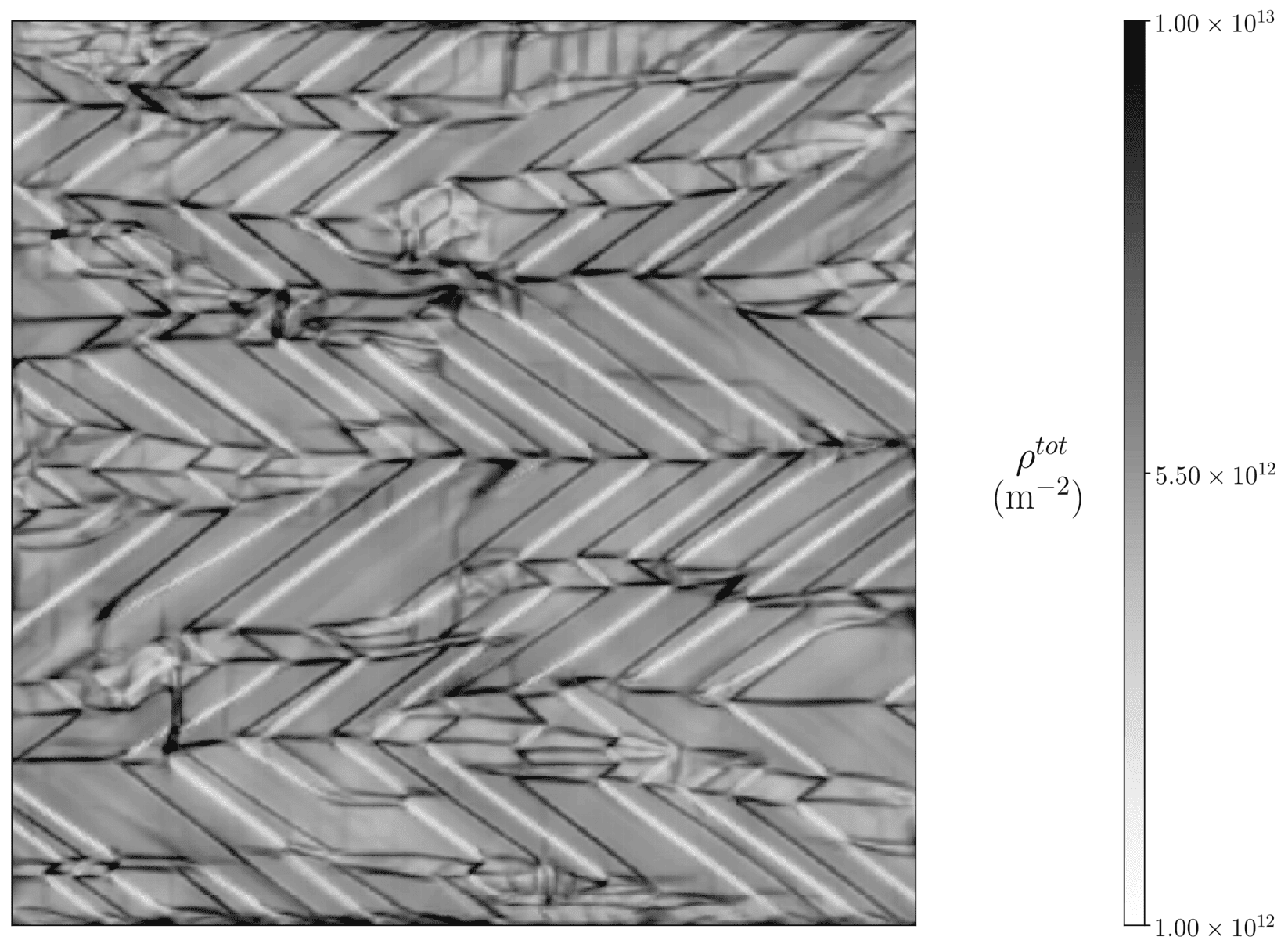}
		\label{subfig:plane_strain_compression_010_gen3f_rho}
	}
	\subfloat[{$100\times100$ hexahedral finite\\ element mesh (C3D20) (Figure~\ref{fig:plane_strain_compression_010_100x100})}]{
		\includegraphics[width=0.48\textwidth,trim=0.075cm 0.075cm 0.cm 0cm,clip]{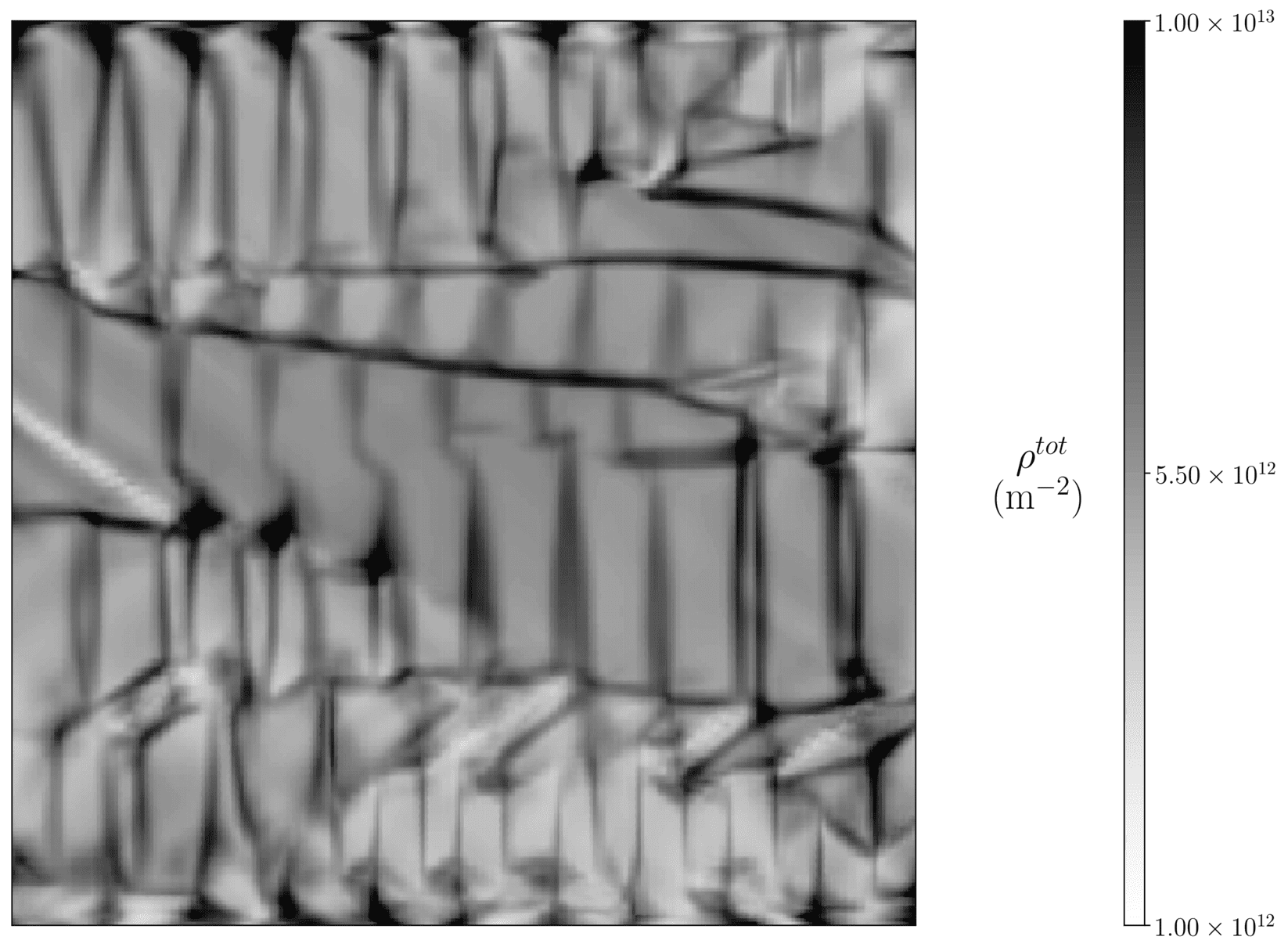}
		\label{subfig:plane_strain_compression_010_100x100_rho}
	}
	\caption{Total dislocation density at 1$\%$ macroscopic strain correspondig to the deformation microstructures shown in Figures~\ref{fig:plane_strain_compression_010_reduced},~\ref{fig:plane_strain_compression_010_tetra},~\ref{fig:plane_strain_compression_010_gen3f},~\ref{fig:plane_strain_compression_010_100x100}.}
	\label{fig:plane_strain_compression_010_rho_num}
\end{figure}

The stress-strain curves corresponding to the deformation microstructures shown in Figures~\ref{fig:plane_strain_compression_010},~\ref{fig:plane_strain_compression_010_reduced},~\ref{fig:plane_strain_compression_010_tetra},~\ref{fig:plane_strain_compression_010_gen3f}, and~\ref{fig:plane_strain_compression_010_100x100} are presented in Figure~\ref{fig:stress_strain_curve}. Despite substantial variations in deformation microstructures arising from differences in finite element type and size, the stress-strain curves remain largely consistent within the range of applied strains, showing only minor deviations. The elastic regime and the yield point are identical across all cases. The strain hardening slope is steepest for quadratic hexahedral elements with reduced integration and shallowest for quadratic hexahedral elements with full integration. Pressure control induces only a slight increase in the strain hardening slope. Compared to hexahedral elements with reduced integration, tetrahedral elements with reduced integration exhibit a softer response. Additionally, increasing the finite element size leads to an increase in the strain hardening rate. These findings can be correlated with the dislocation density fields shown in Figures~\ref{subfig:plane_strain_compression_010_rho} and~\ref{fig:plane_strain_compression_010_rho_num}. The short-wavelength oscillations in the dislocation density field observed with reduced integration hexahedral elements (Figure~\ref{subfig:plane_strain_compression_010_reduced_rho}) result in a higher overall dislocation density and consequently a steeper strain hardening slope. Similarly, the larger mesh size in Figure~\ref{subfig:plane_strain_compression_010_100x100_rho} leads to thicker dislocation walls, contributing to a higher mean dislocation density. In comparison, the dislocation density fields obtained with tetrahedral elements (Figure~\ref{subfig:plane_strain_compression_010_tetra_rho}) and pressure-controlled integration (Figure~\ref{subfig:plane_strain_compression_010_gen3f_rho}) resemble those obtained with fully integrated hexahedral elements (Figure~\ref{subfig:plane_strain_compression_010_rho}), leading to a comparable strain hardening behaviour.
\begin{figure}
	\centering
	\includegraphics[width=0.9\textwidth]{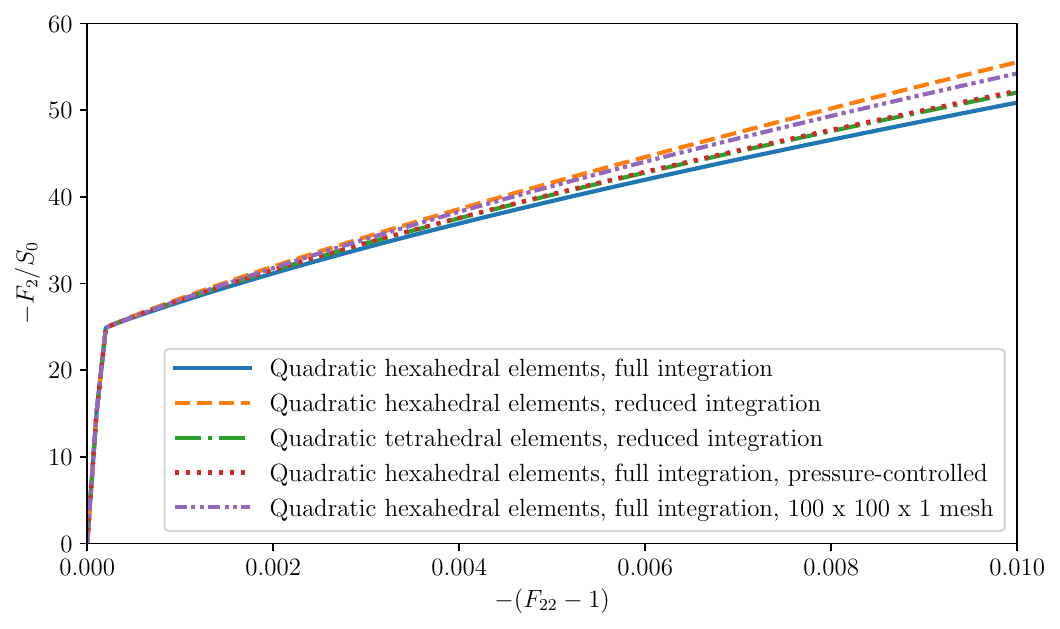}
	\caption{Stress-strain curves corresponding to the deformation microstructures shown in Figures~\ref{fig:plane_strain_compression_010},~\ref{fig:plane_strain_compression_010_reduced},~\ref{fig:plane_strain_compression_010_tetra},~\ref{fig:plane_strain_compression_010_gen3f}, and~\ref{fig:plane_strain_compression_010_100x100}.}
	\label{fig:stress_strain_curve}
\end{figure}

\subsection{Non-uniform simple shear}
\label{subsec:non_uniform_simple_shear}

As in~\cite{rys2024spontaneous}, the crystal is rotated by $\pi/4$ around the $[110]$ axis (see Figure~\ref{subfig:shear_010_superimposed}). Simple shear boundary conditions are applied, which correspond to imposed horizontal and zero vertical displacements on the top boundary, a clamped bottom boundary, and free left and right boundaries. The normal displacement of the front and back faces is set to 0 to enforce plane strain conditions. The top boundary is displaced horizontally to the right at a constant rate such that the applied strain rate is $\dot{u}_{1}(x_2=L)/L = 10^{-3}\, \si{\second}^{-1}$.

As shown in Figure~\ref{fig:shear_010}, we obtain a "butterfly" pattern similar to the one reported in~\cite{rys2024spontaneous}. The pattern consists of four main regions, each located in one of the corners of the domain. Each region exhibits a laminate deformation microstructure, with the activation of slip systems D4--D1 ($\gamma_4-\gamma_5$) or A2--A3 ($\gamma_7-\gamma_9$). The single-slip layers composing the laminate are parallel to their respective slip directions. Thus, the laminate is composed of deformation bands. In~\cite{rys2024spontaneous}, effective in-plane slip systems were used, each combining two real FCC slip systems, and as a result, the subdivision of these regions into parallel layers was not observed. Here, the deformation microstructure is decomposed into a series of single-slip lamellae because the 12 FCC slip systems are considered independently. As in the previous plane strain compression case, the pattern follows from the strong latent hardening interactions (see Table~\ref{tab:interaction_coefficients}). Note that slip systems B5 ($\gamma_3$) and C5 ($\gamma_{10}$) also become active in the four corners (see Figure 15(b) in~\cite{rys2024spontaneous}), but they are not shown here for conciseness. \textcolor{black}{Although the crystal orientation considered in this numerical test is different, our results are in qualitative agreement with the simple shear experiment conducted by Dmitrieva \textit{et al.}~\cite{dmitrieva2009lamination} on single crystal copper which shows a similar laminate microstructure.} The dislocation density field is shown in Figure~\ref{fig:shear_010_rho}. The dislocation density is highest at the four corners of the domain and at the interfaces between the single-slip regions. This dislocation density pattern is consistent with the deformation microstructure observed in Figure~\ref{fig:shear_010}. Similar to Figure~\ref{fig:plane_strain_compression_010_rho}, the dislocation walls form sub-grain boundaries between the single-slip regions. The dislocation density is lower within the interior of the single-slip regions and near the free left and right boundaries of the domain.
\begin{figure}
	\centering
	\subfloat[]{
		\hspace{-1.3cm}
		\includegraphics[width=0.55\textwidth]{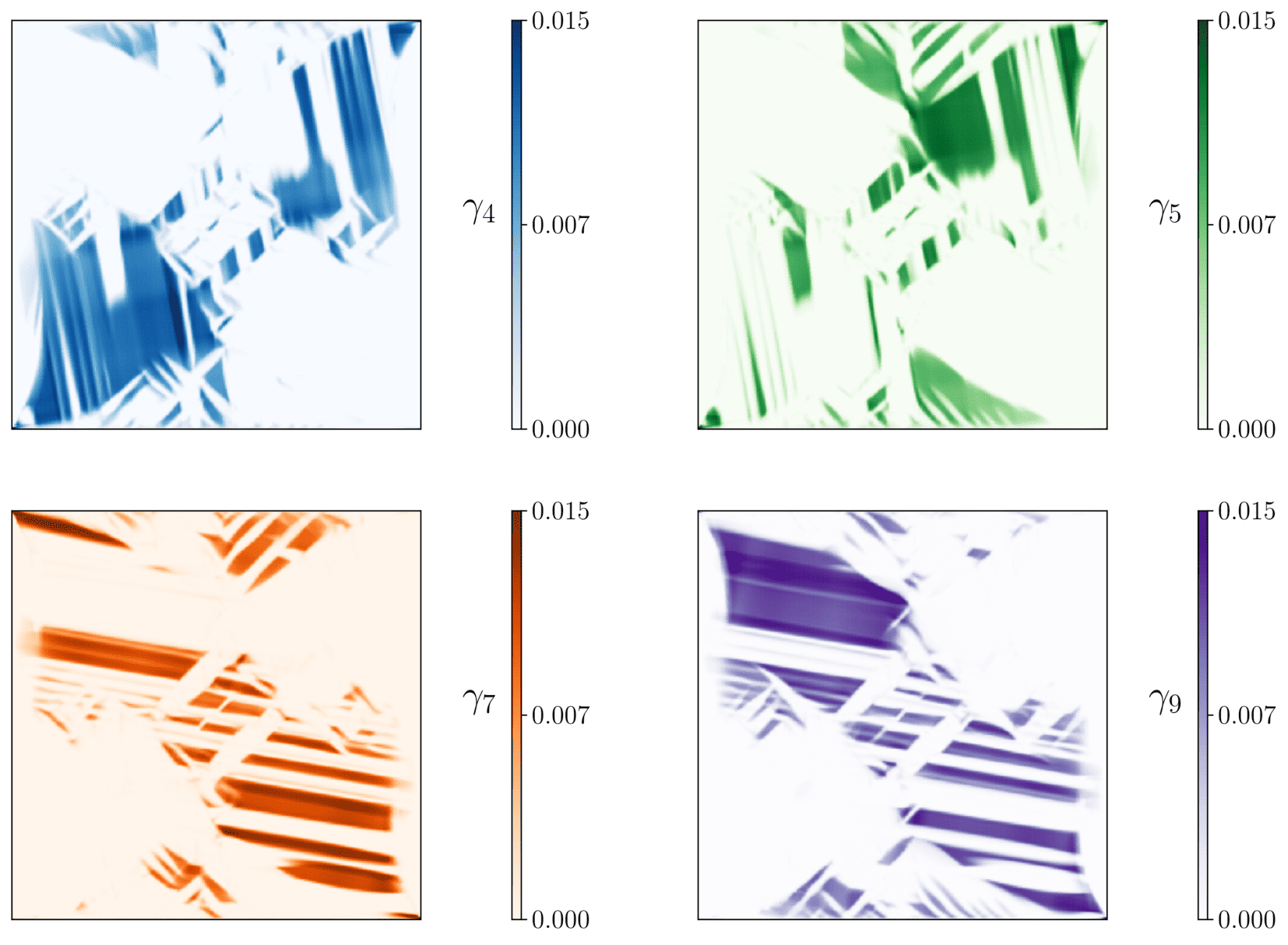}
		\label{subfig:shear_010_grid}
	}
	\subfloat[]{
        \raisebox{5.4cm}{\includegraphics[width=0.13\textwidth]{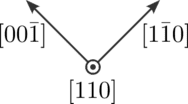}}\hspace{-1.3cm}
        \raisebox{0.cm}{\includegraphics[width=0.3\textwidth, trim=0.115cm 0.15cm 8.5cm 0cm, clip]{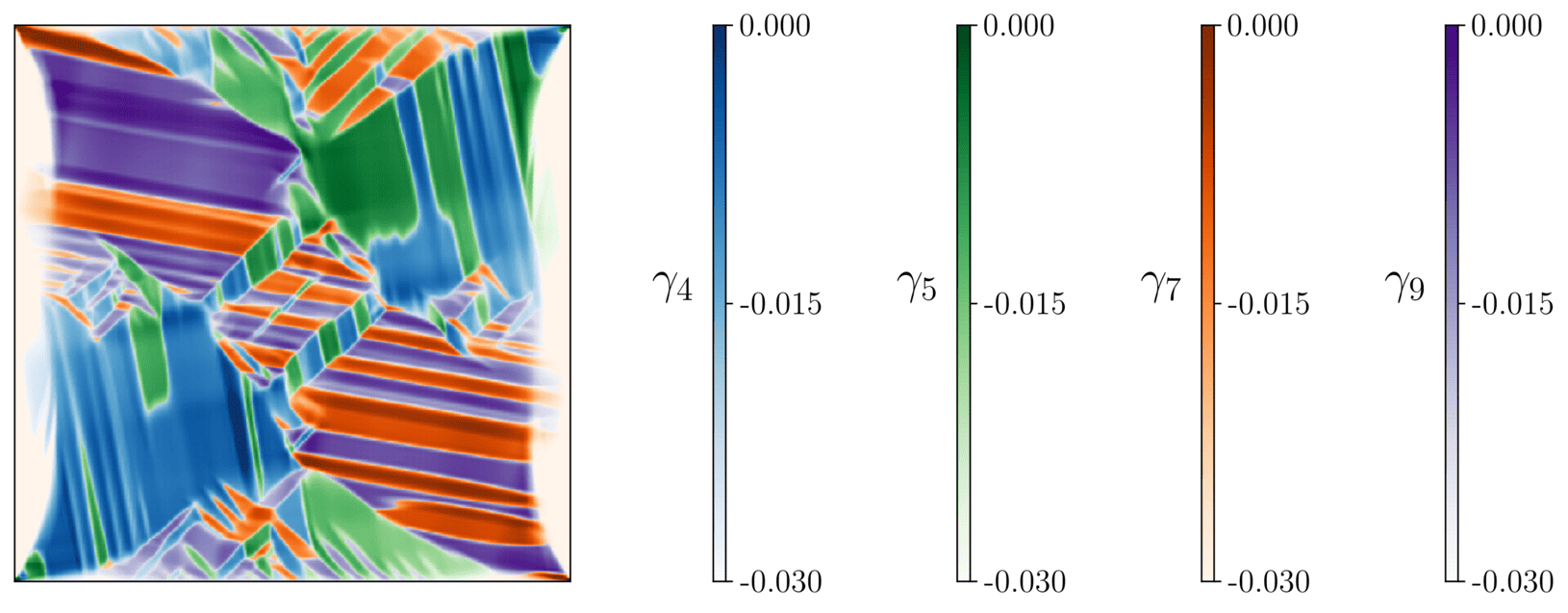}}\hspace{-3.8cm}
		\raisebox{5.4cm}{\includegraphics[width=0.15\textwidth]{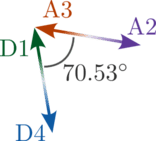}}
		\label{subfig:shear_010_superimposed}
	}
	\caption{Simple shear in the $(110)$ plane, with the crystal rotated by $\pi/4$ compared to Figure~\ref{fig:plane_strain_compression_010}, at 1$\%$ macroscopic strain. (a) Plastic slips on slip systems D4 ($\gamma_4$), D1 ($\gamma_5$), A2 ($\gamma_7$) and A3 ($\gamma_9$). (b) Superimposed plastic slips, where the color is selected according to the slip system with maximum slip activity.}
	\label{fig:shear_010}
\end{figure}

\begin{figure}
	\centering
		\includegraphics[width=0.7\textwidth,trim=0.075cm 0.075cm 0.cm 0cm,clip]{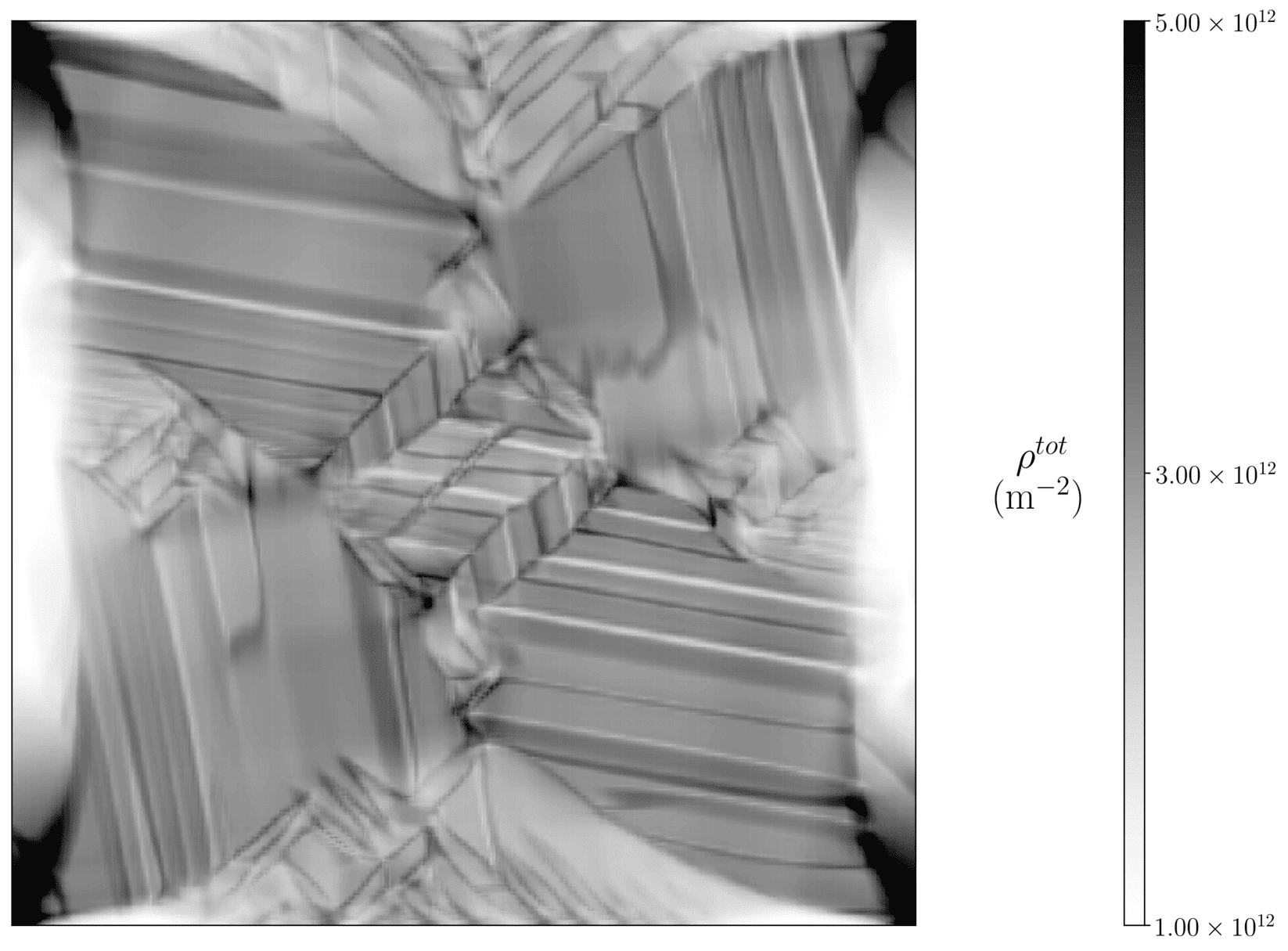}
	\caption{Total dislocation density at 1$\%$ macroscopic strain correspondig to the deformation microstructures shown in Figure~\ref{fig:shear_010}.}
	\label{fig:shear_010_rho}
\end{figure}

\section{Deformation patterns in 3D single crystals}
\label{sec:single_crystal_3D}

The problem of multislip is often analyzed in 0D, at the material point scale, where the deactivation of certain slip systems is studied, or in 2D plane strain cases, as described above, where the formation of deformation patterns is investigated. To the author's knowledge, Dequiedt \textit{et al.}~\citep{dequiedt2015heterogeneous} were the first to study the formation of deformation patterns in 3D simulations, considering single crystal cubic specimens under biaxial tension. This type of loading is known to activate a high number of slip systems for many orientations. Here, we consider simple tension of a single crystal wire oriented along the highly symmetric $[100]$ direction. Along this direction, 8 slip systems are equivalently loaded in tension, each with a Schmid factor equal to $1/\sqrt{6}$.

A cylinder of length $L$ and radius $R$ with a length-to-radius ratio of $L/R = 10$ is considered. The $[100]$ crystal direction is aligned with the cylinder axis. A mesh composed of 960,000 linear hexahedral finite elements is used, resulting in 2,943,243 degrees of freedom. A parallel solver using the Finite Element Tearing and Interconnecting (FETI) domain decomposition method is employed to solve the mechanical equilibrium. The graph centrality-based variant proposed by Bovet~\cite{bovet2023use}, implemented in the finite element software \texttt{Z-set}, is used to solve the equilibrium. The mesh is divided into 64 sub-domains, each assigned to two CPUs of a parallel computer. In total, four Intel Xeon Gold 5118 CPUs at 2.30 GHz with 32 CPUs and 256 GB of RAM each were used.

The cylinder is subjected to a tensile load by imposing a normal displacement on the top face such that the applied strain rate is $\dot{u}_3(x_3=L) / L = 10^{-3}$ s$^{-1}$. The normal displacement on the bottom face is set to 0, and the transverse displacements of the top and bottom faces are free, allowing for the Poisson effect to take place at both ends of the cylinder. The material parameters are listed in Table~\ref{tab:material_parameters}. The plastic slips on the active slip systems are shown in Figure~\ref{fig:cylinder_tension_grid} at a macroscopic strain of 1\%. The plastic slip fields are heterogeneous, with the formation of single-slip regions. Due to strong latent hardening, the single-slip regions do not overlap. The sharp discontinuities between single-slip regions can be seen in Figure~\ref{subfig:cylinder_tension_superimposed}, where all plastic slips are superimposed. The thickness of these boundaries often corresponds to the mesh size. This is confirmed in Figure~\ref{subfig:cylinder_tension_rho}, where the total dislocation density field is shown. The plastic slip fields reveal that the cylinder is predominantly divided into angular sectors, each corresponding to a single-slip region. These sectors extend over a large fraction of the cylinder length. However, the dislocation density field clearly shows that there are not only longitudinal walls, but also transverse walls. These transverse walls are more difficult to identify in the plastic slip fields, as the same slip system is often active on each side of a transverse wall. The dislocation walls are hence predominantly planes containing the cylinder axis and a radius of the cylinder. Close to the bottom face of the cylinder, a unique transverse dislocation wall, normal to the cylinder axis, can be observed.
\begin{figure}
	\centering
	\includegraphics[width=0.9\textwidth]{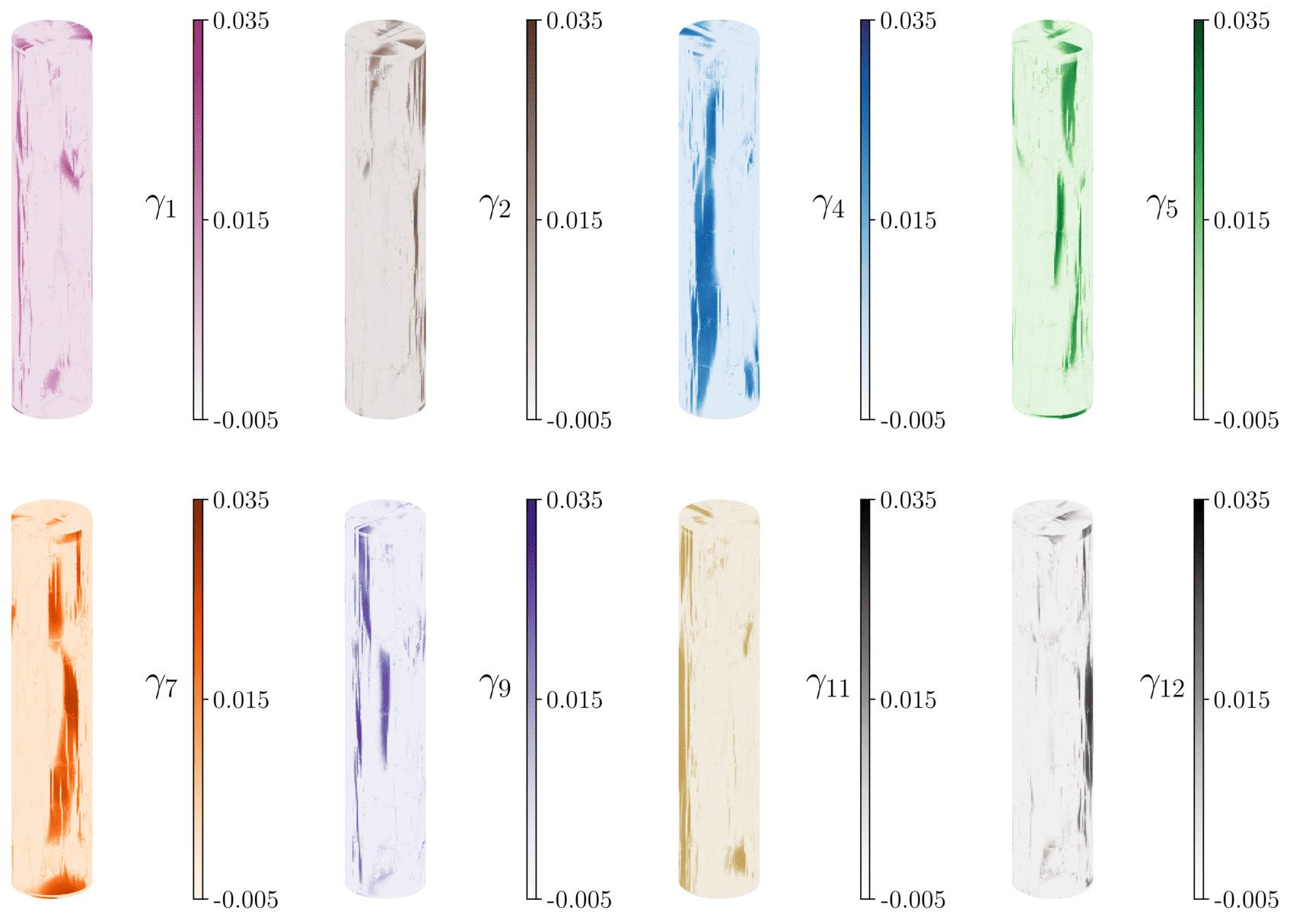}
	\caption{Tension of a $[001]$ wire up to 1$\%$ macroscopic strain. The plastic slips on active slip systems B4 ($\gamma_1$), B2 ($\gamma_2$), D4 ($\gamma_4$), D1 ($\gamma_5$), A2 ($\gamma_7$), A3 ($\gamma_9$), C3 ($\gamma_{11}$) and C1 ($\gamma_{12}$) are shown.} 
	\label{fig:cylinder_tension_grid}
\end{figure}

These dislocation walls can be viewed as low-angle sub-grain boundaries. It was indeed shown in the pole figures of Figure~\ref{fig:pole_figures} that single-slip leads to different rotations of the crystal lattice in each sub-grain. The small orientation jump across the dislocation walls is a necessary precursor for grain fragmentation. Recently, a Cosserat-phase-field model was proposed by Ghiglione \textit{et al.}~\cite{ghiglione2024cosserat} to predict the nucleation of grains in plastically deformed single crystals. Their approach does not require the introduction of seeds to trigger grain nucleation. They demonstrate that a key ingredient for grain nucleation is the presence of orientation gradients, while the sole presence of stored energy gradients is insufficient on its own. However, the stored energy gradients control the mobility of newly formed grain boundaries. In their study of the torsion of a single crystal cylinder, the radial and orthoradial orientation gradients were much weaker than the longitudinal gradients. This resulted in the fragmentation of the specimen into a stack of cylindrical grains with low-angle boundaries. The results presented here suggest that, in tension, accounting for strong latent hardening can lead to radial and orthoradial orientation gradients that are as strong as the longitudinal gradients. This could lead to the formation of more complex grain microstructures when using the Cosserat-phase-field model proposed in~\cite{ghiglione2024cosserat}. This is left for future work.
\begin{figure}
	\centering
	\subfloat[]{
		\raisebox{0.12cm}{\includegraphics[height=.4\textheight, trim=0.115cm 0.15cm 14cm 0cm, clip]{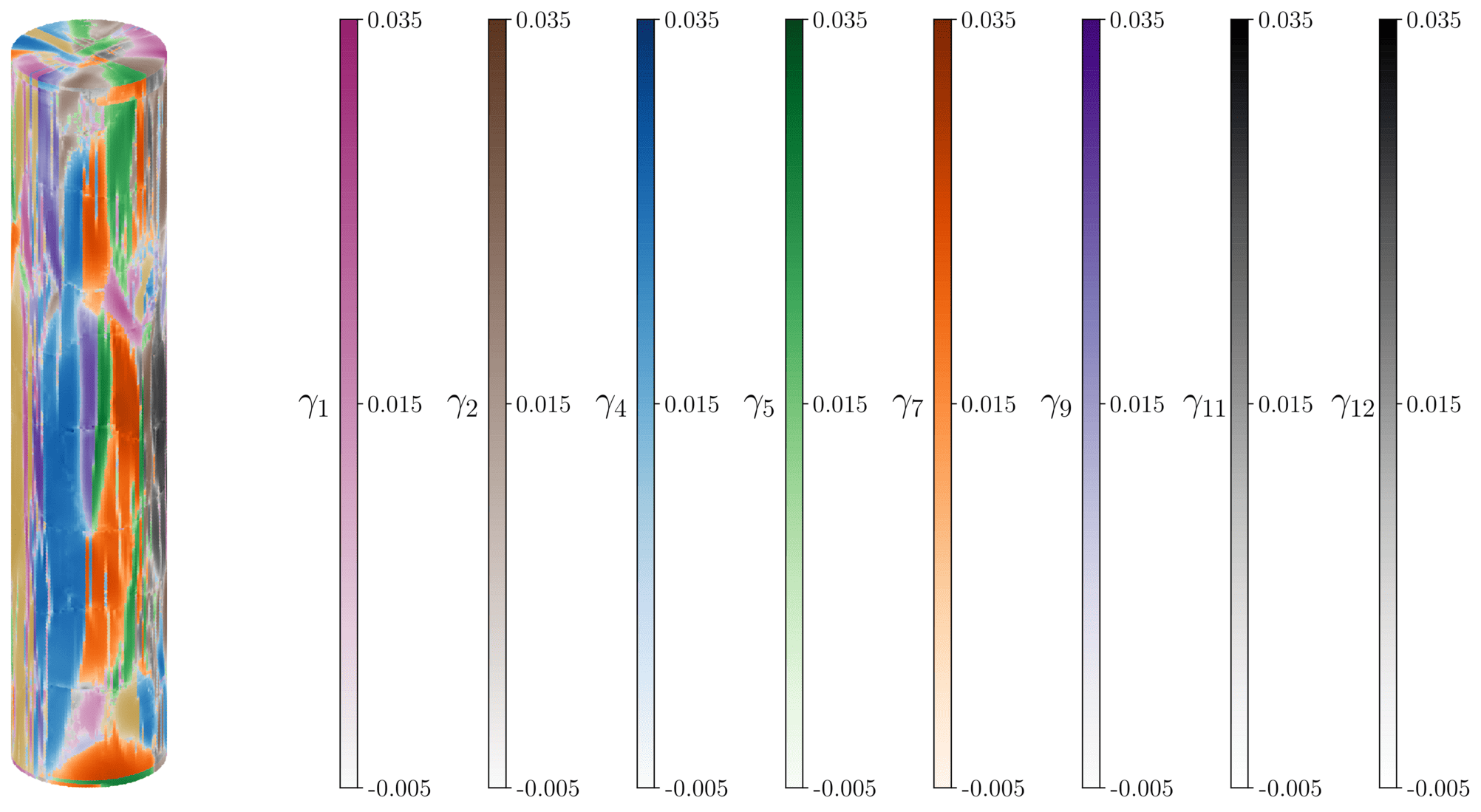}}
		\label{subfig:cylinder_tension_superimposed}
	}
	\subfloat{
		\includegraphics[width=0.2\textwidth]{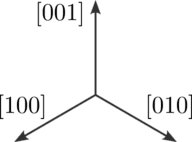}
	}
	\setcounter{subfigure}{1}
	\subfloat[]{
		\hspace{.3cm}
		\includegraphics[height=.4\textheight]{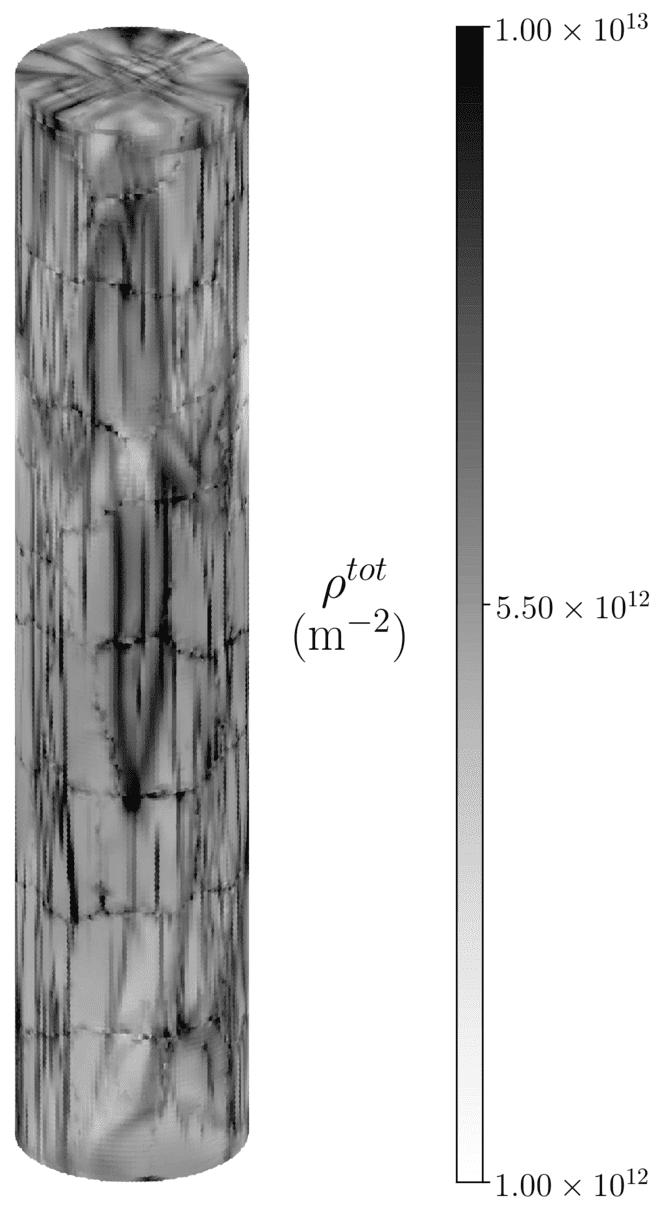}
		\label{subfig:cylinder_tension_rho}
	}
	\caption{(a) Superimposed plastic slips, where the color is selected according to the slip system with maximum slip activity. The colorscales are provided in Figure~\ref{fig:cylinder_tension_grid}. (b) Total dislocation density corresponding to the deformation microstructure shown in (a).}
	\label{fig:cylinder_tension_superimposed_rho}
\end{figure}

\section{Extension to strain gradient plasticity}
\label{sec:strain_gradient}

The plane strain compression of a single crystal along the $[00\bar{1}]$ direction is revisited by introducing a strain gradient plasticity length scale, $\ell$. The main objective is to regularize the characteristic size of the deformation microstructure, as it was shown to depend on the mesh size in Section~\ref{sec:plane_strain_compression}. We use the model presented in Section~\ref{subsec:strain_gradient} with the material parameters from Table~\ref{tab:material_parameters}. The material length scale $\ell$ is defined as follows:
\begin{equation}
	\ell = 2\pi\sqrt{\frac{A}{H_0}}
\end{equation}
where $H_0 = \SI{3500}{\mega\pascal}$ is the initial slope of the strain hardening curve in the absence of strain gradients~\citep{scherer2020lagrange}. This length scale is varied over three orders of magnitude, from $\ell = L/500$ to $\ell = L/0.5$, where $L$ is the side length of the domain. The crystal orientation is \textcolor{black}{such that} $x_1=[1\bar{1}0]$, $x_2=[00\bar{1}]$, and $x_3=[110]$, with plane strain compression applied along $x_2$.

The predicted deformation microstructures at 1$\%$ macroscopic strain are shown in Figure~\ref{fig:plane_strain_compression_010_gradient_l}. The active slip systems are the same as in the case without strain gradients, so the color scales are identical to those in Figure~\ref{fig:plane_strain_compression_010}. For comparison with the strain gradient plasticity model, the deformation microstructure obtained without strain gradients (\textit{i.e.}, $\ell = L / \infty$) is shown again in Figure~\ref{subfig:plane_strain_compression_010_gradient_l_0.0}. The strain gradient plasticity results for $\ell = L/500$, $\ell = L/50$, $\ell = L/5$ and $\ell = L/0.5$ are shown in Figures~\ref{subfig:plane_strain_compression_010_gradient_l_0.002},~\ref{subfig:plane_strain_compression_010_gradient_l_0.02},~\ref{subfig:plane_strain_compression_010_gradient_l_0.2}, and~\ref{subfig:plane_strain_compression_010_gradient_l_2.0}, respectively. The deformation microstructure consists of a second-order laminate of single-slip deformation bands, with their size now being controlled by the strain gradient plasticity length scale $\ell$, rather than the mesh size. As $\ell$ increases, the size of the single-slip regions increases as well. It is interesting to note that both the orientation and shape of the single-slip deformation bands are also affected by the strain gradient plasticity length scale. For $\ell = L/500$, the bands are straight and inclined perpendicular to their slip direction. For $\ell = L/50$, the deformation microstructure adopts a \textit{fence}-like pattern, with mostly vertical deformation bands. For $\ell = L/5$, the bands are again inclined, but the straight boundaries separating two bands are not always strictly perpendicular to the slip direction. For $\ell = L/0.5$, the bands remain inclined, but the boundaries separating two bands are now curved.
\begin{figure}
	\centering
	\subfloat{
		\includegraphics[width=0.2\textwidth]{FiguresReduced_axes.png}\hspace{-1.75cm}
		\raisebox{2.cm}{\includegraphics[width=0.25\textwidth]{FiguresReduced_slip_systems.png}}
	}\hspace{.45cm}
	\setcounter{subfigure}{0}
	\subfloat[$\ell = L/\infty$]{
		\includegraphics[width=0.4\textwidth, trim=0.115cm 0.15cm 8.5cm 0cm, clip]{FiguresReduced_HardeningMatrix_Rys_Redo_step_superimposed_010.png}
		\label{subfig:plane_strain_compression_010_gradient_l_0.0}
	}\\
	\subfloat[$\ell = L/500$]{
		\includegraphics[width=0.4\textwidth, trim=0.115cm 0.15cm 8.5cm 0cm, clip]{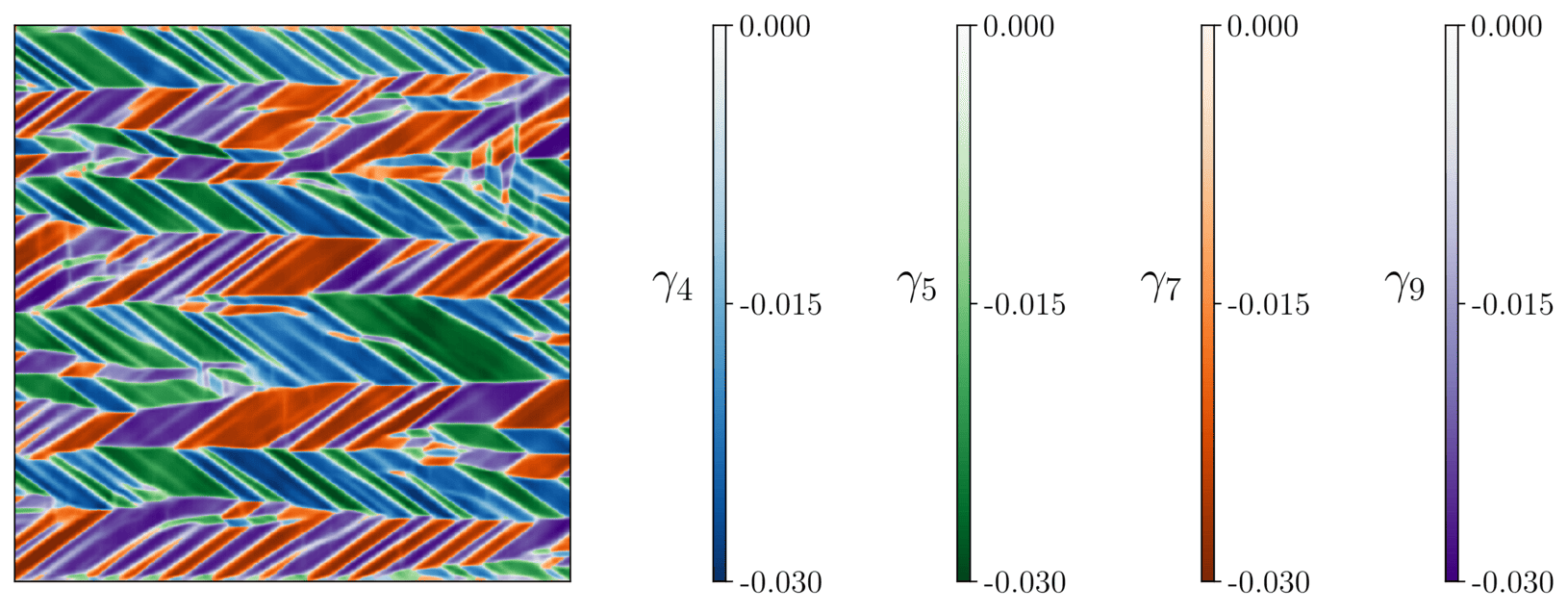}
		\label{subfig:plane_strain_compression_010_gradient_l_0.002}
	}
	\subfloat[$\ell = L/50$]{
		\includegraphics[width=0.4\textwidth, trim=0.115cm 0.15cm 8.5cm 0cm, clip]{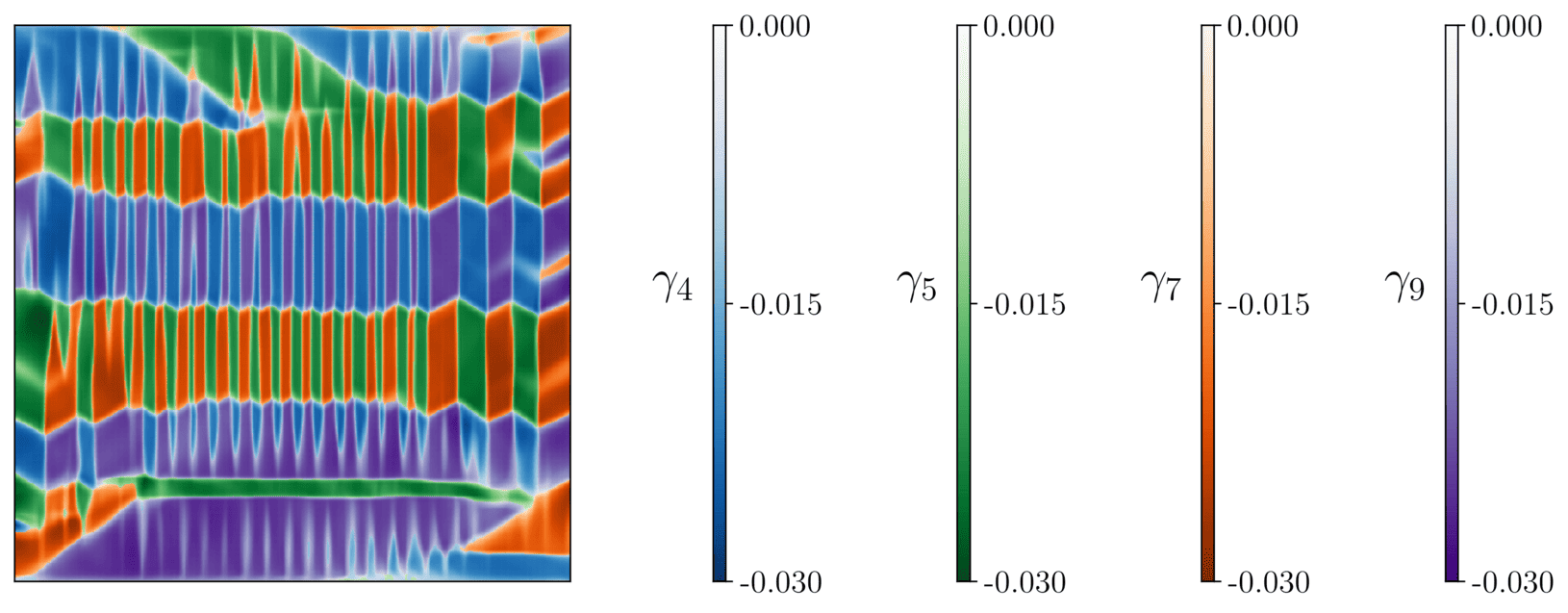}
		\label{subfig:plane_strain_compression_010_gradient_l_0.02}
	}\\
	\subfloat[$\ell = L/5$]{
		\includegraphics[width=0.4\textwidth, trim=0.115cm 0.15cm 8.5cm 0cm, clip]{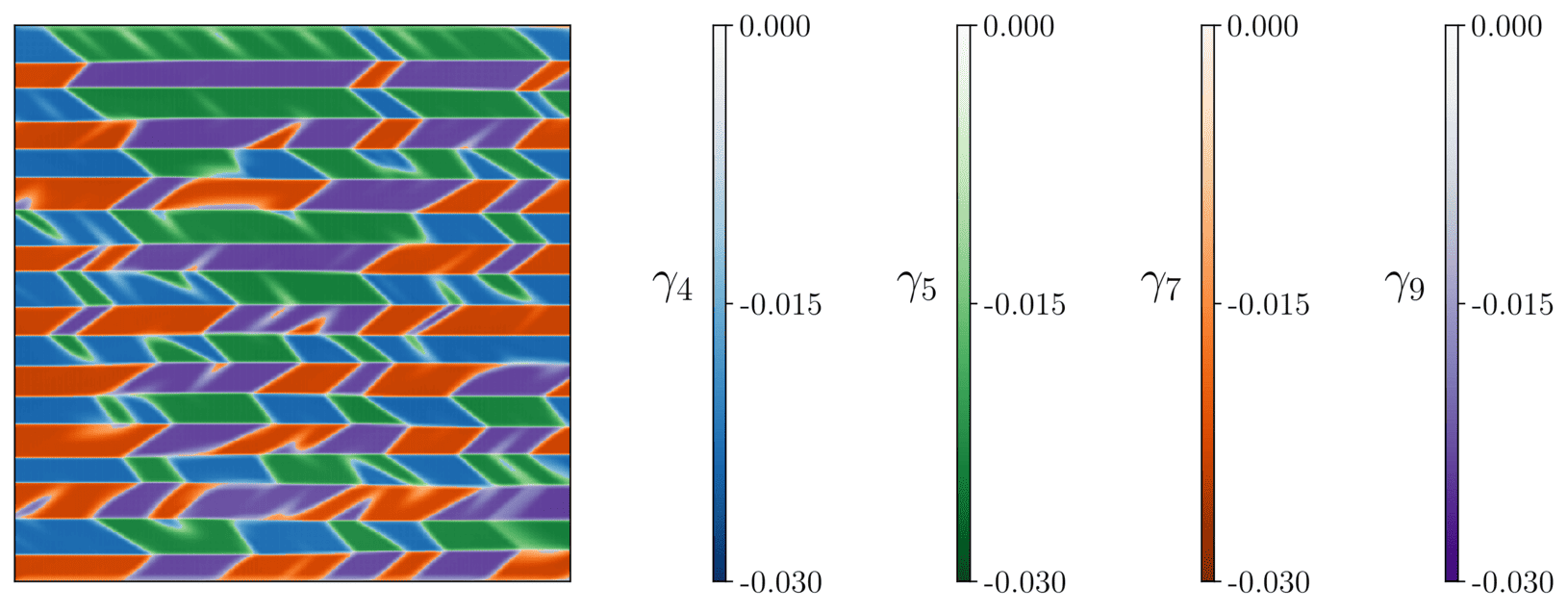}
		\label{subfig:plane_strain_compression_010_gradient_l_0.2}
	}
	\subfloat[$\ell = L/0.5$]{
		\includegraphics[width=0.4\textwidth, trim=0.115cm 0.15cm 8.5cm 0cm, clip]{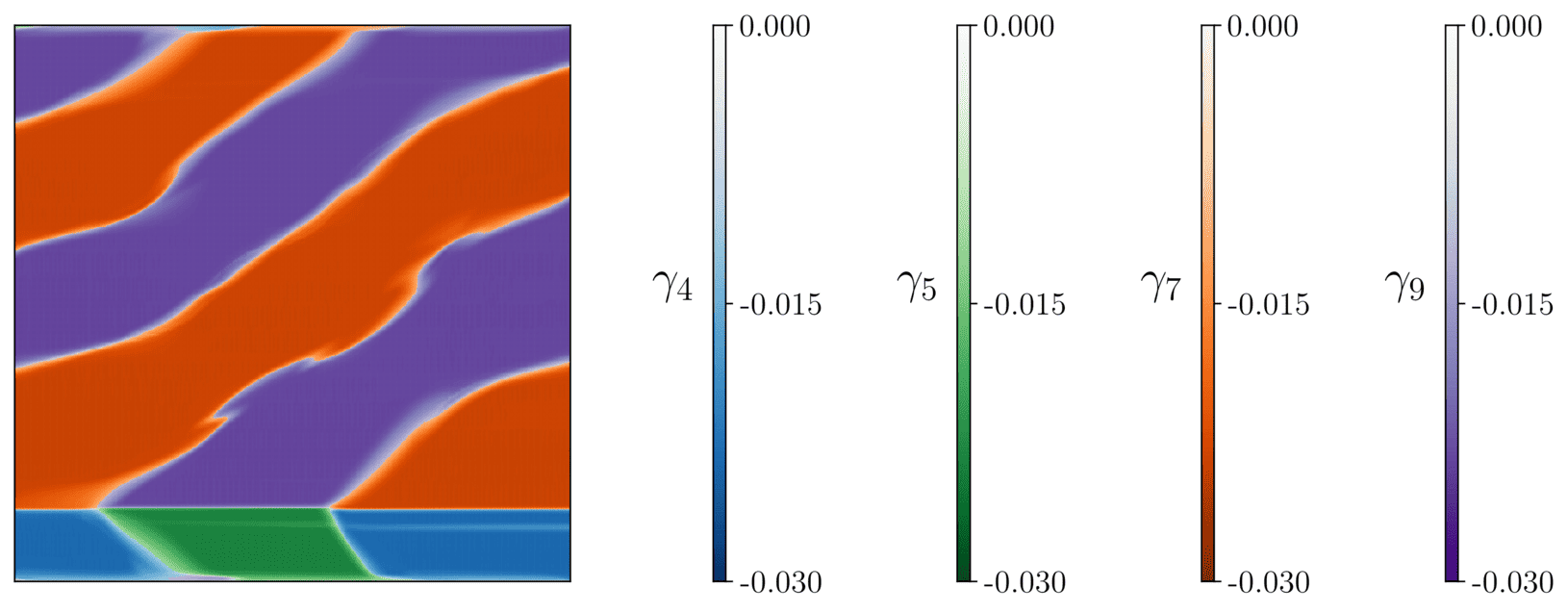}
		\label{subfig:plane_strain_compression_010_gradient_l_2.0}
	}
	\caption{Plane strain compression along $[00\bar{1}]$ in the $(110)$ plane at 1$\%$ macroscopic strain with varying strain gradient plasticity length scale $\ell$. The color scales for the different slip systems are identical to those in Figure~\ref{fig:plane_strain_compression_010}.}
	\label{fig:plane_strain_compression_010_gradient_l}
\end{figure}

The total dislocation density fields corresponding to the deformation microstructures shown in Figure~\ref{fig:plane_strain_compression_010_gradient_l} are displayed in Figure~\ref{fig:plane_strain_compression_010_gradient_l_rho}. The dislocation cell shapes follow the single-slip pattern observed in Figure~\ref{fig:plane_strain_compression_010_gradient_l}. These dislocation density fields are post-processed using thresholding and labeling operations to segment and isolate individual dislocation cells. The area of each cell is computed, and the diameter $\phi$ of a disk with the same area is taken as the characteristic size of the cell. The median, mean, and standard deviation of the characteristic sizes of the dislocation cells are shown in Figure~\ref{fig:plane_strain_compression_010_gradient_l_rho} in a log-log plot. The median size of the dislocation cells $\phi / L$ appears to follow a power law with an exponent close to $0.32$ with respect to the strain gradient plasticity length scale $\ell / L$. \textcolor{black}{This scaling with exponent $0.32$ is close to the scaling law with exponent 0.5 derived by Ortiz \& Repetto~\citep{ortiz1999nonconvex} from a 'principle of similitude'~\citep{kuhlmannwilsdor1962new,kuhlmann1985theory} extended to account for non-local self-energy of dislocations.}
It is interesting to note that, despite the mesh size dependency of the deformation microstructure without strain gradient regularization ($\ell = L/\infty$), the characteristic size of its dislocation cells is still much larger than the mesh size $h/L = 0.005$ (horizontal dash-dotted line in Figure~\ref{fig:plane_strain_compression_010_gradient_l_rho}). This indicates that the size of the dislocation cells is not only governed by the mesh size but also by the domain size $L$. For $\ell = L/0.5$, the characteristic size of the dislocation cells may be somewhat underestimated, as the size of the dislocation cells is close to the size of the domain. In particular, the three cells at the bottom of the domain may be truncated, leading to an underestimate of their size.  The material length scale $\ell$ not only regularizes the characteristic size of the deformation microstructure but also affects the thickness of dislocation cell walls. For $\ell = L / \infty$ and $\ell = L/500$, the cell walls are thin and sharp, while for $\ell = L/50$, $\ell = L/5$ and $\ell = L/0.5$, the cell walls become thicker and more diffuse.
\begin{figure}
	\centering
	\subfloat{
		\raisebox{-.8cm}{
			\includegraphics[width=.4\textwidth]{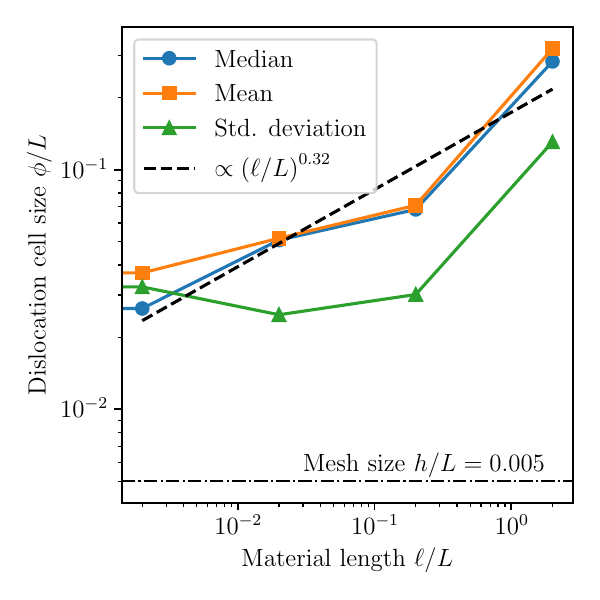}
		}
		\label{subfig:plane_strain_compression_010_gradient_l_median_mean_deviation}
	}
	\setcounter{subfigure}{0}
	\hspace{1.cm}
	\subfloat[$\ell = L/\infty$]{
		\includegraphics[width=0.48\textwidth]{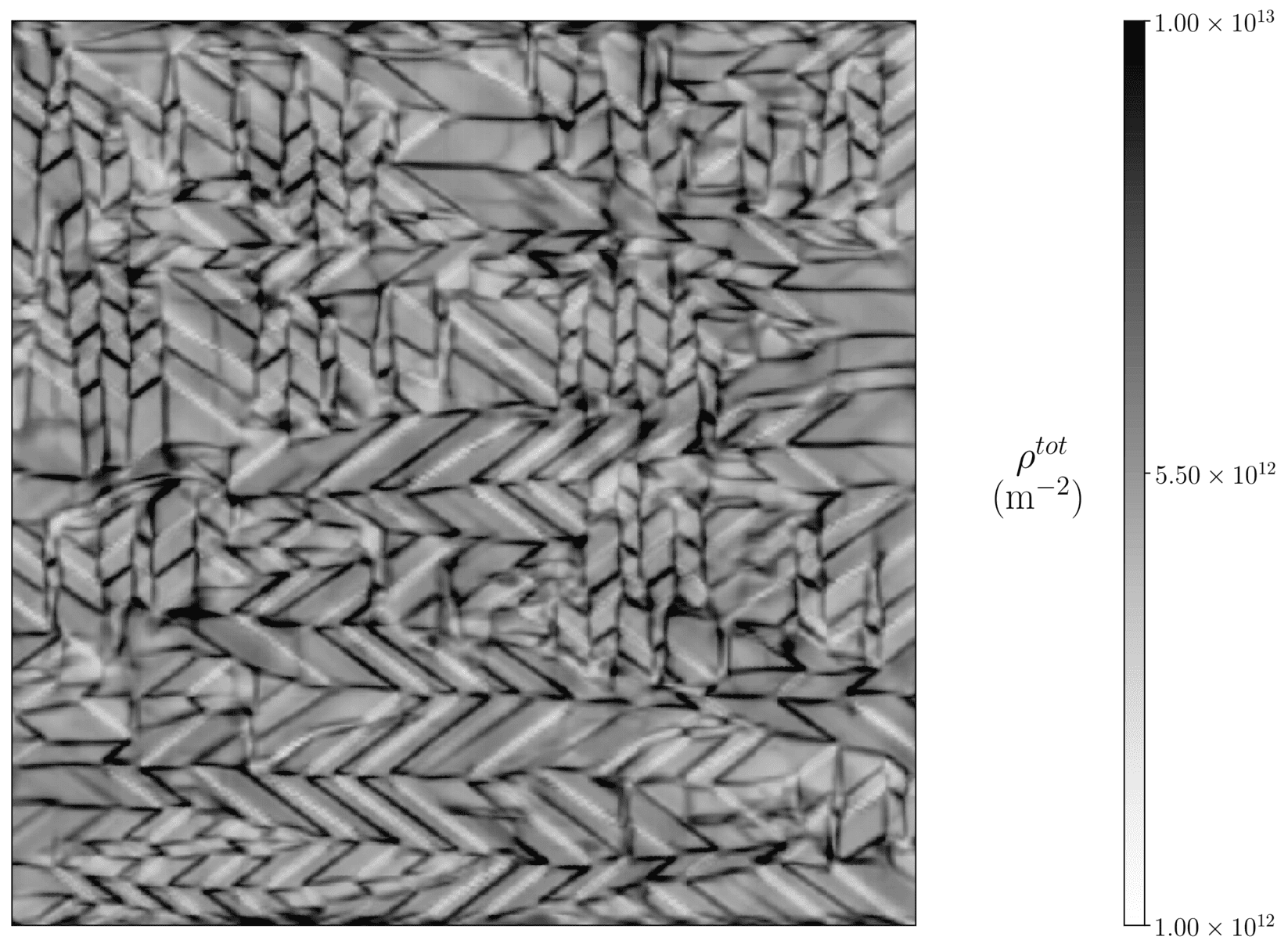}
		\label{subfig:plane_strain_compression_010_gradient_l_0.0_rho}
	}\\
	\subfloat[$\ell = L/500$]{
		\includegraphics[width=0.48\textwidth]{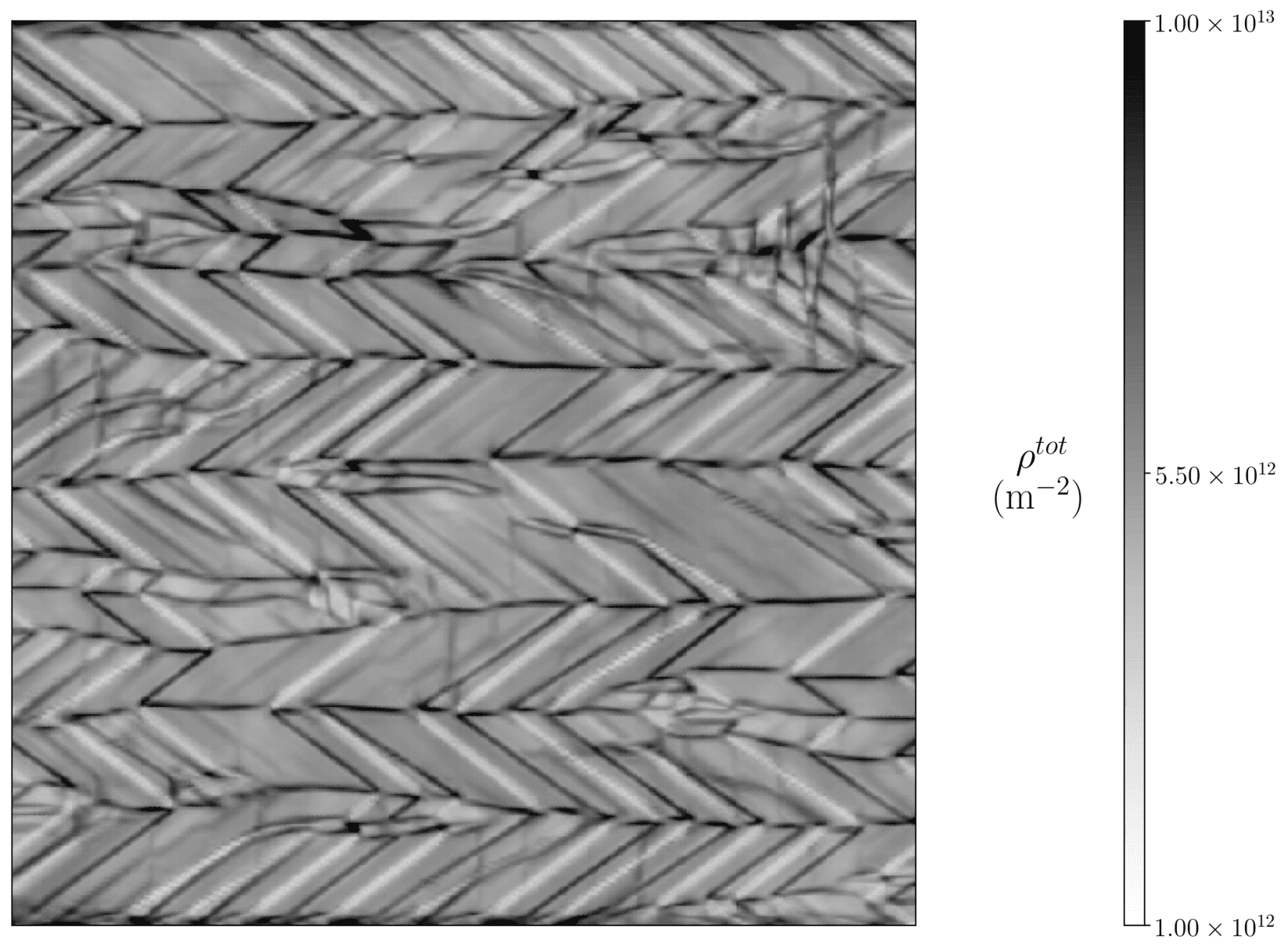}
		\label{subfig:plane_strain_compression_010_gradient_l_0.002_rho}
	}
	\subfloat[$\ell = L/50$]{
		\includegraphics[width=0.48\textwidth]{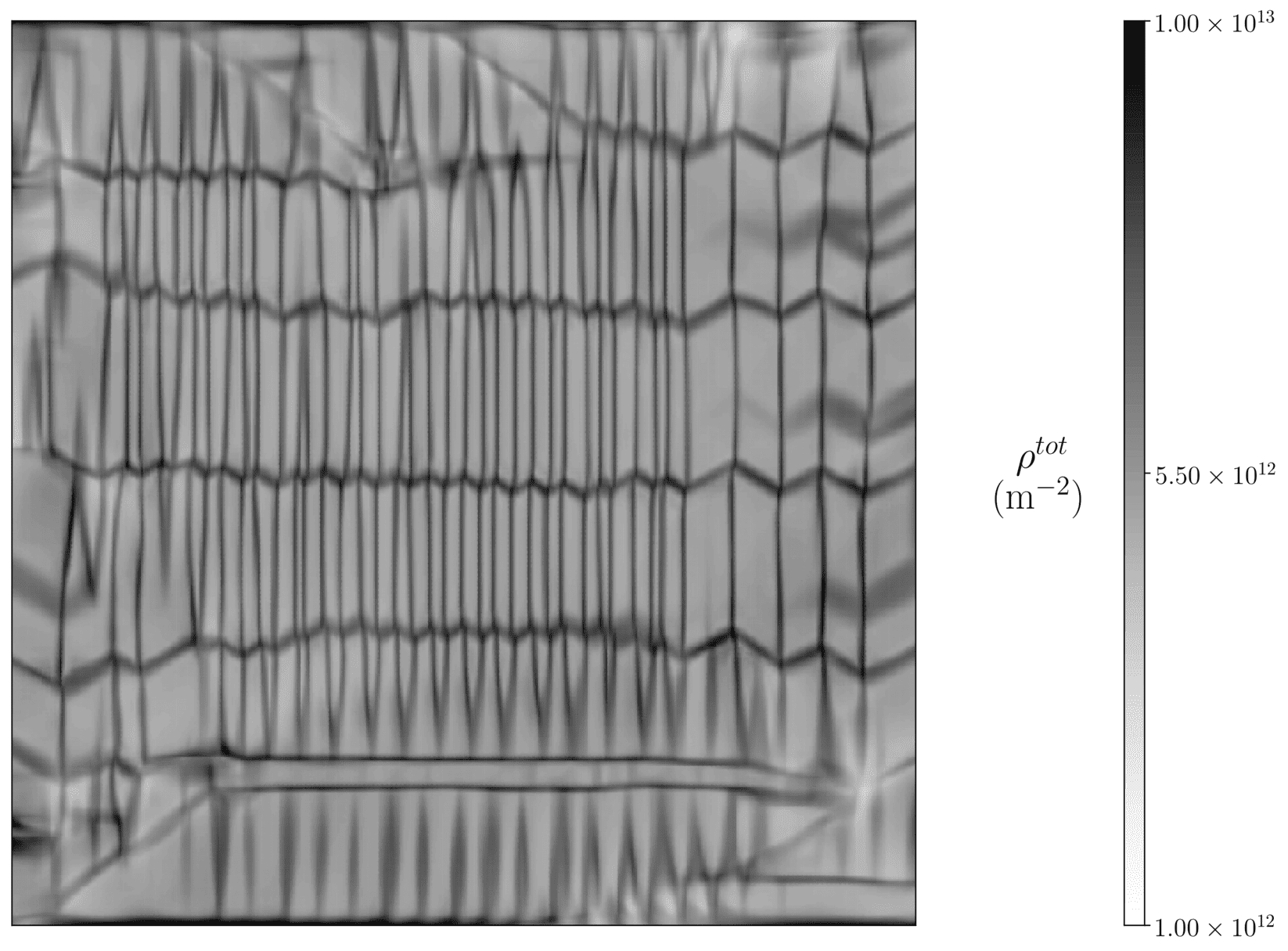}
		\label{subfig:plane_strain_compression_010_gradient_l_0.02_rho}
	}\\
	\subfloat[$\ell = L/5$]{
		\includegraphics[width=0.48\textwidth]{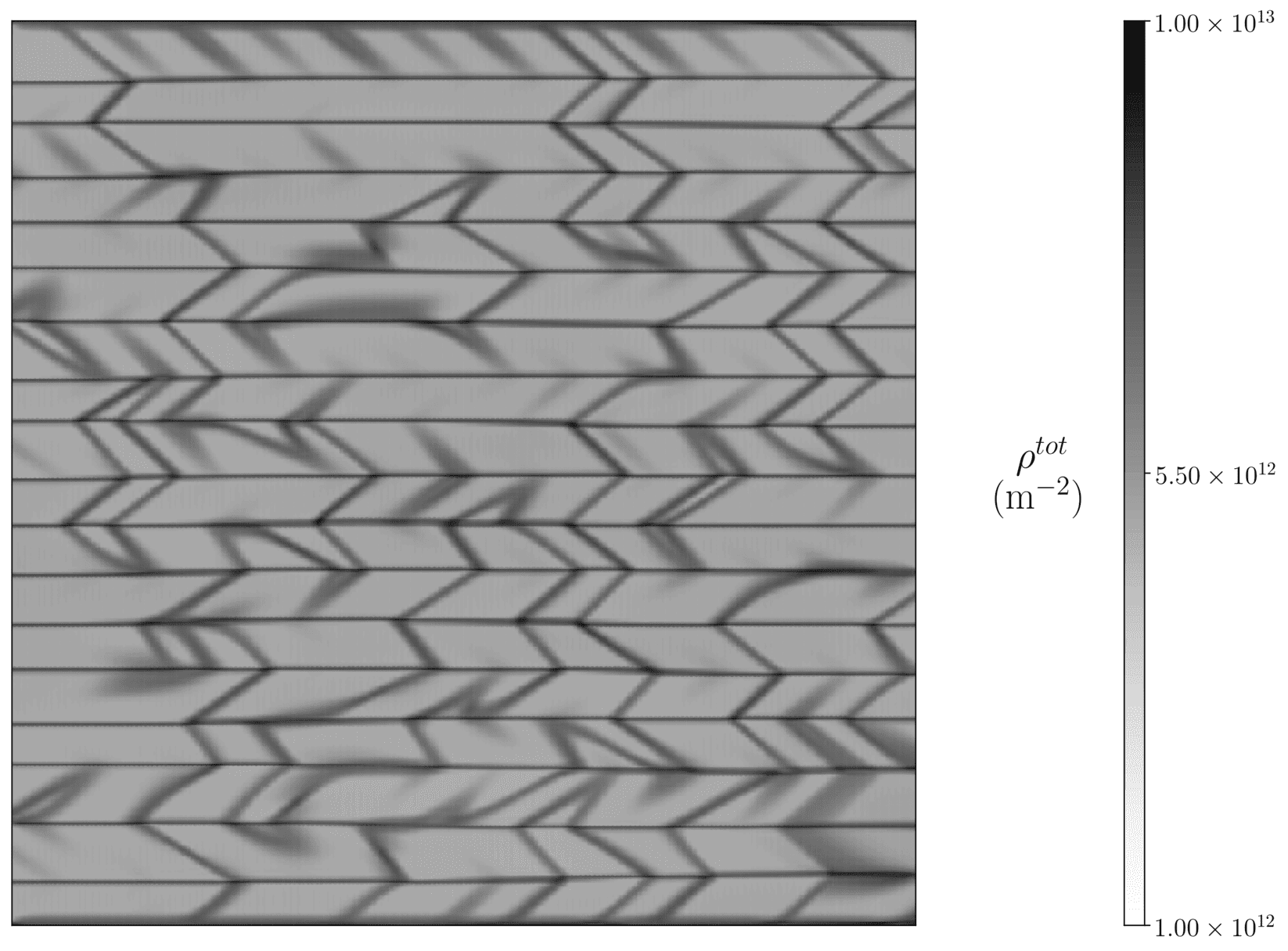}
		\label{subfig:plane_strain_compression_010_gradient_l_0.2_rho}
	}
	\subfloat[$\ell = L/0.5$]{
		\includegraphics[width=0.48\textwidth]{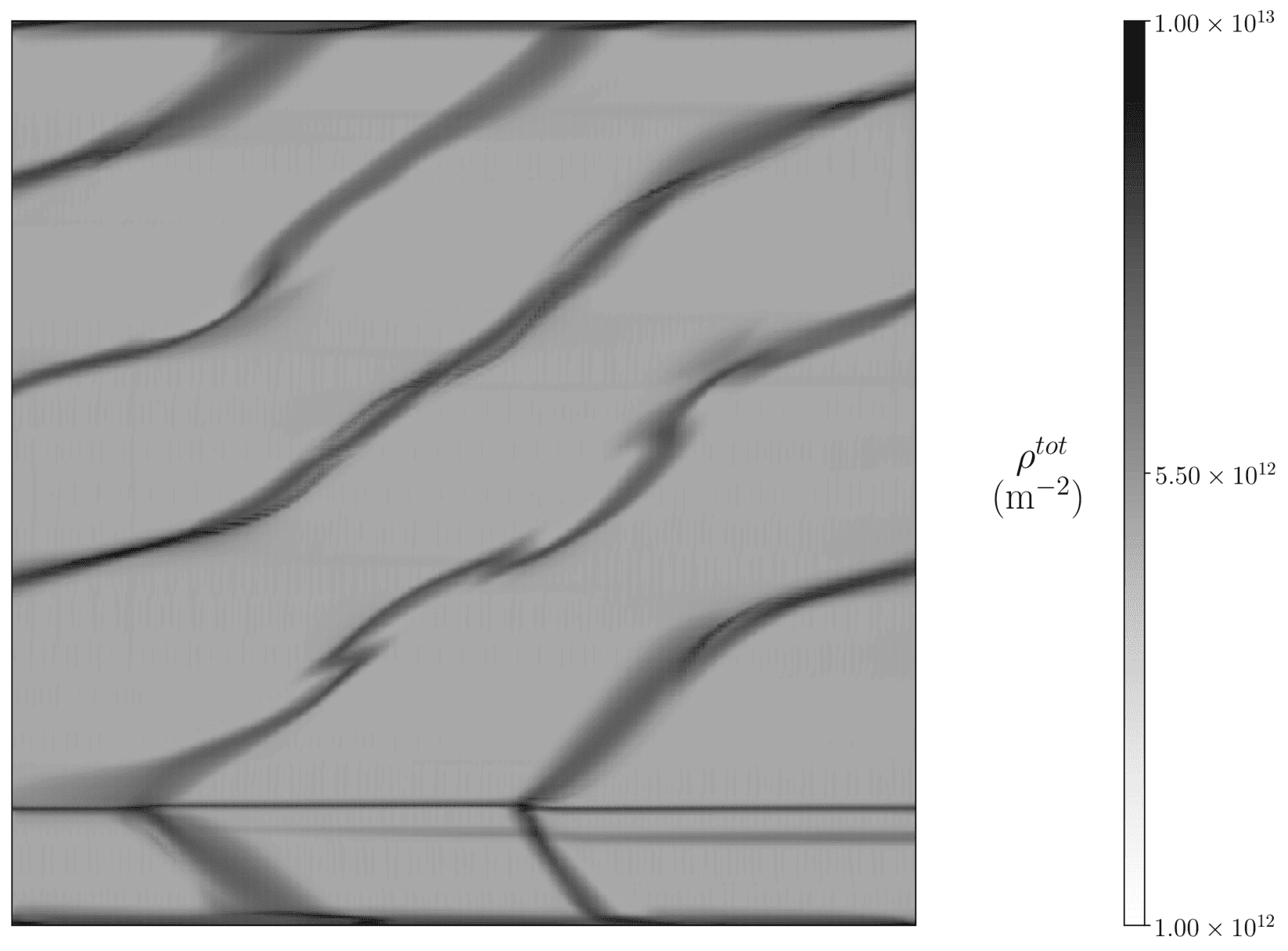}
		\label{subfig:plane_strain_compression_010_gradient_l_2.0_rho}
	}
	\caption{Total dislocation density corresponding to the deformation microstructures shown in Figure~\ref{fig:plane_strain_compression_010}.}
	\label{fig:plane_strain_compression_010_gradient_l_rho}
\end{figure}

\textcolor{black}{A mesh convergence analysis was also performed at a fixed material length scale $\ell = L/5$. Several mesh sizes were considered, with $h/L = 0.01$, $h/L = 0.00667$, $h/L = 0.005$, $h/L = 0.004$ and $h/L = 0.00333$. For mesh sizes below $h/L = 0.005$, the characteristic size of the dislocation cells was found to be independent of the mesh size, hence validating the regularization capacity of the strain gradient plasticity model.} 

\section{Emergence of deformation patterns in 2D and 3D polycrystals}
\label{sec:polycrystal}

In the previous sections, the formation of deformation patterns was studied in single crystals. Crystal orientations of high symmetry were considered due to their propensity to activate a large number of slip systems under uniaxial loading. This naturally raises the question of whether the formation of deformation patterns observed earlier is an artifact of the chosen orientations or if similar patterns can emerge in polycrystals with randomly oriented grains. To explore this, we consider both 2D square and 3D cube polycrystal microstructures, as shown in Figure~\ref{fig:polycrystal_microstructure}. These Laguerre tessellations, each composed of 100 grains, are generated using \texttt{Neper}~\citep{quey2011large,quey2022neper}, with grain orientations following a uniform random distribution. As demonstrated in previous sections, the deformation microstructures consist of very fine details, requiring a very small mesh size to be accurately resolved. To significantly reduce the computational cost, we finely mesh a single grain located at the center of the domain to resolve the deformation patterns. In the other grains, a much coarser mesh is used. A zoomed-in view of the finely meshed grain is shown in Figure~\ref{fig:polycrystal_microstructure}. The coarse mesh size is $h_c/L = 1\times 10^{-2}$ in 2D and $h_c/L = 5\times 10^{-2}$ in 3D, while the fine mesh size is $h_f/L = 4\times 10^{-4}$ in 2D and $h_f/L = 4\times 10^{-3}$ in 3D. The 2D mesh consists of 152,996 quadratic wedge elements with 9 Gauss points (C3D15) and has 2,067,975 degrees of freedom. The 3D mesh is made up of 975,483 quadratic tetrahedral elements with 4 Gauss points (C3D10) and has 3,940,776 degrees of freedom. For the 3D model, the mesh is split into 64 subdomains for the FETI resolution method. The 2D polycrystal is loaded in plane strain compression along the vertical direction, while the 3D polycrystal is loaded in uniaxial compression along the vertical direction. In both cases, the applied strain rate is $\dot{u}_2(x_2=L) / L = 10^{-3}$ s$^{-1}$. The material parameters are provided in Table~\ref{tab:material_parameters}.
\begin{figure}
	\centering
	\includegraphics[width=0.75\textwidth]{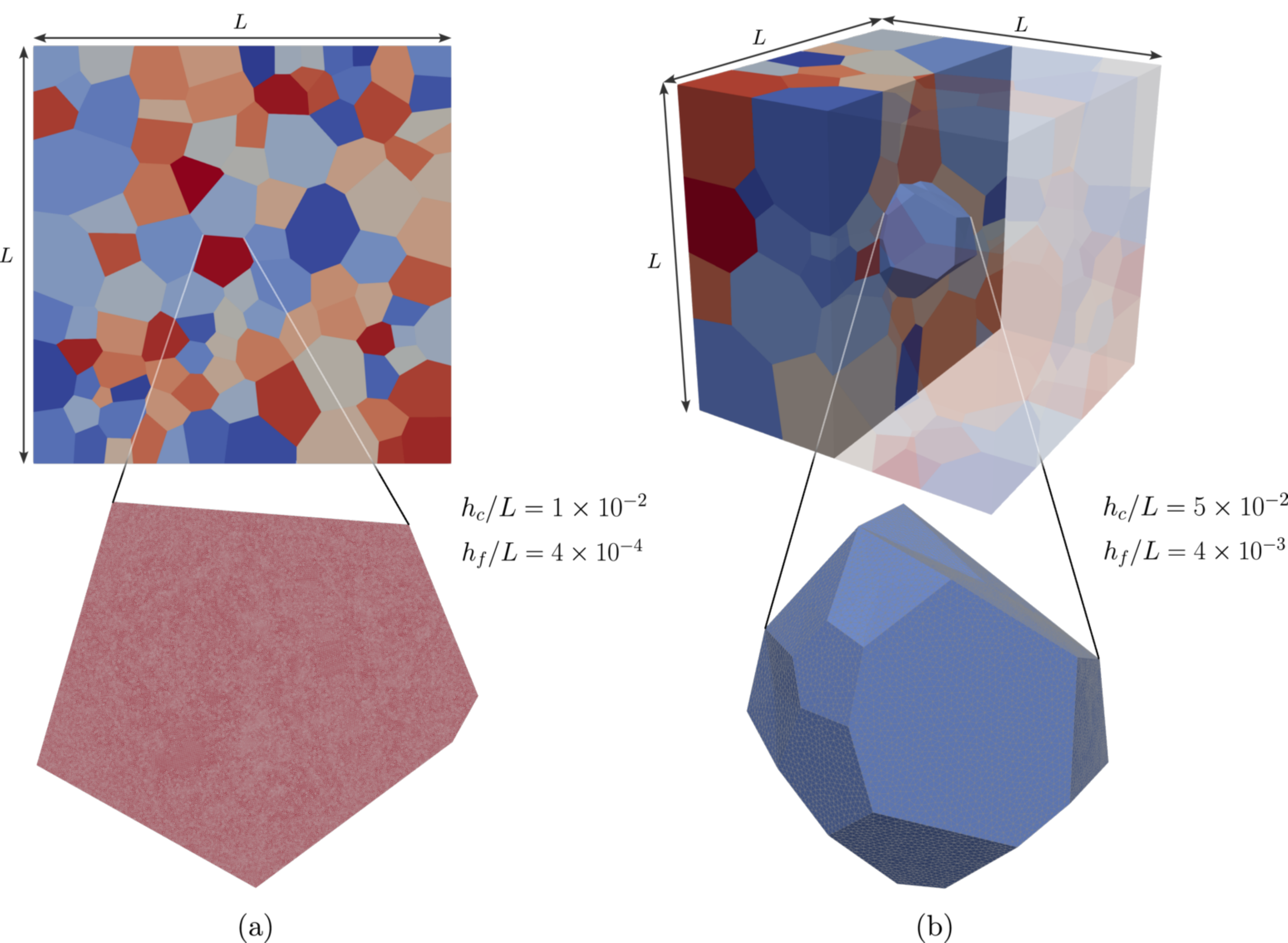}
	\caption{(a) 2D and (b) 3D polycrystal microstructures used in compression simulations, with a single finely meshed grain located at the center of the domain. Both microstructures are composed of 100 grains. The mesh size is $h_c/L = 1\times 10^{-2}$ in the coarsely meshed region in 2D and $h_c/L = 5\times 10^{-2}$ in 3D. The fine mesh size is $h_f/L = 4\times 10^{-4}$ in 2D and $h_f/L = 4\times 10^{-3}$ in 3D.}
	\label{fig:polycrystal_microstructure}
\end{figure}

The plastic fields for the 12 FCC slip systems in the 2D polycrystal are shown in Figure~\ref{fig:polycrystal_2D_slips} at 1\% macroscopic strain. Due to the random distribution of grain orientations, the active slip systems vary across the different grains. In the coarsely meshed grains, multiple slip systems can be simultaneously active at the same location without the formation of a distinct deformation pattern. In contrast, in the finely meshed grain located near the center of the domain, three slip systems become active, resulting in the formation of a deformation pattern. A zoom of the deformation microstructure is presented in Figure~\ref{fig:polycrystal_2D_slips}(b). The pattern consists of a second-order laminate structure, characterized by thick, nearly horizontal deformation bands, each of which is composed of thin, nearly vertical lamellae. Notably, the first type of lamellae is associated with a single active slip system, D4 ($\gamma_4$), whereas the second type involves two simultaneously active slip systems, C5 ($\gamma_{10}$) and C3 ($\gamma_{11}$). Since the strain gradient extension of the model is not applied in this case, the characteristic size of the deformation microstructure is predominantly determined by the mesh size. The coarse mesh size, \( h_c/L \), used in the coarsely meshed grains is approximately 10 times smaller than the grain diameter. However, this level of discretization is insufficient to fully resolve the formation of a detailed deformation microstructure in coarsely meshed grains.
\begin{figure}
	\centering
	\includegraphics[width=0.98\textwidth]{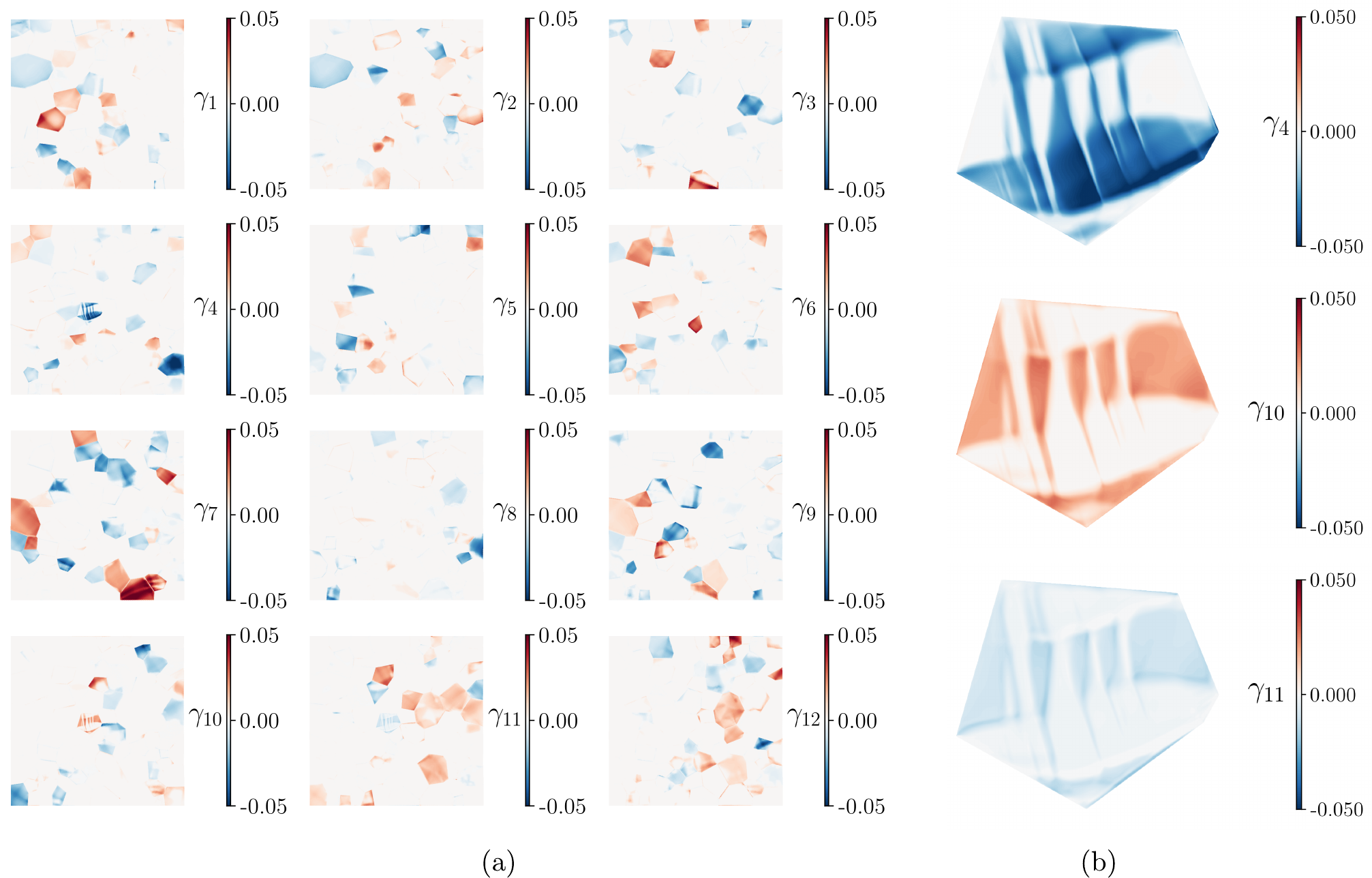}
	\caption{Plane strain compression at 1$\%$ macroscopic strain of the 2D polycrystal. (a) Plastic slip fields on the 12 FCC slip systems. (b) Zoom on the active slip systems in the finely meshed grain located at the center of the domain.}
	\label{fig:polycrystal_2D_slips}
\end{figure}

The total dislocation density field in the 2D polycrystal is presented in Figure~\ref{fig:polycrystal_2D_rho}. The dislocation density reaches its maximum values near stress concentrators, such as grain boundaries and triple junctions, while it is lower within the bulk of the grains. In the coarsely meshed grains, the dislocation density evolves relatively smoothly, without exhibiting sharp discontinuities. In contrast, in the finely meshed grain located at the center of the domain, the dislocation density field displays clear jumps between regions of low and high density. These jumps correspond to the interfaces of the laminate microstructure described in Figure~\ref{fig:polycrystal_2D_slips}.
\begin{figure}
	\centering
	\includegraphics[width=0.98\textwidth]{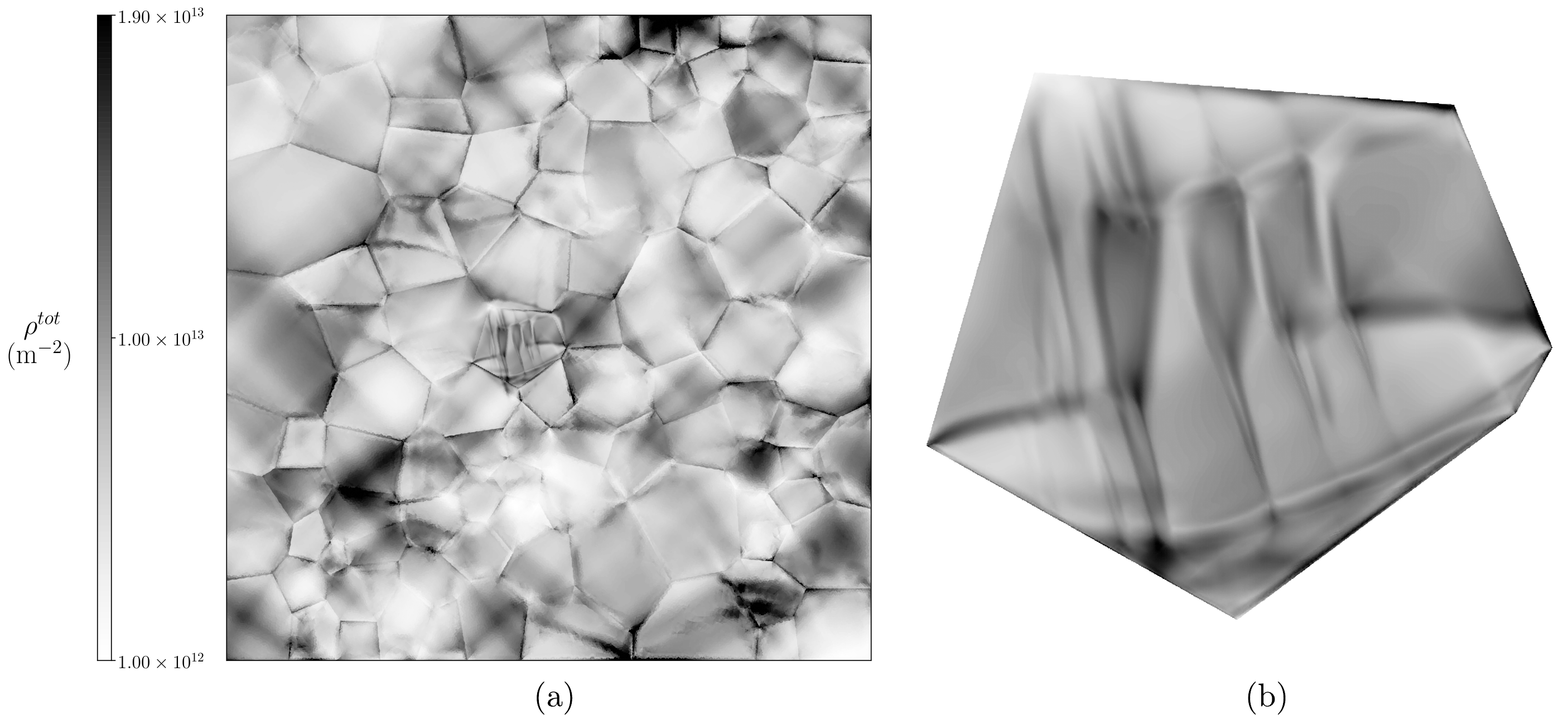}
	\caption{Total dislocation density field in (a) the entire 2D polycrystal and (b) a zoom of the finely meshed grain located at the center of the domain.}
	\label{fig:polycrystal_2D_rho}
\end{figure}

The plastic fields for the 12 FCC slip systems in the finely meshed grain of the 3D polycrystal are shown in Figure~\ref{fig:polycrystal_3D_slips_rho}(a) at 1$\%$ macroscopic strain. Two dominant slip systems, B2 ($\gamma_2$) and B5 ($\gamma_3$), are active in separate regions. Four secondary slip systems, A2 ($\gamma_7$), A3 ($\gamma_9$), C5 ($\gamma_{10}$) and C3 ($\gamma_{11}$), are also active in smaller regions and with lower intensities. The remaining slip systems are also active but in even smaller regions and to even lesser intensities. Although regions with active B2 and B5 slip systems are complementary, a laminate deformation microstructure is not clearly observed in this case. However, the dislocation density field shown in Figure~\ref{fig:polycrystal_3D_slips_rho}(b) reveals the presence of a cellular microstructure. Approximately 25 roughly equiaxed dislocation cells can be observed from the surface of the grain. The dislocation walls are not finely resolved due to the chosen mesh size, but they clearly stand out as their dislocation density is more than twice as large as that in the interior of the cells. The location of the dislocation walls coincides with the interfaces of the 64 subdomains used by the FETI solver. Although the precision of the solver was increased by reducing significantly the convergence tolerance parameter, this did not alter the observed dislocation density structure. In fact, the displacement fields are continuous across the interfaces of the subdomains. Therefore, the coincidence of the dislocation walls with the subdomain interfaces is likely the result of small fluctuations of internal variables induced by the FETI solver. These fluctuations are sufficient to trigger a local instability that leads to the observed pattern. A different decomposition of the microstructure into subdomains would likely result in a different pattern. The presence of these cells could not be easily detected from the plastic slip fields alone. A careful zoom into the regions of dislocation walls reveals that up to four slip systems are active in these regions. Despite the low intensity of plastic slip, the activation of multiple slip systems locally induces strong latent hardening in these walls and causes a rapid increase in dislocation density. Simulations with different mesh refinements (not shown here) revealed that the size of the dislocation cells is governed by the mesh size.
\begin{figure}
	\centering
	\includegraphics[width=0.98\textwidth]{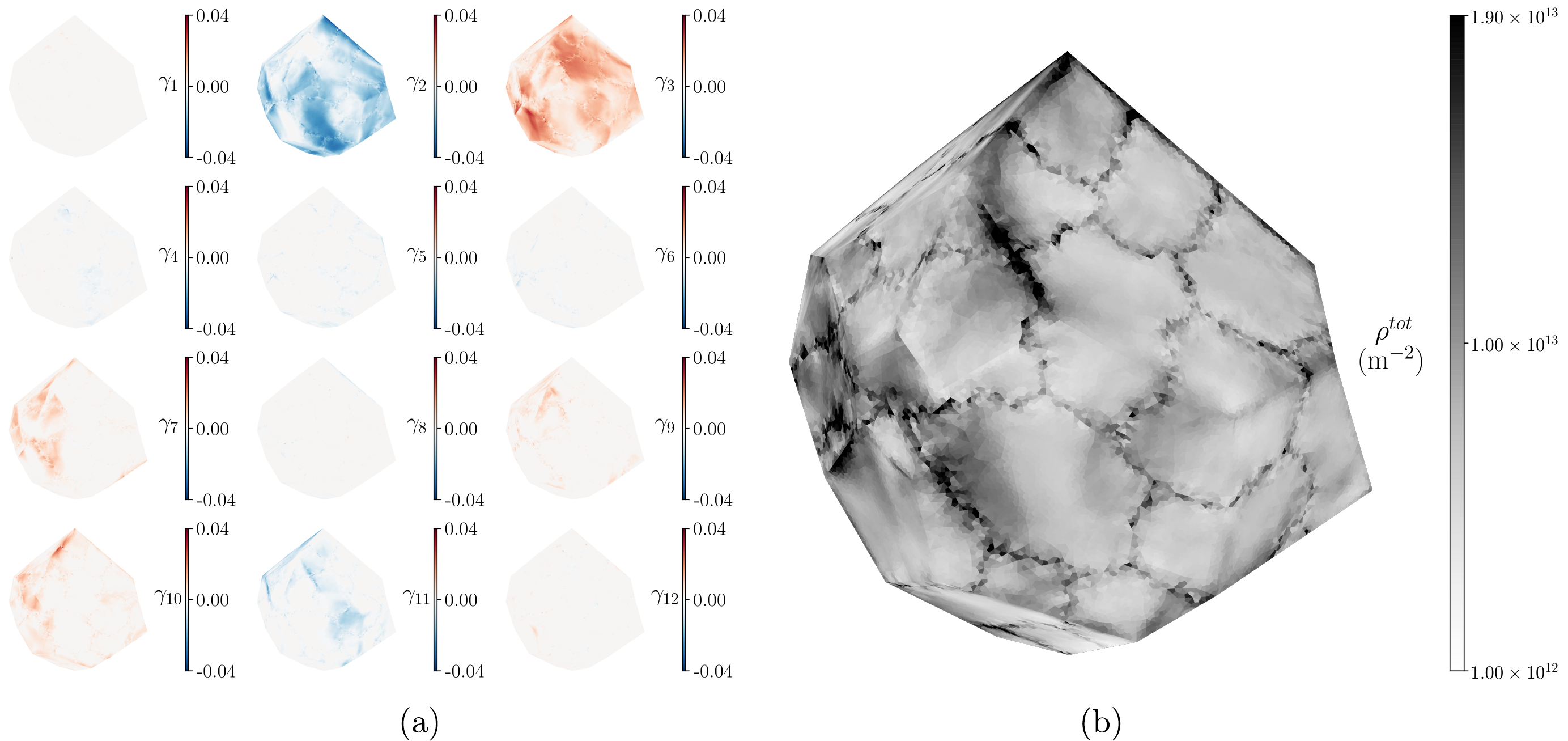}
	\caption{Compression at 1$\%$ macroscopic strain of the 3D polycrystal. (a) Plastic slip fields on the 12 FCC slip systems in the finely meshed grain located at the center of the domain. (b) Total dislocation density field zoomed into the finely meshed grain located at the center of the domain.}
	\label{fig:polycrystal_3D_slips_rho}
\end{figure}

To the author's knowledge, this is the first time that deformation patterning of this kind has been observed in 2D and 3D polycrystal simulations. The results demonstrate that patterning arises when three conditions are met: (i) the latent hardening is strong, thus penalizing the simultaneous activation of multiple slip systems, (ii) the rate-dependence is low, and (iii) the mesh size is sufficiently small to resolve the deformation microstructure. While the first condition is frequently satisfied in the literature (see the relative strength of collinear interactions obtained from Discrete Dislocation Dynamics (DDD) by Madec \textit{et al.}~\cite{madec2003role}), the second and third conditions are rarely met due to their computational cost. A partial solution to this issue is the use of highly efficient spectral solvers~\citep{moulinec1994fast, marano2019intragranular, gelebart2020modified}, which can reduce computation time by one to two orders of magnitude compared to standard finite element solvers~\citep{eisenlohr2013spectral, el2020full}. Although this falls outside the scope of the present study, it will be explored in future work. Additionally, algorithms designed to efficiently integrate nearly rate-independent viscoplastic flow rules, such as those proposed in~\citep{wulfinghoff2013equivalent,rys2024spontaneous}, will also be considered in future studies.

\section{Conclusion}
\label{sec:conclusion}

This study explores the formation of deformation patterns in FCC single crystals using a finite strain crystal viscoplasticity model incorporating strong latent hardening. The model describes the evolution of dislocation densities across 12 slip systems, with latent hardening introduced through a hardening matrix that accounts for the strength of interactions between different slip systems. To identify the minimal conditions necessary for pattern formation, a simplified hardening matrix is considered, where the off-diagonal latent hardening terms are set to 1.6 times the diagonal self-hardening terms. The rate-sensitivity parameters are adjusted to ensure an almost rate-independent response. The model is implemented in the finite element software \texttt{Z-set}~\citep{zset}, and simulations are performed in both 2D and 3D for single and polycrystals.

For single crystals under plane strain compression, deformation patterns emerge early in the deformation process, forming a second-order laminate microstructure composed of single-slip deformation bands. The size of these bands is influenced by both the mesh size and the overall domain size. Pole figures reveal that lattice rotations cause the crystal to fragment into sub-grains, with dislocation walls forming the boundaries between these regions. These walls act as low-angle sub-grain boundaries, marking the segregation of deformation. The morphology of the observed patterns is strongly dependent on the choice of hardening parameters and the initial crystal orientation. This suggests that experimental TEM observations of dislocation structures could provide valuable insight into the relative strength of dislocation interactions, complementing values obtained from discrete dislocation dynamics simulations.

\textcolor{black}{
The patterning phenomena observed in this work are induced by strong latent hardening. Following Franciosi~\cite{franciosi1985concepts} and Feaugas~\cite{feaugas1999origin}, except in the very small energy domain, the latent hardening rate decreases when the stacking fault energy increases. Therefore, it can be expected that FCC materials with low stacking fault energy are more prone to form such patterns. Feaugas~\cite{feaugas1999origin} shows for instance the evolution of dislocation structures in the low stacking fault energy 316L stainless steel ($\sim$30 mJ/m$^2$) as a function of plastic strain. As discussed by the author, low stacking fault energy enhances the extension of the stage I (single-slip dominant). Our results are in agreement with this observation, since most parts of the simulated single crystal domains undergo single slip, while mutli-slip regions are limited to interfaces between patches.
}

In 3D single-crystal wires oriented along the $[100]$ direction and subjected to tension and torsion, the symmetry of the crystal leads to the simultaneous activation of multiple slip systems. However, strong latent hardening forces the deformation to localize into single-slip regions separated by dislocation walls. The characteristic thickness of these dislocation cells and the sub-grain size are primarily determined by the mesh size, highlighting the need for a physically motivated length scale in the model. To address this, a strain gradient plasticity extension is introduced, incorporating a material length scale $\ell$ that regularizes the characteristic size of the deformation microstructure. The resulting microstructure size ($\phi/L$) follows a power-law relationship with an exponent of 0.32 with respect to the material length scale $\ell/L$, demonstrating independence from the numerical discretization.

The formation of deformation patterns is also observed in polycrystals under multiaxial loading conditions. Simulations of 2D and 3D polycrystals with randomly oriented grains show that, despite the arbitrary grain orientations relative to the loading direction, interactions between neighboring grains induce multiaxial stress states that activate multiple slip systems. Strong latent hardening discourages the simultaneous activity of multiple slip systems within the same region, leading to the emergence of deformation patterns. In 2D polycrystals, a finely meshed central grain exhibits a second-order laminate microstructure, whereas coarser meshes fail to resolve it. In 3D polycrystals, a cellular dislocation microstructure appears in a finely meshed grain at the center of the domain, with low-angle sub-grain boundaries indicating grain fragmentation.

These findings demonstrate that latent hardening plays a fundamental role in driving deformation patterning in both single and polycrystals. The link between the types of dislocation structures and the relative strength of the coefficients of the interaction matrix will be explored in future research. Additionally, phase-field models of recrystallization will be employed to investigate the nucleation of new grains from these sub-grain boundaries and their subsequent evolution under continued deformation or heat treatment.

\section{Acknowledgements}
\label{sec:acknowledgements}

The author sincerely thanks Christophe Bovet for his invaluable support in providing and assisting with the setup of the parallel solver for the 3D simulations. Gratitude is also extended to Aubin Geoffre, Lionel Gélébart, Aldo Marano and Samuel Forest for their insightful discussions and constructive feedback, which greatly contributed to this work. Special appreciation is given to the MIT Libraries for facilitating access to the PhD thesis of S. Saimoto.

\appendix

\section{Dislocation density tensor}
\label{sec:nye_tensor}

\textcolor{black}{
The components of Nye's dislocation density tensor, defined as $\pmb{\alpha} = \mathrm{curl}(\pmb{F}^{e^{-1}}) = \frac{1}{J} \mathrm{Curl}(\pmb{F}^p) \cdot \pmb{F}^T$, for the single crystal under plane strain compression along $[00\bar{1}]$ in the $(110)$ plane at 1\% macroscopic strain are shown in Figure~\ref{fig:dislocation_density_tensor_010_redo}. These fields correspond to the deformation band patterns displayed in Figures~\ref{fig:plane_strain_compression_010_redo},~\ref{fig:deformation_gradient_010_redo} and to the pole figures shown in Figure~\ref{fig:pole_figures}. The dominant component is $\alpha_{22}/b$, which corresponds to screw dislocations with the Burgers vector and line vector parallel to the $x_2 = [00\bar{1}]$ direction. Its magnitude reaches peak values above $3\times 10^{13}$ \si{\meter}$^{-2}$, three times the peak value of statistically stored dislocation density shown in Figure~\ref{subfig:ssd_density_010_redo}. Secondary components are $\alpha_{11}/b$, $\alpha_{33}/b$, $\alpha_{23}/b$ and $\alpha_{32}/b$. The two diagonal components correspond to screw dislocations with the Burgers vector and line vector parallel to the $x_1 = [1\bar{1}0]$ or $x_3 = [110]$ directions, respectively. The two off-diagonal components correspond to edge dislocations with the Burgers vector and line vector, perpendicular to each other and parallel to the $x_2 = [00\bar{1}]$ or $x_3 = [110]$ directions. The latter characterize tilt sub-grain boundaries.
}
\begin{figure}
	\centering
	\subfloat{\includegraphics[width=0.32\textwidth]{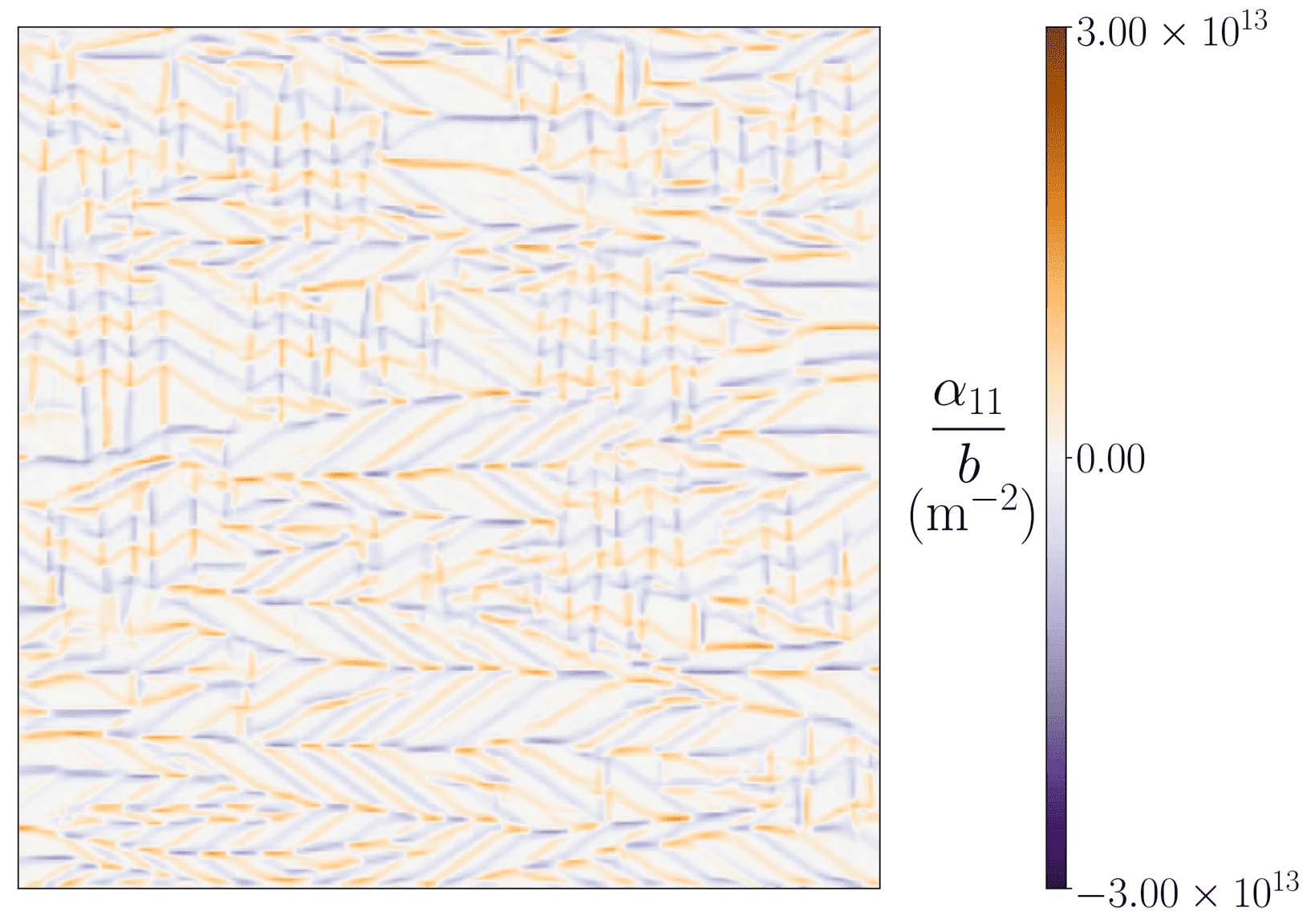}}
	\subfloat{\includegraphics[width=0.32\textwidth]{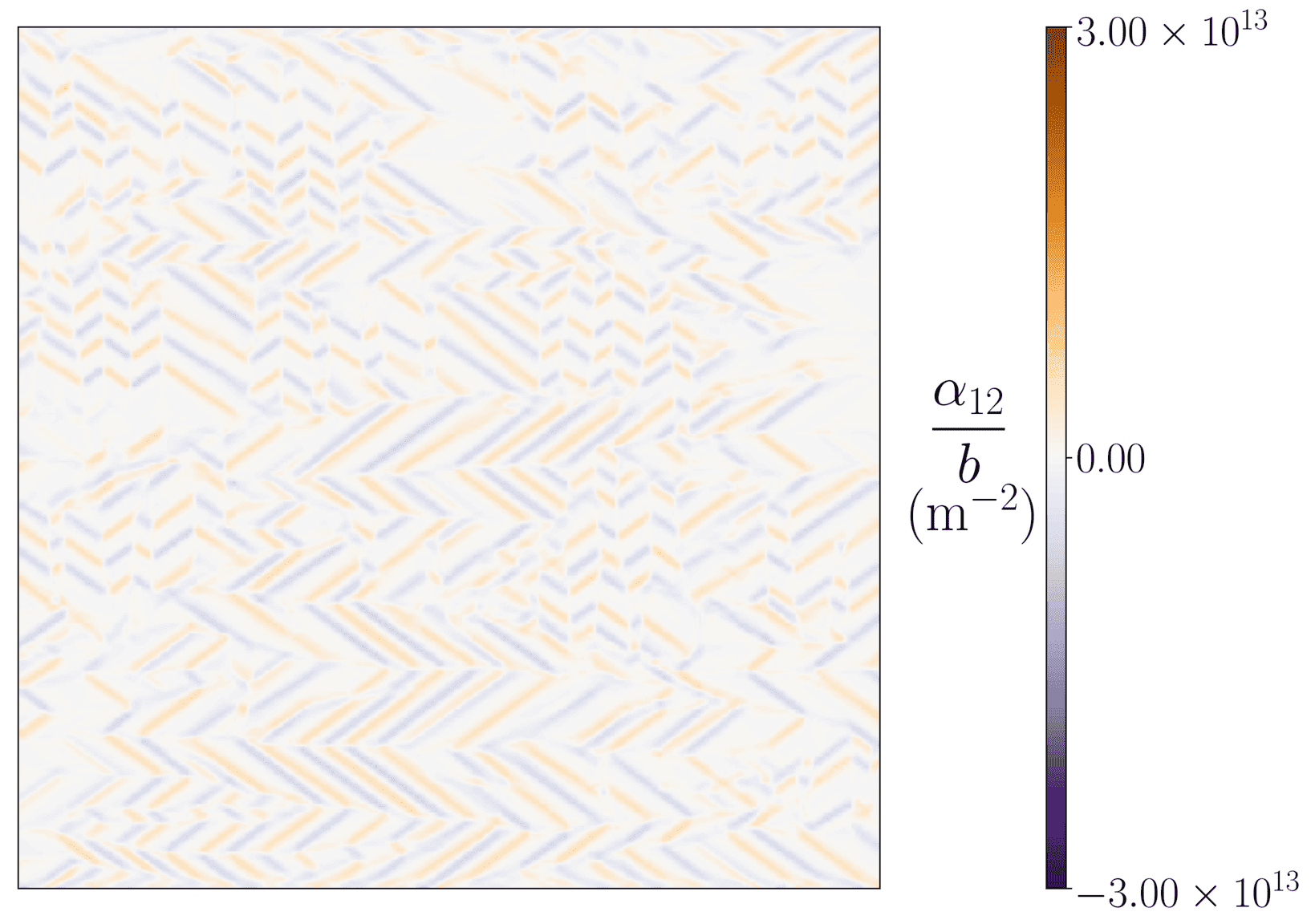}}
	\subfloat{\includegraphics[width=0.32\textwidth]{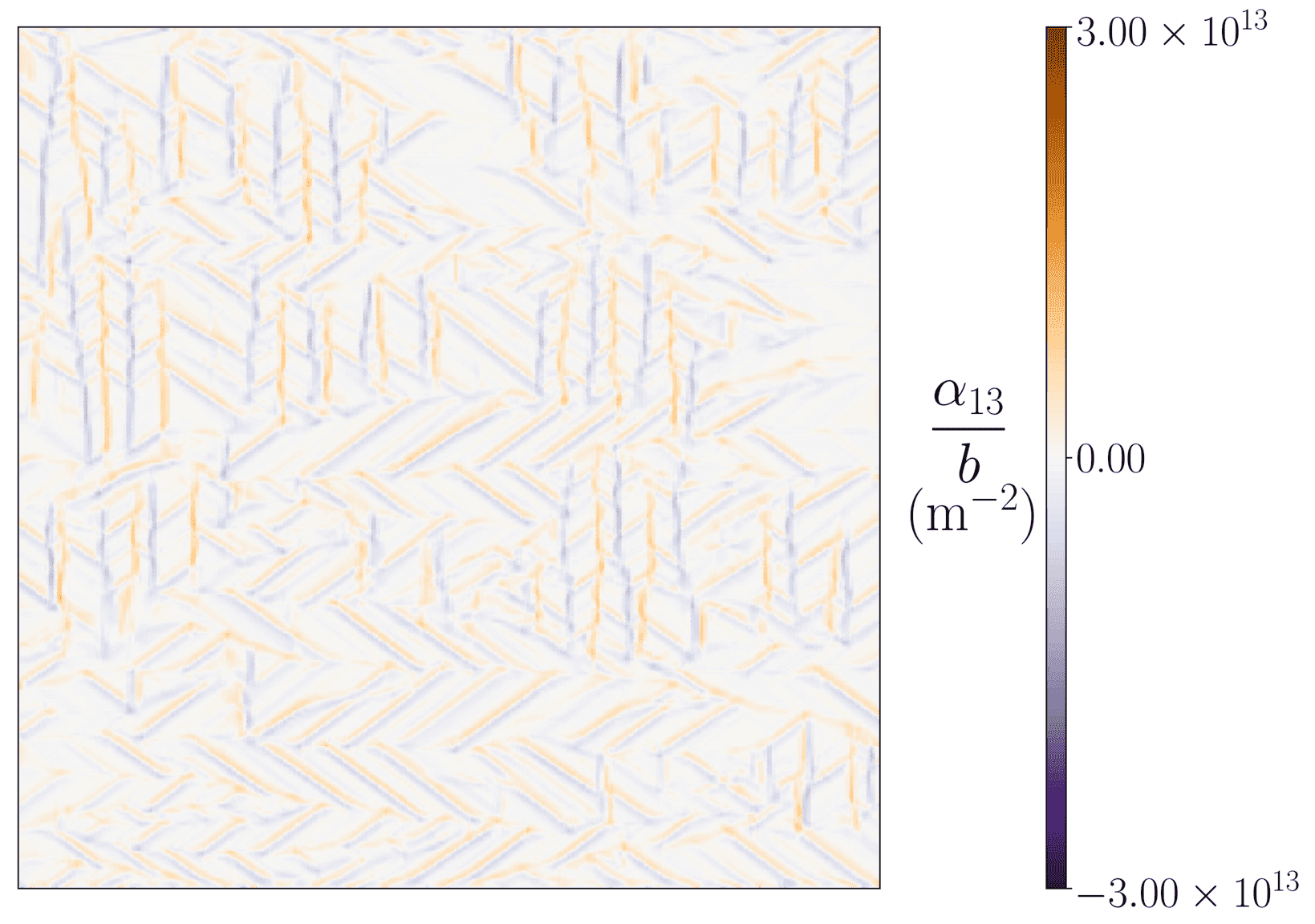}}\\
	\subfloat{\includegraphics[width=0.32\textwidth]{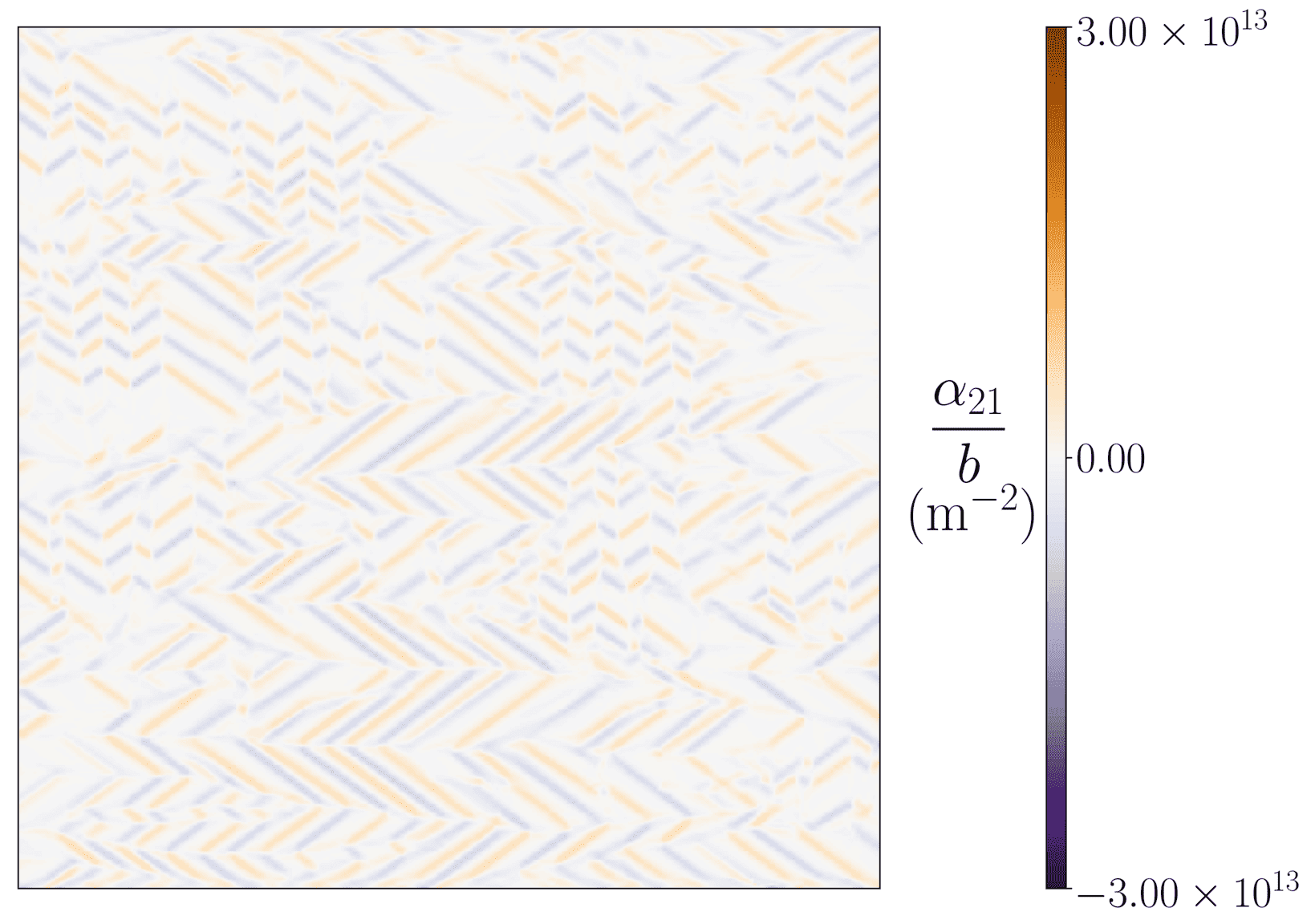}}
	\subfloat{\includegraphics[width=0.32\textwidth]{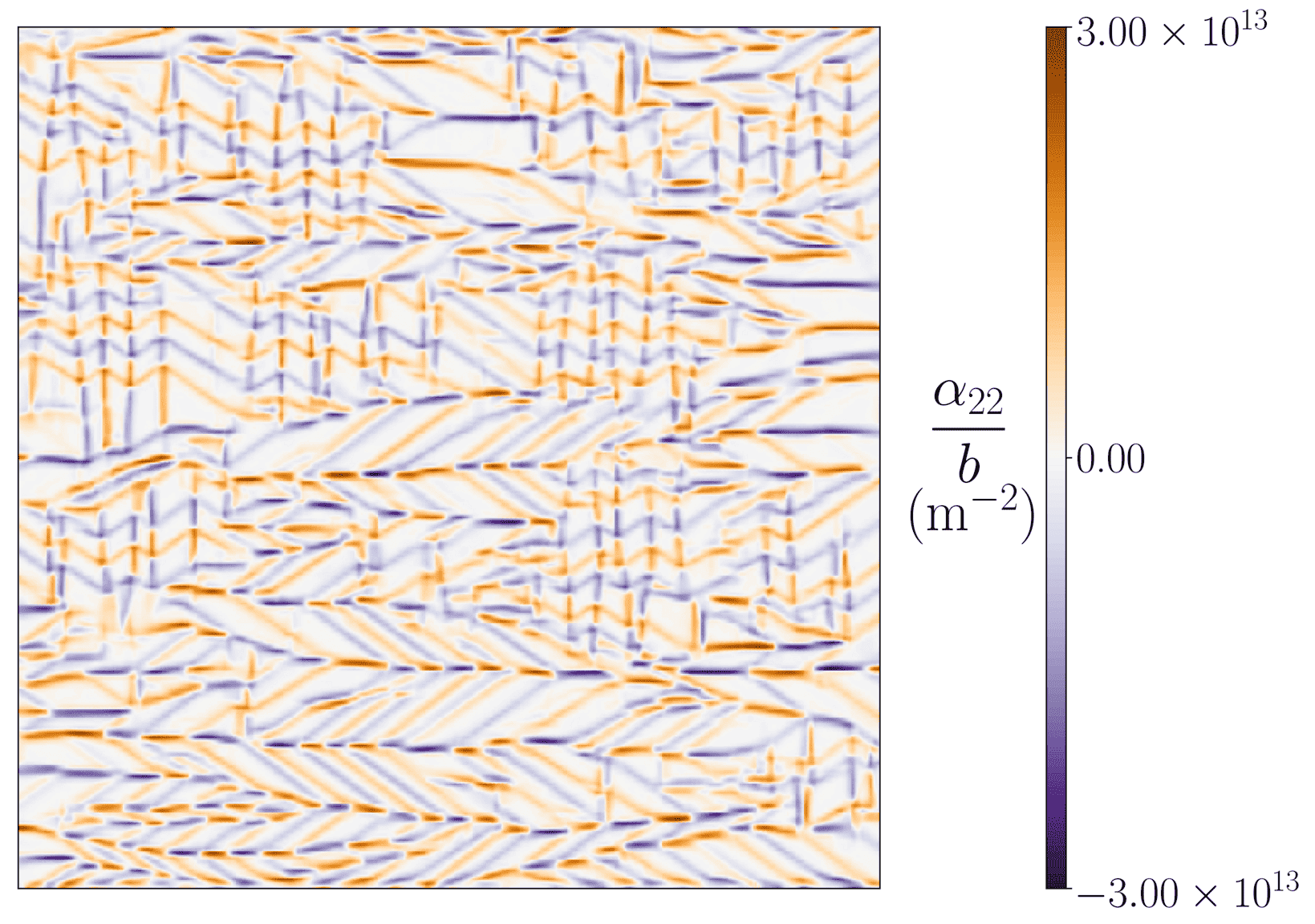}}
	\subfloat{\includegraphics[width=0.32\textwidth]{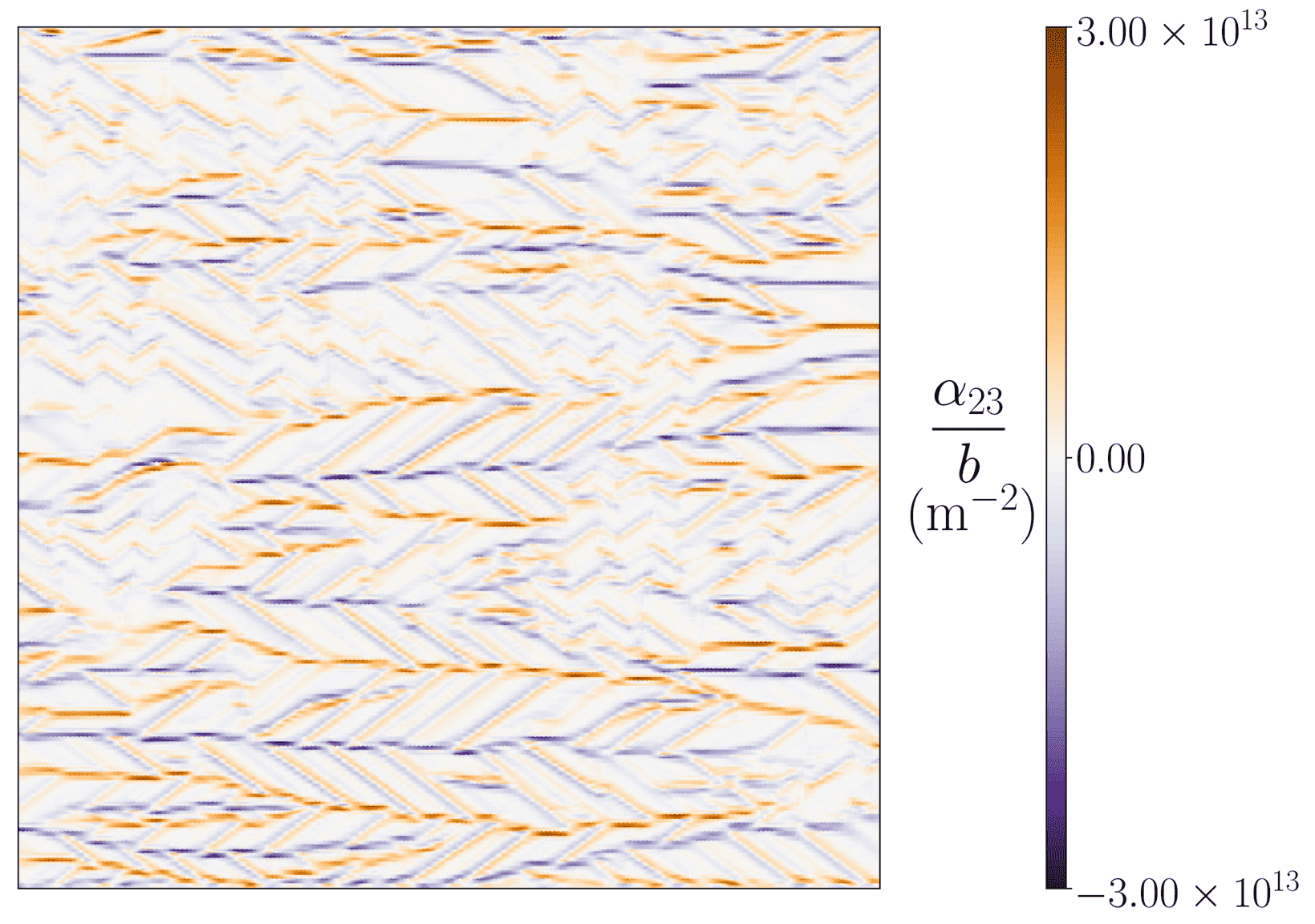}}\\
	\subfloat{\includegraphics[width=0.32\textwidth]{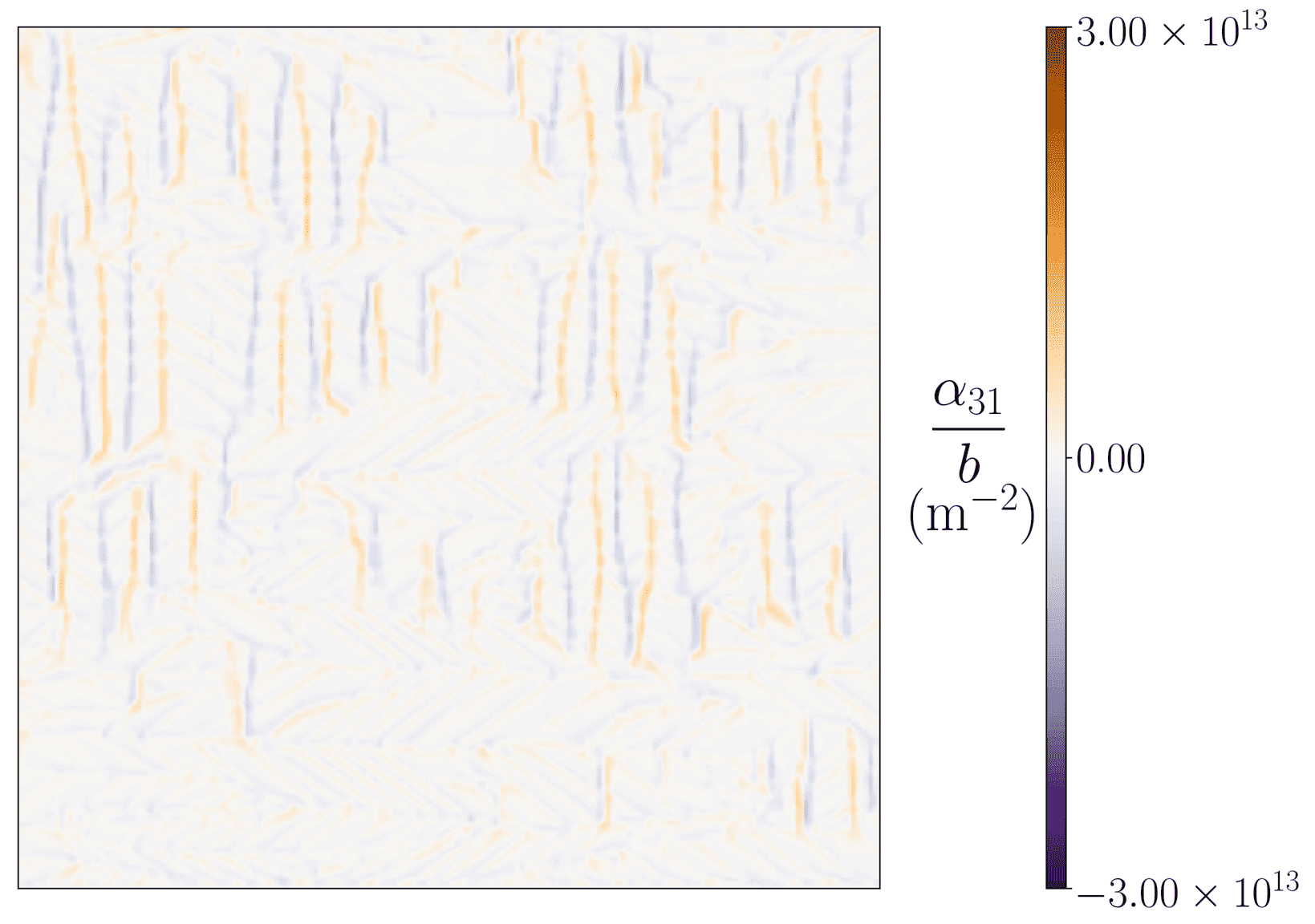}}
	\subfloat{\includegraphics[width=0.32\textwidth]{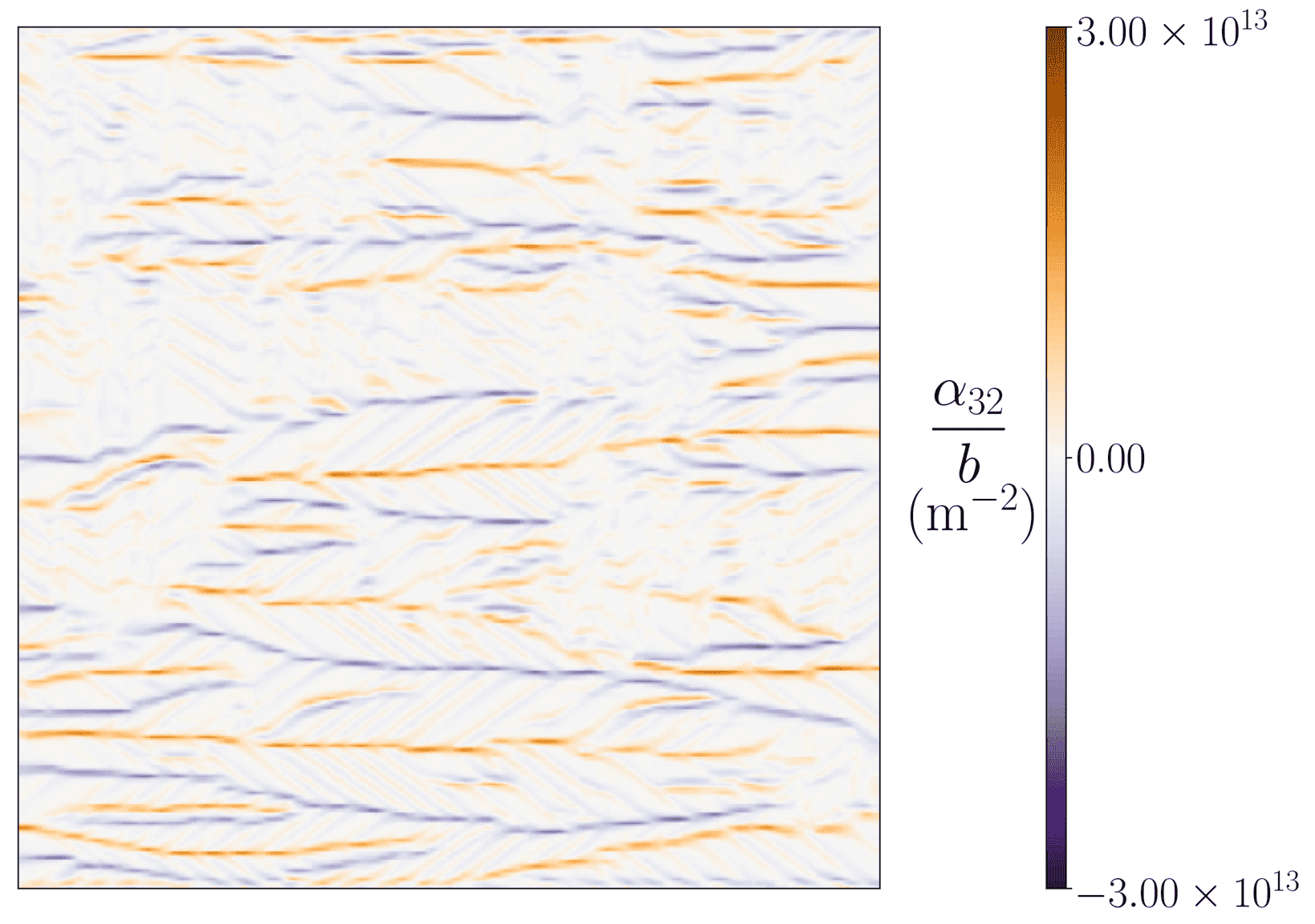}}
	\subfloat{\includegraphics[width=0.32\textwidth]{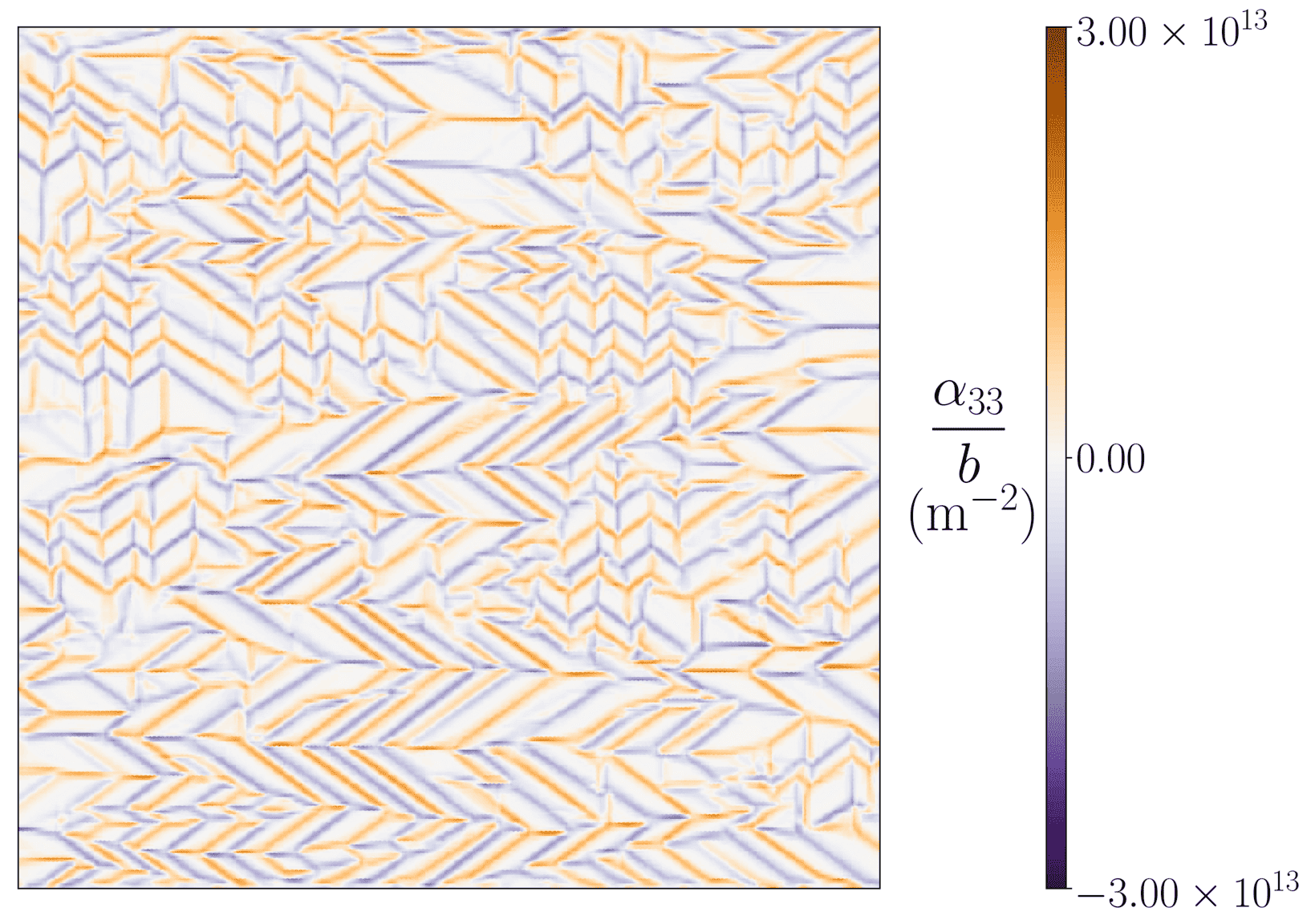}}
	\caption{Components of the dislocation density tensor $\pmb{\alpha}$ at 1\% macroscopic strain under plane strain compression along $[00\bar{1}]$ in the $(110)$ plane.}
	\label{fig:dislocation_density_tensor_010_redo}
\end{figure}

\textcolor{black}{
The total density of statistically stored dislocations (SSD) $\rho^{tot}$ and the norm of the dislocation density tensor $||\pmb{\alpha}||/b$, a measure of geometrically necessary dislocations (GND), are compared in Figure~\ref{fig:dislocation_densities_010_redo}. It appears clearly that both densities are highest in the cell walls. There, the GND density peak values reaches up to 4 times the SSD density peak values. However, it is also interesting to notice that, where the SSD density is lowest (thin band-like light grey regions in Figure~\ref{subfig:ssd_density_010_redo}) the GND density displays significant values, albeit lower than within dislocation cell walls. The GND density is lowest where the SSD density reaches intermediate values (medium grey regions in Figure~\ref{subfig:ssd_density_010_redo}).
}
\begin{figure}
	\centering
	\subfloat{
		\includegraphics[width=0.48\textwidth]{FiguresReduced_HardeningMatrix_Rys_Redo_step_rho_010.png}
		\label{subfig:ssd_density_010_redo}
	}
	\hfill
	\subfloat{
		\includegraphics[width=0.48\textwidth]{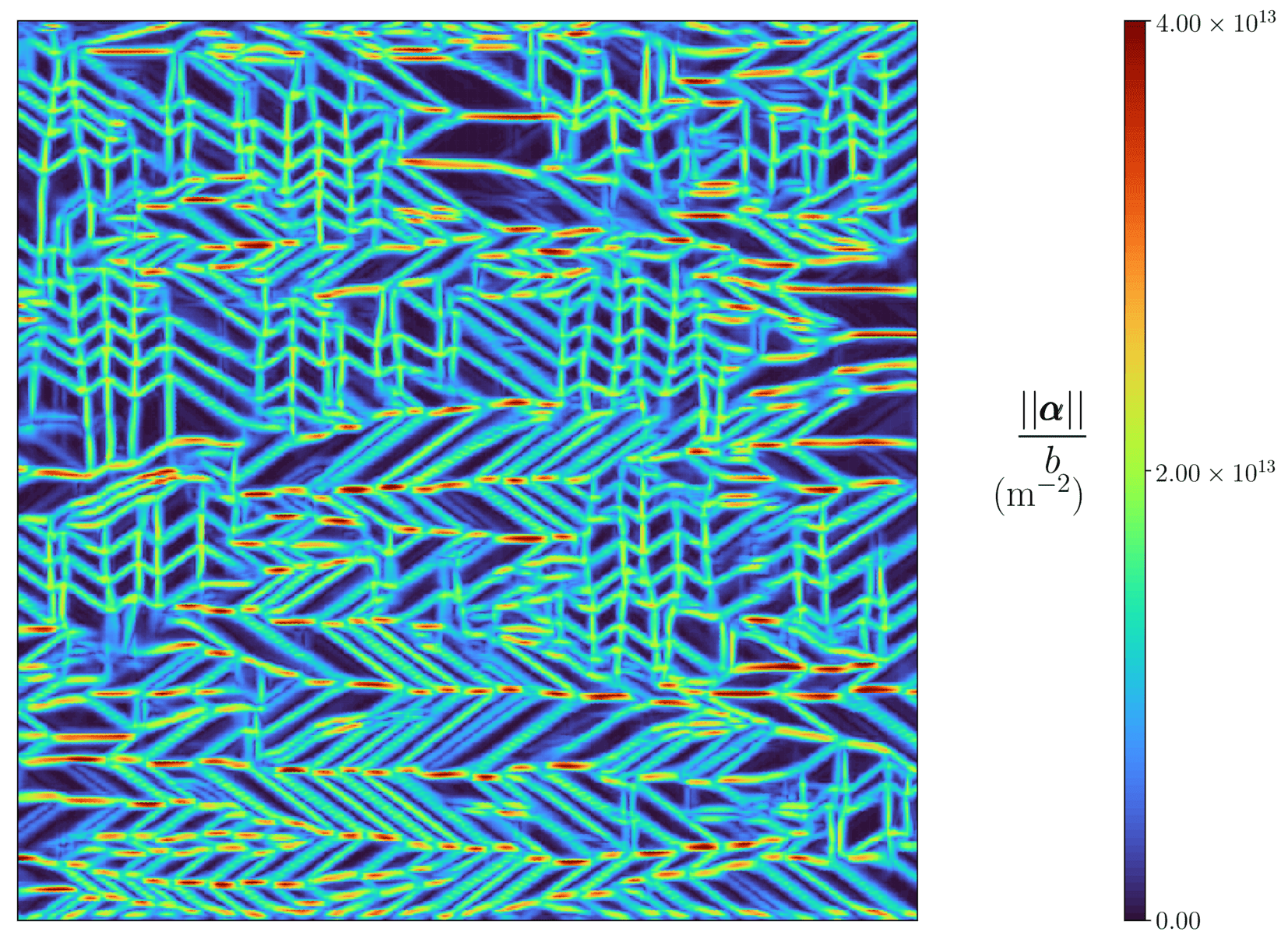}
		\label{subfig:gnd_density_010_redo}
	}
	\caption{(a) Total density of statistically stored dislocations $\rho^{tot}$ and (b) norm of the dislocation density tensor $||\pmb{\alpha}||/b$ at 1\% macroscopic strain.}
	\label{fig:dislocation_densities_010_redo}
\end{figure}

\textcolor{black}{
The regions of high GND and SSD densities are preferential sites for new grain boundary nucleation and growth in recrystallization processes. As discussed by~\cite{ghiglione2024cosserat}, the presence of a strong orientation gradient, thus a high density of GND, is necessary to trigger grain nucleation, while the stored energy, which is correlated to the SSD density, contributes to their mobility. The recrystallization process of the deformation microstructures obtained in this work will be studied in future work.
}




\fi

\bibliographystyle{elsarticle-num.bst}
\bibliography{bibliography.bib}

\end{document}